\documentclass[acmsur]{acmsmall} 

\usepackage{booktabs}
\usepackage{caption}
\usepackage{textcomp}
\usepackage{framed}
\usepackage{slashbox}
\usepackage{makecell}
\usepackage{times}
\usepackage{mathptm}
\usepackage{graphicx}
\usepackage{array}
\usepackage{url}
\usepackage{color}
\usepackage{amsmath}

\begin{document}


\title{RAID Organizations for Improved Reliability and Performance: \\
A Not Entirely Unbiased Tutorial (1st revision}
\author{Alexander Thomasian
\footnote{Thomasian \& Associates, Pleasantville, NY, \url{alexthomasian@gmail.com}}
}
\date{}
\maketitle

\subsection*{Abstract}
RAID proposal advocated replacing large disks with arrays of PC disks,
but as the capacity of small disks increased 100-fold in 1990s 
the production of large disks was discontinued.
Storage dependability is increased via replication or erasure coding.
Cloud storage providers store multiple copies of data obviating for need for further redundancy. 
Varitaions of RAID based on local recovery codes, partial MDS reduce recovery cost. 
NAND flash Solid State Disks - SSDs have low latency and high
bandwidth, are more reliable, consume less power and have a lower TCO than Hard Disk Drives, 
which are more viable for hyperscalers. 

Categories and Subject Descriptors:
B.4.2 [Input/Output and Data Communications]: Input/Output Devices;
B.4.5 [Input/Output and Data Communications]: Reliability, Testing, and Fault-Tolerance;
D.4.2 [Operating Systems]: Storage Management;
E.4 [Data]: Coding and Information Theory.

{\bf TABLE OF CONTENTS}

\begin{small} 
\ref{sec:intro} {\bf Introduction to Storage Systems}.                                                
\ref{sec:HDD} Review of hard disk drives - HDDs.                                   
\ref{sec:companies} Storage companies.                                             
\ref{sec:hyper} Hyperscalers and cloud storage.                                    
\ref{sec:queue} Simple disk performance queueing models.                           
\ref{sec:tail} Dealing with high tail latency.                                     
\ref{sec:compress} Data compression, compaction and deduplication.                 
\ref{sec:UDE} Undetected Disk Errors - UDE and Silent Data Corruption - SDC.       

\ref{sec:RAIDclassify} {\bf RAID classification and extensions}.                   
\ref{sec:Berkeley} Berkeley proposal to replace IBM DASD with inexpensive disks.   
\ref{sec:MDS} {\bf Maximum Distance Separable - MDS} minimum redundancy arrays.    
\ref{sec:RDP} Rotated Diagonal Parity - RDP coded arrays.                          
\ref{sec:PMDS} Partial MDS code.                                                   
\ref{sec:tiger} Disk adaptive redundancy scheme.                                   

\ref{sec:multiD} {\bf Multidimensional coding for higher reliability}.             
\ref{sec:gridfiles} Grid files.                                                    
\ref{sec:HRAID} Hierarchical RAID - HRAID.                                         
\ref{sec:RESAR} Reliable Storage at Exabyte Scale - RESAR.                         
\ref{sec:cubeRAID} Reliability of 2D Square- and 3D Cube-RAID.                      
\ref{sec:2d} 2-d and 3-dimensional parity coding.                                  

\ref{sec:LRC} {\bf Local Recoverable Code - LRC}.                                  
\ref{sec:LRCcomparison} A systematic comparison of existing LRC schemes.           
\ref{sec:widestripe} Wide stripe erasure codes.                                    
\ref{sec:practical} Practical design considerations for wide stripe LRCs.          

\ref{sec:distr} {\bf Reducing rebuild traffic in distributed RAID}.                
\ref{sec:pyramid} Pyramid codes.                                                   
\ref{sec:Xorbas} Hadoop Distributed File System - HDFS-Xorbas.                     
\ref{sec:cidon} Copyset Replication for Reduced Data Loss Frequency.               
\ref{sec:DiskReduce} More efficient data storage: Replication to Erasure Coding.   

\ref{sec:CRAID} {\bf Clustered RAID5 - CRAID}.                                   
\ref{sec:BIBD} Balanced Incomplete Block Design - BIBD.                          
\ref{sec:Thorp} Thorp shuffle implementation of CRAID.                           
\ref{sec:NRP} Nearly Random Permutation - NRP.                                   
\ref{sec:shifted} Shifted parity group placement.                                


\ref{sec:flash} {\bf Flash Solid-State Drives - SSDs}.                            
\ref{sec:HACDFS} Hadoop Adaptively-Coded Distributed File System - HACDFS.        
\ref{sec:FAWN} Fast Array of Wimpy Nodes - FAWN.                                  
\ref{sec:diffRAID} Differential RAID for SSDs.                                    
\ref{sec:RAMCloud} Distributed DRAM-based storage - RAMCloud.                     
\ref{sec:WOM} Write-Once Memory - WOM codes to enhance SSD lifetimes.             
\ref{sec:RAIZN} Redundant Array of Independent Zones - RAIZN.                     
\ref{sec:predictable} NVMe-SSDs and Predictable Microsecond Level Support for Flash. 
\ref{sec:CSD} Computational Storage Drives - CSD.

\ref{sec:IN} {\bf Interconnection Networks}. 

\ref{sec:conclusions} {\bf Conclusions: Cloud Storage}.                                          
 
{\bf Bibliography}  (including references in the Appendices).

Five Appendices provide complementary material in Chapters 10-14.

\ref{sec:perfeval} {\bf Appendix I: RAID performance evaluation}.                   
\ref{sec:reliability} {\bf Appendix II: RAID reliability analysis}.                
%
\ref{sec:simul} {\bf Appendix III: RAID reliability simulation}.                  
%
\ref{sec:tools} {\bf Appendix IV: RAID reliability modeling tools}.               
%
\ref{sec:mirhyb} {\bf Appendix V: Mirrored and hybrid disk arrays}.               

{\bf Abbreviations}  

\end{small}

\section{Introduction to Storage Systems}\label{sec:intro}
\vspace{3mm}


The tutorial is a follow-up to Chen et al. 1994 \cite{Che+94}, 
Thomasian and Blaum 2009 \cite{ThBl09}, and partially updates Thomasian 2021 \cite{Thom21}.
The Berkeley {\it Redundant Array of Inexpensive Disks - RAID} proposal 
foretold a major paradigm shift in magnetic disk technology from large to small 
form factor {\it Hard Disk Drives - HDDs} Patterson, Gibson, and Katz 1988 \cite{PaGK88}.
The term Inexpensive was switched to Independent, 
because the capacity of small form-factor disks grew 100-fold in 1990s Gray and Shenoy 2000 \cite{GrSh00},
replacing large form-factor disks so that the production of IBM 3390s ceased (1989-1993).
\footnote{\url{https://www.storagenewsletter.com/2019/10/22/history-1989-ibm-3390-hdds-finally-arrived/}}
HDDs have dominated the storage market. A list of IBM disk drives 
\footnote{\url{https://en.wikipedia.org/wiki/History_of_IBM_magnetic_disk_drives}}
and disk drives in general.
\footnote{\url{https://en.wikipedia.org/wiki/History_of_hard_disk_drives}}

There is a shift from HDDs to Flash {\it Solid State Drive - SSD} drives in consumer products such as laptops,
while HDDs are mainly used in hyperscalers,
hence a coexistence of these two technologies is expected for the foreseable future.
Emerging {\it Storage Class Memories - SCM} Freitas and Wilcke 2008 \cite{FrWi08}
serve specific roles, but some like {\it Micro-ElectroMechanical - MEMS}
\footnote{\url{https://en.wikipedia.org/wiki/MEMS}}
are fragile due to tiny mechanical parts.
\footnote{\url{https://www.symmetryelectronics.com/blog/is-mems-technology-fragile-a-guide-to-caring-for-your-mems-products-symmetry-blog/}}

The prevalence of large disk arrays makes frequent disk failures inevitable.
Data loss is preventable by tape or disk backup. 
\footnote{\url{https://www.nakivo.com/blog/how-tape-backup-software-works-short-overview/}}
While this is a timeconsuming task for a large number of disks
it is necessary to deal with catastrophic failures (fire, flood, earthquake, large-scale power failure, 
and ransomware attacks which holds data hostage by encrypting it 
\footnote{\url{https://en.wikipedia.org/wiki/Ransomware}}). 
While automated tape library recovery does not require human intervention,
\footnote{\url{https://en.wikipedia.org/wiki/Tape_library}}
even short periods of unavailability incur a high cost.
\footnote{\url{https://trilio.io/resources/cost-of-downtime/}}
Companies providing disk backup also provide data deduplication (see Section \ref{sec:compress}).
\footnote{\url{https://www.trustradius.com/disk-based-backup}}
Veritas Back Exec is an example.
\footnote{\url{https://en.wikipedia.org/wiki/Backup_Exec}}

Replication or coding both provide high data resiliency 
by allowing on demand recovery and systematic reconstruction of 
the contents of a failed HDD while processing external disk requests. 

While flash SSDs are more expensive per GB than HDDs 
they provide lower latency, higher throughout, lower {\it Annual Failure Rate - AFR}
and lower {\it Total Cost of Ownership - TCO} due to reduced power consumption. 
\footnote{\url{https://www.snia.org/sites/default/files/SNIA_TCOCALC_Workpaper_Final_0.pdf}}

We discuss analytical performance modeling using queueing theory in Appendix I and
simulation in Appendix III to evaluate the performance of HDDs
with delays in milliseconds not microseconds in SSDs. 
References to SSD performance evaluation are provided.

Reliability analysis of RAID arrays with HDD failures 
is first undertaken {\it Continuous Time Markov Chains - CTMCs},
which are then extended with probabilistic analysis to take into account 
the effect of disk scrubbing and {\it IntraDisk Redundancy - IDR} 
in dealing with {\it Latent Sector Errors - LSEs} in Appendix II.

Simulation for reliability evaluation of disk arrays is discussed in Appendix III,
while tools for this purposes are discussed in Appendix IV.

Mirrored and hybrid disk arrays which instead of replicas store 
{\it eXclusive-OR - XOR}ed data  are discussed in Appendix V. 
A shortcut method for reliability comparison is presented. 




Sections 1:9 are the main body of the paper.
Sections 10-14 are Appendices I:V on journal's web site.
Sections indicated by asterisks in the Appendices require background.
in queueing theory and reliability, which can be studied together
by expanding computer science and engineering curricula 
Hardin 2102 \cite{Hard12} and SEFI 2013 \cite{SEFI13}.

\subsection{Review of Hard Disk Drives}\label{sec:HDD}

The volume of data growing exponentially is to reach 175 ZettaBytes - ZB=$10^{21}$B by 2025.
\footnote{\url{https://www.networkworld.com/article/3325397/}}
Rate of data collected by organizations tends to exceed the rate at which it can be analyzed,
but it is projected that nearly 30\% of the data generated will be consumed in real-time by 2025.
IBM estimate that roughly 90\%  data generated by sensors never gets used. 
Data not used to derive insights or for decision making is referred to as dark data.
\footnote{\url{https://en.wikipedia.org/wiki/Dark_data}}
Ephemeral or transitory data exists briefly 
as temporary, transient files held in containers and is deleted periodically.
\footnote{\url{https://www.mongodb.com/databases/ephemeral-storage}}

HDDs are the prevalent data storage medium for holding high data volumes,
because of low cost per GigaByte - GB and access times in a few milliseconds (ms).
SSDs are more expensive per GB than HDDs, e.g., \$0.20 versus \$0.03.
\footnote{\url{https://nexstor.com/hdd-vs-flash-storage/}} 
Data access from robotic tape libraries (2 cents per GB) takes several seconds
\footnote{\url{https://en.wikipedia.org/wiki/Tape_library}} 
and from flash SSDs microseconds ($\mu$s).
Flash memory densities and capacities are increasing over times as price per GB is dropping.
\footnote{\url{https://jcmit.net/flashprice.htm}}

HDDs have grown 100,000x in capacity, 100x in bandwidth, 
but only 10x in {\it I/Os Per Second - IOPS}, due to their mechanical nature.
HDDs can be classified as {\it Personal Storage - PS} versus {\it Enterprise Storage - ES}
or {\it Advanced Technology Attachment - ATA} versus {\it Small Computer System Interface - SCSI} 
Anderson et al. 2003 \cite{AnDR03}.
\footnote{\url{https://en.wikipedia.org/wiki/SCSI}}
{\it Serial ATA - SATA} replacing parallel ATA is less expensive and better suited for desktop storage.
\footnote{\url{https://en.wikipedia.org/wiki/SATA}}
More expensive SCSI drives tend to provide higher performance via faster seeking 
and 10K and 15K {\it Rotations Per Minute - RPM} versus 7200 RPM SATA drives.
IBM's {\it Serial Storage Architecture -- SSA} in 1990
\footnote{\url{https://en.wikipedia.org/wiki/Serial_Storage_Architecture}}
but was taken over by the {\it Fibre Channel - FC} protocol for SANs on optical fibers allowing switching. 
\footnote{\url{https://en.wikipedia.org/wiki/Fibre_Channel}}
{\it Serial Attached SCSI - SAS} is more expensive than SATA and better suited for use in servers.
Storage types are: object, file, block. 
\footnote{\url{https://www.backblaze.com/blog/object-file-block-storage-guide/}}
{\it Amazon Web Service - AWS} provides block storage.
\footnote{\url{https://aws.amazon.com/what-is/block-storage/}}
HDDs are discussed in Part III in Jacob et al. \cite{JaNW08} and below.
\footnote{\url{https://en.wikipedia.org/wiki/Hard_disk_drive}}

{\it Online Transaction Processing - OLTP} workloads are known 
to generate accesses to small randomly placed disk blocks. 
This was ascertained by analyzing disk I/O traces Ramakrishnan et al. 1992 \cite{RaBK92}. 
Disk positioning time is the sum of seek time to move the {\it Read/Write - R/W} head 
attached to the disk arm to the target track and 
rotational latency for the block on the target track to reach the R/W head.
For small block accesses the mean delay is one half of disk rotation time,
e.g., $T_R/ 2= 4.167$ ms for 7200 RPM HDDs. 
The slowly improving mechanical seek time takes few ms.
The 4 KB or 8 KB block transfer time is negligible with respect to rotational latency,
since 8 or 16 512 B sectors constitute a small fraction of sectors per track.

{\it Zoned Bit Recording - ZBR} or zoning increases HDD capacity 
by maintaining about the same linear recording density across disk tracks.
\footnote{\url{https://en.wikipedia.org/wiki/Zone_bit_recording}}
Utilizing a fraction of outer tracks, 
which leads to lower seek times due to short stroking Jacob et al. 2008 \cite{JaNW08}.
Outermost tracks whose capacity exceeds innermost tracks possibly by a factor of $\approx$1.7 
have a higher data transfer rate by this factor.

Rotational latency is shortened via two disk arms, which are 180$^\circ$ apart, 
or just replicating records 180$^\circ$ apart on a track, 
but this adversely affects sequential accesses. 
Seagate's 18 TB disks have dual actuators achieving 554 MB/s. 
\footnote{\url{https://www.tomshardware.com/news/seagate-launches-2nd-gen-dual-actuator-hdds-18-tb}} 
based on {\it Heart Assisted Magnetic Recording - HAMR} technology, 
but even higher capacity HDDs have been introduced,
\footnote{\url{https://www.anandtech.com/show/21100/seagate-unveils-exos-x24-24tb-and-28-tb-setting-the-stage-for-hamr-hdds}}
Seagate announced a 30 TB {\it Heat Assisted Magnetic Recording - HAMR} HDDs in April 2023. 
\footnote{\url{https://blocksandfiles.com/2023/04/21/seagate-30tb-hamr/}}
A dated but influential presentation on areal density is as follows.
\footnote{\url{http://www.thic.org/pdf/Mar03/tarnotek.gtarnopolsky.030304.pdf}}
While Flash NAND SSDs tend to be smaller in capacity than HDDs Pure Storage proposed 300 TB flash drive due in 2026.
\footnote{\url{https://blocksandfiles.com/2023/03/01/300tb-flash-drives-coming-from-pure-storage/}}

Disk accesses contribute heavily to transaction response time 
{\it Transaction Processing Council - TPC}'s TPC C benchmark 
determines maximum transactions per minute - tpm at a certain threshold for response times.
\footnote{\url{https://www.tpc.org/tpcc/detail5.asp}}

A brief review of the components of a memory hierarchies is given in Table \ref{tab:hierarchy},
which is based on Patterson and Hennessey 2017 \cite{HePa17}. 
NAND Flash memory can be used as the secondary store or caches for HDDs Yu et al. 2012 \cite{Yu++12}.
Flash wear can be reduced by first caching into an SCM, 
e.g., {\it Magnetorestrictive RAM - MRAM},
where data can be updated several times before downloading to Flash SSDs. 
\footnote{\url{https://www.digikey.com/en/articles/how-to-use-mram-to-improve-edge-computing}}
Storage technologies are described in Section 2.3.14 in Thomasian 2021 \cite{Thom21}.

\begin{table}[t]
\centering
\begin{footnotesize}
\begin{tabular}{|c|c|c|c|}\hline
Hierarchy        &Memory technology       &Access time           & Cost \$ per GB   \\ \hline\hline
CPU cache        &Static RAM - SRAM       &0.5-2.5 ns            & 500-1000         \\ \hline
Main memory      &Dynamic RAM - DRAM      &50-70 ns              & 10-20            \\ \hline
Cache for disks  &Flash storage           &5-50 $\mu$s           &0.05-0.1          \\ \hline 
Secondary store  &Magnetic disk           &5-20  ms              &0.5-0.10          \\ \hline 
\end{tabular}
\end{footnotesize}
\caption{\label{tab:hierarchy}Technologies associated with memory hierarchy.} 
\end{table} 

\subsection{Storage Companies}\label{sec:companies}


{\it Western Digital - WD} producing HDDs  
\footnote{\url{https://en.wikipedia.org/wiki/Western_Digital}}
started producing SSDs after acquiring SanDisk Flash memory manufacturer in 2016.
\footnote{\url{https://en.wikipedia.org/wiki/SanDisk}}
WD is now splitting along product lines because of differences in HDD and SSD markets.
\footnote{\url{https://www.techtarget.com/searchstorage/news/366557538/Western-Digital-to-split-flash-HDD-business-units-in-two}}
SSDs are intended for client devices, such as laptops, while HDDs for hyperscalers.
\footnote{\url{https://www.techradar.com/news/ssd-vs-hdd-which-is-best-for-your-needs}}

Examples of HDDs and SSDs based on the following footnote are as follows:
\footnote{\url{https://www.enterprisestorageforum.com/hardware/ssd-vs-hdd-speed/}} 
\noindent
(1) {\bf Toshiba MG Series} with SATA interface, model MG06ACA800E, 
8 TB capacity: up to 241 MB/s, sustained transfer rate.   
(2) {\bf Seagate EXOS 2X18} with SATA or SAS interface, 18 TB capacity: up to 554 MB/s sustained data rate. 
(3) {\bf Micron 5200} ECO with SATA interface, 7.68 TB capacity: 540/520 MB/s sequential read/write.   
(4) {\bf Western Digital PC SN720} with {\it NonVolatile Memory Express - NVMe} interface,  
one TB capacity: 3,400/2,800 MB/s, sequential read/write.
\footnote{\url{https://www.theverge.com/2023/10/30/23938334/western-digital-separating-hard-drive-flash-business}}


IBM Corp. (1911-) in 1964 introduced the S/360 family of computers,
whose more powerful members of the family are known as mainframes.
\footnote{
\url{https://www.academia.edu/25337525/One_Architecture_Fits_All_IBM_Mainframe?email_work_card=abstract-read-more}}
mainframes are still important since 92/100 of top banks, 
23/25 top airlines, 
all top 10 insurance companies, 
and 71\% of Fortune 500 companies use them.
\footnote{\url{https://www.precisely.com/blog/mainframe/9-mainframe-statistics}}
IBM like many other early computer companies 
was a vertically integrated company producing most computer components,
\footnote{\url{https://en.wikipedia.org/wiki/Vertical_integration}}
but this paradigm has changed across the industry. 

M. Yanai was responsible for the EMC Symmetrix in 1992 
\footnote{\url{https://en.wikipedia.org/wiki/EMC_Symmetrix}} 
whose {\it Disk Array Controller - DAC} could emulate IBM 3390 DASD on fixed block disks.
\footnote{\url{https://siliconangle.com/2020/09/29/dell-emc-sees-30-years-of-symmetrix-storage-array-innovation-cubeconversations/}}
EMC was acquired by Dell in 2015.
Yanai's Nextra with self-healing and self-tuning and dynamic scaling capabilities 
was acquired by IBM and renamed xiv storage in 1998, 
\footnote{\url{https://en.wikipedia.org/wiki/IBM_XIV_Storage_System}}
but has been replaced by IBM FlashSystem.
\footnote{\url{https://en.wikipedia.org/wiki/IBM_FlashSystem}}
Yanai's Infinidat provides protection against cyberattacks using advanced machine-learning models 
that provide 99.5\% confidence in detecting cyber threats.
\footnote{
\url{https://www.cybersecuritydive.com/press-release/20230615-infinidat-enhances-cyber-storage-resilience-with-infinisafe-cyber-detection/}}

Cleversafe was renamed IBM Object Storage after being acquired by IBM in 2016 
\footnote{\url{https://en.wikipedia.org/wiki/IBM_Cloud_Object_Storage}}
It relies on M. Rabin's {\it Information Dispersal Algorithm - IDA}.
\footnote{\url{https://www.eecs.harvard.edu/~michaelm/TALKS/RabinIDA.pdf}}
Data is encrypted, and sliced with twelve slices dispersed to separate disks, 
storage nodes and/or geographic locations.
Multiple failure scenarios can be handled.
\footnote{\url{https://www.spectrumscale.org/wp-content/uploads/2016/12/IBM-Cleversafe-Object-Storage-Overview-022016.pdf}}


The IBM 2105 emulates a large number of control unit 
and 3390 disk drives on SCSI or SSA - Serial Storage Architecture drives.
\footnote{\url{https://www.ibm.com/docs/en/zos-basic-skills?topic=concepts-mainframe-hardware-disk-devices}}
It was replaced by IBM TotalStorage DS8000 series (2004-06) and IBM System Storage DS8000 series (2006-19)
\footnote{\url{https://en.wikipedia.org/wiki/IBM_DS8000_series}}
IBM DS8880 supports SSDs and HDDs with disk scrubbing.
\footnote{\url{https://www.redbooks.ibm.com/redbooks/pdfs/sg248323.pdf}}
The following configurations are supported for HDDs.
(5+P+Q+S),(6+P+Q),(6+P+S),(3+3,2S),(4+4)RAID10s. 
The 12/29/2021 expired RDP patent is used to implement RAID6 (Fig. 3.7 on  p. 93). 
\footnote{\url{https://patents.google.com/patent/US20060107135}}
RAID6 and RAID10 for Flash SSDs have no spares.

Tandem Computers (1974-1997) was an early multi-minicomputer providing high availability through duplexing.  
\footnote{\url{https://en.wikipedia.org/wiki/Tandem_Computers}}
Tandem was acquired by PC maker Compaq (1982-2002), 
which also acquired {\it Digital Equipment Corp. - DEC} (1957-1998),
known for its PDP computers, VAX cluster, and finally Alpha pocessor. 
\footnote{\url{https://en.wikipedia.org/wiki/Digital_Equipment_Corporation}}
before Compaq was acquired by HP (2013-).
\footnote{\url{https://en.wikipedia.org/wiki/Hewlett-Packard}}
Teradata shipped its {\it Data Base Computer DBC}/1012 to Wells-Fargo before 1984.
\footnote{\url{https://en.wikipedia.org/wiki/DBC_1012}}
was acquired by {\it National Cash Register - NCR} Corp. (1884-),
\footnote{\url{https://en.wikipedia.org/wiki/NCR_Voyix}}
which was acquired by AT\&T. Both acquisitions were nulled in 1997.

NCR is one of five early computer companies competing with IBM known as the BUNCH,
\footnote{\url{https://en.wikipedia.org/wiki/BUNCH}}
i.e., Burroughs, UNIVAC (joined forces in 1986 as Unisys), 
\footnote{\url{https://en.wikipedia.org/wiki/Unisys}}
NCR, {\it Control Data Corp - CDC}, and {\it Honeywell}.

In 2010 Oracle acquired Sun Microsystems 
\footnote{\url{https://en.wikipedia.org/wiki/Sun_Microsystems}}
which had developed SPARC microprocessors, the Java programming language, and Solaris OS and acquired StorageTek.
\footnote{\url{https://en.wikipedia.org/wiki/Oracle_Solaris}}
Oracle's Exadata provides a sophisticated storage system to run Oracle databases.
\footnote{\url{https://en.wikipedia.org/wiki/Oracle_Exadata}}
As companies move to the cloud they reduce hardware development. 
\footnote{url{https://www.mercurynews.com/2017/01/20/oracle-lays-off-450-employees/}}

Gartner groups 2-dimensional chart of ''ability to execute'' versus ''completeness of vision" for storage systems.  
Starting with the NE corner company rankings are roughly as follows:
Pure Storage, 
\footnote{\url{https://www.purestorage.com/resources/gartner-magic-quadrant-primary-storage.html}}
NetaApp, 
\footnote{\url{https://en.wikipedia.org/wiki/NetApp}}
HPE, 
\footnote{\url{https://en.wikipedia.org/wiki/Hewlett_Packard_Enterprise}}
Dell, 
\footnote{\url{https://en.wikipedia.org/wiki/Dell}}
IBM, 
\footnote{\url{https://en.wikipedia.org/wiki/IBM}}
Huawei, 
\footnote{\url{https://en.wikipedia.org/wiki/Huawei}}
Infinadat, 
\footnote{\url{https://en.wikipedia.org/wiki/Infinidat}} 
Hitachi Ventura, 
\footnote{\url{https://www.hitachivantara.com/en-us/home.html}}
IEIT Systems, 
\footnote{\url{https://en.ieisystem.com/}}
and 
{\it DataDirect Networks - DDN} (Tinitri). 
\footnote{\url{https://www.ddn.com/}}  

A list of storage startups in 2023 is as follows.
\footnote{\url{https://www.crn.com/news/storage/the-10-hottest-storage-startups-of-2023}}

\subsection{Hyperscalers and Cloud Storage}\label{sec:hyper}

Hyperscalers are data centers with a very large number of disks.
4/12 of hyperscaler platforms which represent 78\% of capacity in this category and consume 13,177 MegaWatts 
are AWS (Amazon Web Services), Google Cloud, Meta, and Microsoft Azure. 
\footnote{\url{https://www.datacenterknowledge.com/manage/2023-these-are-world-s-12-largest-hyperscalers}}   
The combined storage capacity of HDDs and SSDs has exceeded an ExaByte EB$=10^{18}$
\footnote{\url{https://home.cern/news/news/computing/exabyte-disk-storage-cern}}
at European Organization for Nuclear Research - CERN. 
\footnote{\url{https://en.wikipedia.org/wiki/CERN}}

Microsoft is developing Silica glass storage that could make ransomware attacks 
impossible in the data center and hyperscalers for its Azure customers.
\footnote{\url{https://www.microsoft.com/en-us/research/publication/project-silica-towards-sustainable-cloud-archival-storage-in-glass/}}
Microsoft Azure's {\it Local Redundant Storage -LRS} stores data thrice 
in the same primary region yielding eleven 9's durability for objects over one year.
All three copies should be written synchronously before the write is considered completed.
{\it Zoned Redundancy Storage - ZRS} replicates data in three zones.
{\it Geo Redundant Storage - GRS} in addition to writing data thrice in primary region,
writes it again in the secondary region which is miles away. 
{\it Geo-Zone Redundancy Storage - GZRS} combines LRS with GRS. 
\footnote{\url{https://learn.microsoft.com/en-us/azure/storage/common/storage-redundancy}}
This info is summarized in Table \ref{tab:summary},
Modeling techniques in Appendix II can be used to find configurations meeting user reliability requirements.

\begin{table}[h]
\begin{footnotesize}
\begin{center}
\begin{tabular}{|c|c|c|c|c|}\hline \hline
Config.          &LRS     &ZRS      &RA-GRS        &RA-ZRS  \\ \hline
Durability               &11 9s   &12s      &16 9's        &16 9's  \\ \hline
copies/ node             &3       &3        &6             &6       \\ \hline
\end{tabular}
\end{center}
\end{footnotesize}
\caption{Key parameters for redundancy options in Microsoft Azure storage.\label{tab:summary}}
\end{table}

HDDs due to their lower cost per GB than SSDs are the preferred medium for hyperscalers. 
Brewer et al. 2016 \cite{Brew16,Bre+16} identified the following metrics: 
(1) Higher throughput.
(2) Higher capacity.
(3) Lower tail latency in Section \ref{sec:tail}. 
(4) Security requirements. 
(5) TCO.  
Higher throughput can be attained at a lower IOPS by transferring larger blocks, since:
$\mbox{throughput} = \mbox{IOPs} \times \mbox{blocksize}\mbox{ or }\lambda_i < 2^{-i} / \bar{x} \leq 1 , i \geq 0.$ 
Replication is implemented at various levels (geographic, data center, servers) Brewer 2016 \cite{Brew16}
Reliability, performance, and cost (pay-as-you-go)
are identified as key cloud storage parameters in Huang 2013 \cite{Huan13}. 

HDDs save power to various degrees by halving RPM, stopping rotations, turning off electronics.
{\it Massive Array of Idle Disks - MAID} powered down the majority of data center disks 
Colarelli and Grunwald 2002 \cite{CoGr02}.
MAID's {\bf nearline storage} can be made online quickly without human intervention,
\footnote{\url{https://en.wikipedia.org/wiki/Nearline_storage}}
while {\bf offline storage} requires human intervention to become online.

Copan for archival storage combines MAID based on RAID5 requires at least two disks to be powered up for updating.  
\footnote{\url{http://www.thic.org/pdf/Jul05/copansys.aguha.050720.pdf}}
Schemes to save disk power are discussed in Chapter 7 in Thomasian 2021 \cite{Thom21}. 
Lower power consumption is attainable by NAND Flash SSDs as discussed in Section \ref{sec:flash}.

Google datacenters use clusters, e.g. 15,000 commodity PCs with fault-tolerant software.
Power usage is a major consideration.
{\it Power Usage Effectiveness - PUE} is a metric used to determine data center energy efficiency.
It is the ratio of total power consumption (cooling plus power conversion) to the power used to run IT equipment. 
Variation of PUE over (2008-22) across different Google sites is as follows. 
\footnote{\url{https://www.google.com/about/datacenters/efficiency/}}

\subsection{Simple Disk Performance Queueing Models}\label{sec:queue}

RAID5 performance was successfully evaluated in Menon 1994 \cite{Meno94}
by modeling disks as single server M/M/1 queues with Poisson arrivals with rate $\lambda$
and exponential service times with mean $\bar{x}=1/\mu$ Kleinrock 1975 \cite{Klei75}.
The disk utilization factor $\rho=\lambda \bar{x} \leq 1$ is the fraction of disk busy time,
hence the maximum sustainable arrival rate is $\lambda_{max} = [\bar{x}_{disk}]^{-1}$,
e.g., for $\bar{x}=10$ ms $\lambda_{max}=[\bar{x}]^{-1}=100$ IOPS.  
The mean response and waiting time for M/M/1 queues is:
$R=\bar{x}/(1-\rho) = [\mu-\lambda]^{-1},\hspace{3mm} W = R - \bar{x} = \rho \bar{x}/(1-\rho), \hspace{3mm}
\rho=\lambda \bar{x} \leq 1.$

At $W=\bar{x}$ $\rho=0.5$ and at $\rho=1/3$ rule-of-thumb 
for mainframe HDDs $R \leq 1.5 \bar{x}$ or $W=\bar{x}/2$.
For a target mean response time: $\lambda_{target} < 1/\bar{x} - 1/R_{target},$     \newline
e.g., $\bar{x}=10$ ms and $R_{target} =20$ ms then ${\lambda}_{target}=50 = \lambda_{max}/2$.  
Disk scheduling can be used to shorten disk service time are discussed in Section \ref{sec:sched} in Appendix III.

If the offered load to disk files $\lambda \bar{x} > 1$ 
then if all files fit on one disk then one solution is to replicate files on $m$ disks, 
so that $\rho = \lambda \bar{x} / m < 1$,
which is an M/M/m queueing system where $m$ disks have a shared queue Kleinrock 1975 \cite{Klei75}.
$m$ should be selected to be sufficiently large to ensure an acceptable response time, 
e.g., for $m=2$ and $\rho=0.9$ $R_{(2)} = \bar{x}/(1-\rho^2) \approx 5 \bar{x}$.

Replicated disks can be modeled as $m$ M/M/1 queues,
which with uniform routing and balanced loads are inferior in mean response time to M/M/m queues,
e.g., $R^{(2)} < R_{(1)}$.
This is because there may be pending requests at some disks, while others are idle,
a principle known as resource sharing in Section 5.1 in Kleinrock 1976 \cite{Klei76}.

Response time can be shortened by initiating requests at $m$ disks with replicated data concurrently.  
In fact hedged requested are sent after a certain delay Dean and Barroso 2013 \cite{DeBa13}.
Response times in M/M/1 queues are exponentially distributed Kleinrock 1075 \cite{Klei75}  

\vspace{-2mm}
\begin{eqnarray}\label{eq:Rdistr}
R(t) = 1 - e^{-t/R(\rho)}\mbox { where } R(\rho)=\bar{x}/(1-\rho).
\end{eqnarray}

Provided disk accesses are initiated concurrently at $m$ disks then the expected value of the minimum is:
$R_m^{min} = R (\rho)  / m$ Trivedi 2001 \cite{Triv01}.    
Once there is a response the request at the other $m-1$ disks may be cancelled 
except during seeks Thomasian 1995 \cite{Thom95}.
 
The arrival rate ($\lambda_p$) to achieve a certain percentile ($p$) of response time $R_{(p)}$ is: 
\vspace{-1mm}
$$R_{(p)}  =  - R  \mbox{ln} (1-p), \mbox{ so that }\lambda_p = 1/\bar{x} + \mbox{ln} (1-p) / R_{(p)}.$$ 

More accurate estimates of mean disk response time can be obtained using the M/G/1 queueing model,
with Poisson (M) arrivals and general (G) service times,
which is utilized in Section \ref{sec:normalanal} in Appendix I.
Given the mean ($R$) and variance ($\sigma^2_R$) of response time the 90$^{th}$ and 95$^{th}$ percentiles are: 
$R_{(90)} \approx R + 1.3 \sigma_R$ and $R_{(95)} \approx R + 2 \sigma_R.$
Approximations and bounds for GI/G/1 queues are given in Chapter 2 Kleinrock 1976 \cite{Klei76}. 
Request arrivals to a disk may be bursty and exhibit long tails, e.g., occasional very long file transfers from disk.
\footnote{\url{https://en.wikipedia.org/wiki/Long-tail_traffic}}
Self-similarity has been observed in disk arrival process.
\footnote{\url{https://en.wikipedia.org/wiki/Self-Similarity_of_Network_Data_Analysis}}


The problem of allocating files on $n$ devices 
to minimize mean response time is considered in Piepmeier 1975 \cite{Piep75}. 
Given the total rate of requests $\Lambda$ the problem at hand is to determine the rate of disk requests: 
$\lambda_i \geq 0 , 1 \leq i \leq n, \sum_{i=1}^n= \Lambda$ to minimize the mean overall response time 
\vspace{-1mm}
$$R= \sum_{i=1}^n \frac{\lambda_i}{\Lambda} R_i \mbox{  s.t.  }\sum_{i=1}^n \lambda_i = \Lambda, \hspace{2mm} 
R_i= \bar{x}_i+ W_i \mbox{  where } W_i =\frac{\lambda_i \overline{x^2_i}}{2(1-\rho_i)},\hspace{2mm}
0 \leq \lambda_i \leq [\bar{x}_i]^{-1}, $$
where $\overline{x^j_i}, 1 \leq i \leq n$ is the $j^{th}$ moment of service time at the $i^{th}$ device
and $W_i$ the mean waiting time in an M/G/1 queue Kleinrock 1975 \cite{Klei75}. 
The Lagrange multiplier method is used to determine the optimal $\underline{\lambda}$.
\footnote{\url{https://en.wikipedia.org/wiki/Lagrange_multiplier}} 
In allocating files on heterogeneous drives the mean response time 
is possibly minimized by not utilizing the slowest devices.

A similar observation is made in Georgiadis et al. 2004 \cite{GeNT04},
where the allocation criterion is minimizing the expected maximum response time
using the min-max policy, see e.g., Ibaraki and Katoh 1988 \cite{IbKa88}.
Heterogeneous arrays combining SSDs and HDDs to accommodate hot and cold data are viable by reducing cost, 
while meeting response time requirements for certain applications as discussed in Section \ref{sec:optimal}.


\subsection{Dealing with High Tail Latency}\label{sec:tail}

Tail latency is high percentile latency of requests 
that are longer than say 99\% of all requests Dean and Barroso 2013 \cite{DeBa13}.        
Reasons for high latency are given as follows: 
{\bf 1. Shared resources:}  CPU cores, processor caches, memory and network bandwidth.                 
{\bf 2. Daemons:} They are periodically activated and hence use limited resources, 
but cause delays when active. 
{\bf 3. Global resource sharing:} Network switches, shared file systems.                               
{\bf 4. Maintenance:} Examples are data reconstruction in RAID, periodic disk defragmentation on disk,
\footnote{\url{https://en.wikipedia.org/wiki/Defragmentation}}
storage reclamation in {\it Log-Structured File Systems - LFS} Rosenblum and Ousterhout 1992 \cite{RoOu92}. 
In the case of SSDs there may be a 100-fold increase in random read access time for a modest write activity.      
{\bf 5. Queueing delays} are introduced in Section \ref{sec:queue} and used in Appendix I for RAID analysis.

CPUs which temporarily overrun their power limits, throttle to lower temperature slower CPUs.
Disk spindown in laptops incurs extra power and latency for spinup.

The effect of subsecond response times in timesharing computer systems 
on programmer productivity is observed in Thadhani 1981 \cite{Thad81}. 
An extra benefit is that programmers react faster and this reduces the total time to complete a task. 
The effect of response time on successful conversion in e-commerce is quantified as follows: 
\footnote{\url{https://queue-it.com/blog/ecommerce-website-speed-statistics/}}

\subsection{Data Compression, Compaction and Deduplication}\label{sec:compress}

More efficient storage utilization is attainable by preserving storage capacity 
by using data deduplication, compaction, and compression.
Data compression methods including Huffman, Lempel-Ziv - LZ
and {\it Arithmetic Coding - AC} are discussed in Sayood 2017 \cite{Sayo17}.
Deflate
\footnote{\url{https://en.wikipedia.org/wiki/DEFLATE}}
and zip/gzip use lz77 + Huffman code, 
AC whose patent expired in 2008 
\footnote{\url{https://patents.google.com/patent/US4905297A/en}}
has been implemented 
\footnote{\url{https://www.proquest.com/openview/62157a4dcd72d24d0038014e888c7af6/1?pq-origsite=gscholar&cbl=1936364}}
but not used in common tools,
Zstd 
\footnote{\url{https://en.wikipedia.org/wiki/Zstd}}
uses lz77 + tANS (t: table) as {\i Asymmetric Numeral Code - ANS} is much faster than AC, 
\footnote{\url{https://en.wikipedia.org/wiki/Asymmetric_numeral_systems}}
and open.
\footnote{\url{https://www.theregister.com/2022/02/17/microsoft_ans_patent/}}
(Private communication from Lihao Xu at Wayne State Univ.)

Magnetic tapes are a high density inexpensive medium for storing data with 148 GB/inch$^2$ achieved by Sony.
\footnote{\url{https://en.wikipedia.org/wiki/Magnetic_tape}}
A 2.5 improvement in magnetic tape capacity is achieved by applying compression.
\footnote{
\url{https://www.techtarget.com/searchdatabackup/news/252495598/Potential-magnetic-tape-storage-capacity-surges-in-renaissance}}
Tape libraries with robotic arms are attractive in that they do not require human intervention. 
An automated tape library is described and analyzed in Section \ref{sec:tapelibrary} in Appendix I.

LFS data compression yields different file sizes when they are modified, 
but this is not a problem since files are not written in place.
LFS collects a cylinders worth of modified files and 
writes them to disk in batches, one cylinder at a time 
or as full-stripe writes in  RAID5/6 {\it Log Structured Array -LSAs} in Section \ref{sec:LFS} in Appendix I.
Hardware and software data compression techniques double disk capacity.
While disk space and data transmission time is reduced there is an associated CPU processing cost.
\footnote{\url{https://en.wikipedia.org/wiki/Disk_compression}} 

Main memory compression was attractive in mid-1990s when DRAM prices were high.
Compressed data in main memory is decompressed 
as it is loaded into the lowest level CPU cache (L3) and vice-versa.
Data decompression time is in the critical path of instruction execution and 
hence more important than compression time.
IBM's {\it Memory eXpansion Technology - MXT} was applied to its x86-based servers Tremaine et al. 2001 \cite{Tre+01}.
Lenovo acquired IBM's x86 server business in 2014 following sale of its PC business in 2005, 
which made Lenovo the \#1 PC maker.
Specifications for Lenovo ThinkSystem SR630 V3 Server in 2023 are as follows.
\footnote{\url{https://lenovopress.lenovo.com/lp1600-thinksystem-sr630-v3-server}}

The complication in memory management due to variability of compressed segment sizes
is addressed using {\it Linearly Compressed Pages - LCP}, 
i.e., all cache lines within a page are compressed to the same size Pekhimento et al. 2013 \cite{PSK+13}. 
This avoids performance degradation without requiring energy-inefficient hardware.

Intel 4th Gen Xeon Scalable processors use {\it Quick Assist Technology - QAT} 
to carry out main memory data compression and encryption.
\footnote{https://www.intel.com/content/www/us/en/content-details/784036}
Memory compression and encryption has been applied by IBM z16 mainframes.
\footnote{CP Assist for Cryptographic Functions - PACF} coprocessor performs 
symmetric key encryption operations and calculates message digests in hardware. 
{\it Advanced Encryption Standard - AES}, 
\footnote{\url{https://en.wikipedia.org/wiki/Advanced_Encryption_Standard}.}
Data/Triple Data Encryption Standard (DES/TDES), Secure Hash Algorithm (SHA)-1,2,3 are supported
\footnote{\url{https://www.redbooks.ibm.com/redbooks/pdfs/sg248951.pdf}}   
{\it American MicroDevices - AMD} also provides encryption support.
\footnote{
\url{https://www.amd.com/content/dam/amd/en/documents/epyc-business-docs/white-papers/memory-encryption-white-paper.pdf}}
Main memory compression studies are discussed in Section 2.14 in Thomasian 2021 \cite{Thom21}. 

Deduplication is applied at block, page, or file level 
to archival and backup data to reduce storage and data transmission costs. 
Duplication can be checked quickly by applying hash functions such as SHA-2 
and in the case of a match a byte-by-byte comparison can be applied.  
\footnote{\url{https://en.wikipedia.org/wiki/SHA-2}}
Data compaction reclaims storage held by old versions of objects 
or defunct objects based on their {\it Time-to-Live - TTL}.

Data deduplication is surveyed in Mohammed and Wang 2021 \cite{MoWa21}
and classified in Paulo and Pereira 2014 \cite{PaPe14}.
(1) granularity, (2) locality, (3) timing, (4) indexing, (5) technique, and (6) scope.
\footnote{\url{https://en.wikipedia.org/wiki/Data_deduplication}}
Deduplication is discussed in Section 2.15 in Thomasian 2021 \cite{Thom21} 

\subsection{Undetected Disk Errors and Silent Data Corruption - SDC}\label{sec:UDE}


{\it Silent Data Corruption - SDC} manifests itself as {\it Undetected Disk Errors - UDEs}.
The causes of UDEs and their effects on data integrity is discussed by Hafner et al. 2008 \cite{HDBR08}.
Techniques to address the problem at various software layers in the I/O stack
and solutions that can be integrated into the RAID subsystem are discussed here.     
\footnote{\url{https://en.wikipedia.org/wiki/Data_corruption}}

{\it Undetectable Write Errors - UWEs} are in the form of dropped or phantom writes and 
off-track or misdirected writes resulting in stale data.
{\it Undetected Read Errors - UREs} are in the form of ECC miscorrects or off-track reads, 
which result in reading of stale or corrupted data.

The work presents technologies for addressing data corruption problems caused by UDEs,
as well as other causes of data corruption.
Technologies are divided into classes depending on the software layer and the type of problem.

\begin{description}

\item[Checking at middleware or application software layers:]
Some file systems checksum 4-8 KB data chunks and store them separately.
When the data chunk is read the checksum is recomputed and compared.
A mismatch indicates corruption of either the data or the checksum,
but the former is more likely because of its larger size.

This method was utilized by ZFS Karlsson 2006 \cite{Karl06}
and the {\it IRON - Internal RObustNess} file system Prabhakaran et al. 2005 \cite{Pra+05},
The IRON file system prototype which incurs minimal time and space overheads via in-disk checksumming, 
replication, and parity is shown to greatly enhance file system robustness.

\item[Detection of Errors by the Storage System:]
Errors introduced along the datapath between the storage system and the application may not be detected.
This class has several subclasses with respect to different powers of detection and correction 
with associated performance impacts and performance/cost tradeoffs.

\end{description}

By itself data scrubbing cannot detect UDEs.
A parity scrub provides an additional step beyond a data scrub.

\begin{description}

\item[Data scrubbing.]
A data scrub in addition to reading all of the data blocks recomputes the parity
and compares this computed parity with the stored parity on disk.
A miscompare implies a data corruption error.
Data scrubs and parity scrubs do not always ensure recovery,
and a comparison does not always provide a guarantee that no data corruption has occurred.

\item[Metadata options.]
When metadata is colocated with data and metadata contains checksums,
a read of the data and metadata can detect data corruption errors,
especially those introduced by firmware, software, or the memory bus.
When the metadata contains the data address, a read can detect off-track writes, 
but only at the offset target location, where data was incorrectly overwritten.
These methods are usually combined with a data scrub 
to enhance its effectiveness as discussed in Sundaram 2006 \cite{Sund06}. 
\footnote{\url{https://atg.netapp.com/wp-content/uploads/2018/08/TECH_ONTAP-Private_Lives_of_Disk_Drives.pdf}}

\item[SCSI Write Verify Disk Command.]
This command rereads data after it is written to ensure that it was written correctly.
This works for dropped write errors but cannot necessarily detect off-track writes,
because the same possibly incorrect track is accessed.
This is an obvious performance impact, since a full disk rotation is required.
This method has a simple recovery algorithm, 
which rewrites the data to the same or a different location on disk.

\item
[Verification handled by fresh data in write cache.]
When the data needs to be evicted from the cache,
the disk is first read and compared with the cache copy
and the eviction occurs only if the data comparison is correct.
This has the same simple recovery algorithm as the write-verify approach.

\item[\bf Idle Read After Write - IRAW.]
proposed by Riedel and Riska 2008 \cite{RiRi03}.
is an improvement over {\it Read After Write - RAW} to reduce this performance penalty. 
The idea is to retain the written content in the disk cache and verify it once the disk drive becomes idle.
Trace-driven evaluation of IRAW show that disk idleness
can be utilized for WRITE verification with minimal effect on user performance.

\item[\bf Metadata as checksum or version number.]
The second copy of the metadata is stored in memory for fast comparison and access,
but must be flushed to disk to maintain detection capabilities after system crashes.

\item[\bf Metadata stored as RAID.]
This provides a locality of reference that can be used 
to mitigate some of the disk I/O overheads of separate locations.
The advantages of this approach are lower I/O and bandwidth penalties;
the disadvantage is potential loss of detection under certain failure scenarios.

\end{description}


SDCs may go undetected until a system or application malfunction occurs.
A major problem with SDC is that data errors propagate during data rebuild.
A reconstruction method in the presence of silent data corruption is developed in Li and Shu 2010 \cite{LiSh10}, 
which outperforms other methods when periodic validation is carried out.



\section{RAID classification and Extensions}\label{sec:RAIDclassify}
\vspace{3mm}

Two main categories of RAID redundancy are replication, e.g., mirrored disks,
and erasure coding, e.g., parity-based protection.
In erasure coding $k$ out of $n$ disks are data disks and $m=n-k$ disks serve as check disks.
The {\it Maximum Distance Separable - MDS} code is optimal, 
where $m$ is the minimum number of disks to recover as many failed disks known as erasures, 
which are failures whose location is known Thomasian and Blaum \cite{ThBl09}.
 
Striping partitions files into fixed size {\it Stripe Units - SUs} or strips, 
which are placed in round-robin manner across $N$ disks modulo $N$.
Full stripe writes have the advantage that $m$ strips are computed  
on-the-fly as data strips are being transmitted to be written.
The effect of striping on disk load balancing is ascertained in Ganger et al. 1996 \cite{GWHP96}.
A striped RAID with no check strips is classified as RAID0.

Chen et al. 1994 \cite{Che+94} report on studies to determine the optimal strip size. 
Most files which are small may be held in a single strip and hence accessed by a single disk access.
Default 128 KB and 256 KB strip sizes are adopted by some operating systems. 
Readahead size should be matched to strip size to minimize the number of disks accessed.  
\footnote{\url{https://ioflood.com/blog/2021/02/04/optimal-raid-stripe-size-and-filesystem-readahead-for-raid-10/}}

RAID4 and RAID5 dedicate one strip per stripe to parity,
while RAID3 based on synchronized disks uses small parity blocks,
so that disks are read/written concurrently providing a high bandwidth.
The parity are used to an be used to check the integrity of data 
and computed on the fly before writing. 

Parity blocks are computed as follows.
$P_N = D_1 \oplus D_2 \oplus D_3 \ldots \oplus D_{N-1}$.
The one disk dedicates to parity in RAID4 may become a bottleneck if the workload is write intensive.
RAID5 with distributed parities has an advantage over RAID4 in utilizing all disks for read processing. 
The left symmetric organization places parity strips in repeating left to right diagonals
allows the maximum parallelism in reading full stripes ($N=4$ consecutive strips).
\vspace{-1mm}
\begin{scriptsize}
$$
(D_0,D_1,D_2,D_3,P_{0:3}),
(D_5,D_6,D_7,P_{4:7},D_4),
(D_{10},D_{11},P_{8:11},D_8,D_9),
(D_{15},P_{12:15},D_{12},D_{13},D_{14}),
(P_{16:19},D_{16},D_{17},D_{18},D_{19})
$$
\end{scriptsize}
Parity striping proposed in Gray et al. 1990 \cite{GrHW90} preserves the original data layout, 
which has the advantage of obviating accesses to multiple disks 
when the accessed block size exceeds the strip size. 
It has he disadvantage of possibly unbalanced disk loads, known as access skew, 
which can be dealt with via disk load balancing support.
Placing parity blocks at the middle disk cylinders will reduce seek distances for updates. 

This results in a significant increase in RAID5's disk loads and favors parity striping, 
To show that load balanced disks provide a lower response time
consider two disks with $\bar{x}=1$ and $\Lambda=1.2$.
Let $f_1$ and $f_2$ denote the fractions of requests to two disks,
so that $\rho_i=f_i \lambda \bar{x}$ and $R= \sum_{i=1}^2 f_i \bar{x}_i / (1-\rho_i)$.
If loads are balanced: $f_1=f_2=0.5$ then $R_b = 1/(1-0.6)=2.5$
and if the loads are unbalanced, e.g., $f_1=1/3$ and $f_2=2/3$, 
then $R_u=(1/3)/(1-0.4)+(2/3)/(1-0.8)=3.89$.

{\it Zettabyte File System - ZFS} was developed at Sun Microsystems 
and acquired by Oracle in 2010 operates with either disk controllers or RAIDZ software RAID, 
which handles multiple RAID levels including RAID5.
\footnote{\url{https://en.wikipedia.org/wiki/ZFS}}
ZFS dRAID is the software RAID version available with Linux. 
\footnote{\url{https://pve.proxmox.com/wiki/ZFS_on_Linux}}
{\it Software-Defined Storage - SDS} provides policy-based provisioning 
and management of data storage independent of the underlying hardware.
\footnote{\url{https://en.wikipedia.org/wiki/Software-defined_storage}}

RAID(4+k) disk arrays have $k$ check strips per stripes with $N$ strips.
$N-i$ disks need be accessed to recover from $i \leq k$ failures, possibly just unreadable sectors.
Single disk failures are dealt with parity strips, 
which are the least expensive computationally.
The use of both P and Q codes to reduce page accesses is discussed in Section \ref{sec:singlediskrebuild}.

Disk arrays with Hamming codes are classified as RAID2 Fujiwara 2006 \cite{Fuji06}. 
\footnote{\url{https://en.wikipedia.org/wiki/Hamming_code}}
The Hamming code was adopted by Thinking Machines for its Connection Machine CM-5.  
\footnote{\url{https://en.wikipedia.org/wiki/Connection_Machine}}
Disks are read and written in parallel allowing parity strips updates. 
With the positions of failed disks known two rather than 
just one failed disk can be recovered by reading disks holding check blocks.
Parity bits $p_i,0 \leq i \leq 4$ are XORs of data blocks with a bit in that position. 
$$
\begin{matrix}
\mbox{Bit position} &1 &2 &3 &4 &5 &6 &7 &8 &9 &10 &11 &12 &13 &14 &15 &16 &17 &18 &19 &20 \\ \hline
\mbox{Encoded data bits}$$ &p_1 &p_2 &d_1 &p_4 &d2 &d_3 &d_4 &p_8 &d_5 &d_6 &d_7 &d_8 &d_9 &d_{10} &d_{11} &p_{16} &d_{12} &d_{13} &d_{14} &d_{15}
\end{matrix}
$$
If the parity bits in positions 1, 2 and 8 indicate an error, then bit 1+2+8=11 is in error.

{\it RAI Memories - RAIM} Meaney et al. 2012
\footnote{\url{https://en.wikipedia.org/wiki/Redundant_array_of_independent_memory}}               
and {\it RAI Libraries - RAILs} (tapes) in Ford et al. 1998 \cite{FoMB98} utilize the RAID4 paradigm.
Quantum has a product based on the RAIL paradigm.                          
\footnote{\url{https://blocksandfiles.com/2021/10/08/quantums-exabyte-munching-scale-out-modular-tape-library/}}

\subsection{Berkeley Proposal to Replace IBM DASD with Inexpensive Disks}\label{sec:Berkeley}

The RAID paper in addition to a classification of disk arrays also advocated replacing expensive,  
large form-factor, reliable disks known as {\it Direct Access Storage Devices - DASD}, 
used by IBM mainframes and its {\it Plug-Compatibles Mainframes - PCMs}
with arrays of inexpensive small form-factor, 
less reliable disks used in {\it Personal Computers -PCs} in 1980s. 
\footnote{\url{https://en.wikipedia.org/wiki/Plug_compatible}}
Amdahl Corp. (1970-1997) founded by S/360 architect G. Amdahl was an early PCM, 
which was later taken over by Fujitsu.
\footnote{\url{https://en.wikipedia.org/wiki/Amdahl_Corporation}}
Hitachi exited the mainframe hardware business but runs its software on IBM hardware.
\footnote{\url{https://www.prnewswire.com/news-releases/hitachi-to-deliver-new-mainframe-based-on-ibm-z-systems-in-japan-300461913.html}}

DASD used variable block sizes with the {\it Extended Count Key Data - ECKD} format,
same as CKD layout, but with five additional {\it Channel Command Words -CCWs}. 
\footnote{\url{https://en.wikipedia.org/wiki/Count_key_data}}
ECKD is supported by IBM's {\it Multiple Virtual Storage - MVS} OS and its descendant z/OS, 
whose I/O routines deal with {\it virtual DASD - v-DASD} with user files mapped to 3390 HDDs.
User applications via MVS's I/O routines issue I/O requests to v-DASD, which are enqueued dor v-DASD's, 
although is blocks may reside on different physical HDDs. 
Given the high cost of modifying z/OS and legacy software DASD 
are emulated on {\it Fixed Block - FB} disks by DACs - Disk Array Controllers.
\footnote{\url{https://en.wikipedia.org/wiki/Disk_array_controller}}

Early IBM file systems such as the {\it Index Sequential Access Method - ISAM} had a cylinder and track index,
whose benefits are lost when files are mapped onto FB disks.
Topics such as physical database design and tuning also lose their relevance, 
see e.g., Ramakrishnan and Gehrke 2002 \cite{RaGe02} 

ECKD format disks are mapped onto FB disks with 512 B (byte) and 4096 B disk Advanced Format sectors since 2010.
\footnote{\url{https://en.wikipedia.org/wiki/Disk_sector} }
The {\it Error Correcting Code - ECC}                      
\footnote{\url{https://en.wikipedia.org/wiki/Error_correction_code}}
in the latter case is 100 rather 50 bytes with an efficiency of $4096/(4196)=97.3\%$ versus $512/562= 88.7\%$.
\footnote{\url{https://en.wikipedia.org/wiki/Advanced_Format}}
CCWs addresses issued to DASD, e.g. HHTTRR, are translated to LBAs on FB disks by DAC.

Section 3.1 in Gibson 1992 \cite{Gibs92} presents three cases  
of replacing IBM 3390 HDDs with arrays of smaller capacity disks. 
Advances in disk technology since 1988 by 1993 invalidated the premise 
upon which the 1988 RAID hypothesis was based,
that with a large number of disks a higher storage bandwidth is achievable.
The availability of high capacity small form-factor: 2.5 and 3.5 inch less expensive disks 
obviated the need for large form factor disks and DASD production was discontinued in 1993. 
\footnote{\url{https://en.wikipedia.org/wiki/List_of_disk_drive_form_factors}}
DASD were used by {\it Plug Compatible Mainframes - PCMs}
Such disks consume less power than 10\begin{footnotesize}7/8\end{footnotesize} inch 3390s, 
but even less power is consumed by SSDs. 
This is the reason for change in terminology: Inexpensive to Independent.

\subsection{Maximum Distance Separable - MDS Minimum Redundancy Arrays}\label{sec:MDS}

RAID(4+k) {\it Maximum Distance Separable - MDS} erasure-coded arrays 
use the capacity of $k$ disks to tolerate $k$ disk failures,
which is the minimum redundancy according to the Singleton bound. 
\footnote{\url{https://en.wikipedia.org/wiki/Singleton_bound}}
RAID(4+k) kDFT arrays for $k \geq 1$ can be implemented 
using the Reed-Solomon code and its variations or parity codes for $k=2,3$.
\footnote{\url{https://en.wikipedia.org/wiki/Reed-Solomon_error_correction}}

StorageTek's Iceberg was an early RAID6 product with RS coding 
which used the LFS paradigm extended to disk arrays coined LSA - Log-Structured Array Menon 1995 \cite{Meno95}.
Check strips were computed as data was transmitted for full stripe writes.
LSA is not suitable for OLTP applications with frequent record updates,
since extra processing is required to update all the indices of updated records not written in place.

Iceberg under the name {\it RAMAC Virtual Array - RVA} 
temporarily served as an IBM product IBM Corp. 1997 \cite{IBMC97}. 
StorageTek was acquired by Sun Microsystems, itself acquired by Oracle Corp.  
\footnote{\url{https://www.oracle.com/it-infrastructure/}}

Iceberg's {\it Thin Provisioning - TP} method for optimizing storage utilization has been adopted by others. 
TP relies on on-demand allocation of blocks of data versus the traditional method of allocating all the blocks in advance. 
\footnote{\url{https://en.wikipedia.org/wiki/Thin_provisioning}}
3PAR, an HPE subsidiary, advocates TP using {\it Dedicate on Write - DoW} 
rather than {\it Dedicate on Allocate - DoA}, to reduce required disk space.          

Three {\it 2-Disk-Failure-Tolerant - 2DFTS} arrays which are also MDS are as follows:
(1) EVENODD by Blaum et al. 2002 \cite{Bla+02}, 
(2) X-code by Xu and Bruck 1999 \cite{XuBr99}, and 
(3) {\it Rotated Diagonal Parity - RDP} by Corbett et al. 2004 \cite{Cor+04}. 
EVENODD has been extended to $k=3$ most notably the STAR code Huang and Xu 2005 \cite{HuXu05},
whose decoding complexity is  lower than comparable codes. 
RDP has similarly been extended to $k=3$ Thomasian and Blaum 2009 \cite{ThBl09}. 

RAID7.3 is an MDS 3DFT, whose coding details are not specified.
It is argued that 3DFT is required to deal with additional disk failures in view of long rebuild times with HDDs.
Given 12 TB drives 540 TB capacity is achieved by 12 RAID6 arrays with six disks each
or RAID7.3 with data disks and 24 P and Q check disks,
but it is claimed that the same {\it Mean Time to Data Loss - MTTDL} is achieved by 45 data and only 3 check disks.      
\footnote{\url{https://www.hyperscalers.com/jetstor-raidix-1185mbs-storage-nas-san-record}}

RM2 is a non-MDS 2DFT parity code where 1-out-m rows are dedicated to parity strips Park 1995 \cite{Park95}.
Data strips in the first $m-1$ rows are protected by two parities in the $m^{th}$ row, 
RM2 performance is analyzed in Thomasian et al. 2007 \cite{ThFH07}. 

\subsection{Rotated Diagonal Parity - RDP Coded Arrays}\label{sec:RDP}

RDP invented at NetApp and adopted as its RAID6 array Corbett et al. 2004 \cite{Cor+04} 
There are $p+1$ blocks in each row (stripe) where the first $p-1$ blocks hold data,
the $p^{th}$ block holds a horizontal parity and $p+1^{st}$ block holds the diagonal parity.
The controlling parameter $p > 2$ should be a prime number.
When  a block is updated the horizontal parity is updated first and 
is used in updating the corresponding diagonal parity,  
RDP with $p=5$ is shown in Fig.~\ref{fig:RDP}.

\begin{figure}[t]
\begin{footnotesize}
\begin{center}
\begin{tabular}{|c|c|c|c|c|c|}   \hline
Disk$_0$       &Disk$_1$       &Disk$_2$       &Disk$_3$       &Disk$_4$       &Disk$_5$    \\ \hline \hline
$d_{0,0}^0$    &$d_{0,1}^1$    &$d_{0,2}^2$    &$d_{0,3}^3$    &$d_{0,4}^4$    &$d_{0,5}$   \\ \hline
$d_{1,0}^1$    &$d_{1,1}^2$    &$d_{1,2}^3$    &$d_{1,3}^4$    &$d_{1,4}^0$    &$d_{1,5}$   \\ \hline
$d_{2,0}^2$    &$d_{2,1}^3$    &$d_{2,2}^4$    &$d_{2,3}^0$    &$d_{2,4}^1$    &$d_{2,5}$   \\ \hline
$d_{3,0}^3$    &$d_{3,1}^4$    &$d_{3,2}^0$    &$d_{3,3}^1$    &$d_{3,4}^2$    &$d_{3,5}$   \\ \hline
\end{tabular}
\end{center}
\end{footnotesize}
\caption{\label{fig:RDP}Storage system with RDP code with $p=5$.
The superscripts are the diagonal parity groups on Disk$_5$. The parity of diagonal 4 is not stored.}
\end{figure}

Horizontal parity blocks on Disk$_4$ hold the even parity of the preceding data blocks in that row. 
Diagonal parity blocks on Disk$_5$ hold the even parity of data and row parity blocks in the same diagonal.
\vspace{-1mm}
$$d^{i,4} = d^{i,0} \oplus d^{i,1} \oplus d^{i,2} \oplus d^{i,3}, i=0,3. \hspace{5mm}
d_{0,5} = d_{0,0} \oplus d_{3,2} \oplus d_{2,3} \oplus d_{1,4} $$
\vspace{-4mm}
$$d_{1,5} = d_{0,1} \oplus d_{1,0} \oplus d_{3,3} \oplus d_{2,4}  \hspace{5mm}
d_{2,5} = d_{0,2} \oplus d_{1,1} \oplus d_{2,0} \oplus d_{3,4}    \hspace{5mm}
d_{3,5} = d_{0,3} \oplus d_{1,2} \oplus d_{2,1} \oplus d_{3,0}    $$


RDP protects $(p-1)^2$ data blocks using $2p^2 - 6 p + 4$ XORs.
Setting $n=p-1$ we have $2n^2 -2n$ XORs, hence RDP requires $2-2/n$ XORs per block.
It is shown that EVENODD requires $2-2/(n-1)$ XORs per block and 
is hence outperformed by RDP in the number of required XORs.  


If the block size in RDP and EVENODD was a strip then a full stripe write 
would require updating the four diagonal parity strips.
This inefficiency is alleviated by selecting the prime number $p=2^n+1$,
which allows defining diagonal parities within a group of $2^n$ stripes.
This allows the block size for RDP to be the usual system block size 4 KB divided by $2^n$.
If the system's disk block size is 4 KB for $p = 17$ 
there are sixteen 256 B blocks per strip allowing efficient processing of full stripe writes.
EVENODD developed a similar scheme, but used $p=2^{10}+1=257$.

It is shown in Blaum and Roth 1999 \cite{BlRo99} that the minimum update complexities 
of MDS codes for an $m \times n$ array with $k < n $ data and $r=n-k$ check columns are:  
$$
r=2:  \hspace{2mm} 2 +\frac{1}{m} ( 1 - \frac{1}{k})     \hspace{5mm}
r=3:  \hspace{2mm} 3 + \frac{3}{m} (\frac{2}{3} - \frac{1}{k}).
$$

\subsection{Partial Maximum Distance Separable Code}\label{sec:PMDS}

{\it Partial MDS - PMDS} relies on local and global parities Blaum et al. 2013 \cite{BlHH13}.
Given an $m \times n$ array of sectors $r$ erasures can be corrected in a horizontal row.
The case of RAID5 $r=1$ with $p_i, 0 \leq i \leq 3$
and $s=2$ with two global parities $g_1$ and $g_2$ in Fig.~\ref{fig:PMDS}.
Given six check sectors as many failed sectors can be corrected.

\begin{figure}[b]
\begin{center}
\begin{footnotesize}
\begin{tabular}{|c|c|c|c|c|c|c|}\hline
Column 1  & Column 2   & Column 3  &  Column 4  & Column 5   & Column 6  & Column 6  \\ \hline
$d_{0}$   & $d_{1}$    & $d_{2}$   &  $d_{3}$   & $d_{4}$    & $d_{5}$   &$p_0$      \\ \hline
$d_{6}$   & $d_{7}$    & $d_{8}$   &  $d_{9}$   & $d_{10}$   & $d_{11}$  &$p_1$      \\ \hline
$d_{12}$  & $d_{13}$   & $d_{14}$  &  $d_{15}$  & $d_{16}$   & $d_{17}$  &$p_2$      \\ \hline
$d_{18}$  & $d_{19}$   & $d_{20}$  &  $d_{21}$  & $g_1$      & $g_2$     &$p_3$      \\ \hline
\end{tabular}
\end{footnotesize}
\end{center}
\caption{\label{fig:PMDS}Layout of the local and global parities for PMDS with $r=1$ and $s=2$.}
\end{figure}

Recovery is possible in the following cases:
\begin{description}
\item[Recoverable case I:] One drive failure: $(d_3, d_9, d_{15}, d_{21})$ 
and two additional sector failures: $d_2$ and $d_{12}$
\item[Recoverable case II:] $r=1$ failures per row $(d_2, d_11, d_{12}, d_{19}$
and two additional failures anywhere: $d_4$ and $d_19$.
\item[Recoverable case III:]
$d_2$ and $d_4$ in row zero and $d_{12}$ and $d_{14}$ in row 2.
$d_{11}$ and $d_{19}$ in row 1 and three.
$d_1$ ad $d_{12}$ can be recovered by $g_1$ and $g_2$ 
and then remaining blocks can be recovered using row parities.
\end{description}

RAID6 is an overkill to tolerate the failure of one disk and one sector 
according to Plank and Blaum 2014 \cite{PlBl14}.   
{\it Sector Disk - SD} is an erasure code based on the PMDS concept,
whose decoding process using SD is based on linear algebra. In brief:


\begin{description}

\item[PMDS:] Given a stripe defined by the parameters $(n,m,s,r)$ 
a PMDS code tolerates the failure of any $m$ blocks per row, and any additional blocks in the stripe. 
PMDS codes are maximally fault-tolerant codes that are defined with $m \times r$ local parity equations 
and $s$ global parity equations. 
However, PMDS codes make no distinction for blocks that fail together because they are on the same disk.

\item[SD:] Given a similarly defined stripe as PMDS 
SD code tolerates the failures of any $m$ disks (columns of blocks) plus any additional $s$ sectors in the stripe

\end{description}

For a given code construction, the brute force way to determine whether 
it is PMDS or SD is to enumerate all failure scenarios and test to make sure that decoding is possible. 
Given $n$ disk, $m$ decoding disks, $s$ sectors per stripe dedicated to coding, and $r$ rows per stripe
a set of $m  \times r + s$ equations are set up each of which sums to zero.
For SD codes the number is larger: $ \binom{n}{m} \binom{r(n-m)}{s}$.

There are many constructions which are valid as SD codes, but not as PMDS codes.
SD codes can recover the failure of a column 4 $(d_3, d_9, d_{15}, d_{21})$ and $d_3$ and $d_12$,
but not aforementioned case II
Since this abbreviated discussion is hard to follow and readers should refer to the main text.


\subsection{Tiger: Disk Adaptive Redundancy Scheme}\label{sec:tiger}

Tiger {\it Disk-Adaptive Redundancy - DAR} scheme dynamically tailors 
itself to observed disk failure rates Kadekodi et al. 2022 \cite{Kad+22}. 
This scheme has been observed to reduce the space overhead by up to 20\%.
Tiger results in significant savings with respect to existing DAR schemes, 
which are constrained in that they partition disks into subclusters with homogeneous failure rates.  

Pacemaker DAR scheme is less desirable from the viewpoint of the fraction of viable disks for higher n-of-k's 
as shown in Fig. 2a for 7-of-9, 14-of-17, 22-of-25, and 30-of-33.
$n-k$ parity are added to $k$ data chunks.
Pacemaker reduces intra-stripe diversity and is more susceptible to unanticipated changes 
in a make or model's failure rate and only works for clusters committed to DAR.

Tiger avoids constraints by introducing eclectic stripes 
in  which redundancy is tailored to diverse disk failure rates. 
Tiger ensures safe per-stripe settings given that device failure rates change over time. 
Evaluation of real-world clusters shows that Tiger provides better space-savings, 
less bursty IO to change redundancy schemes and better robustness than prior DAR schemes.

Appendix A in \cite{KSCM23} derives MTTDL for eclectic stripes 
using the Poisson binomial distribution for disks which have different failure rates.  
\footnote{\url{https://en.wikipedia.org/wiki/Poisson_binomial_distribution}}

\section{Multidimensional Coding for Higher Reliability}\label{sec:multiD}
\vspace{3mm}

Several schemes to attain higher reliability are proposed in Hellerstein et al. 1994 \cite{Hel+94}.
2-dimensional arrays are analyzed in Newberg and Wolf 1994 \cite{NeWo94},
but the analysis of Full-2 codes took longer Lin et al. 2009 \cite{LZW+09}.
Simulation was used to investigate the effect of varying frequency of repairs 
to replacing broken disks on the MTTDL \cite{Hel+94}

\subsection{Grid Files}\label{sec:gridfiles}
\vspace{2mm}

Grid files by Mingqiang Li et al. 2009 \cite{LiSZ09} 
is an instance of {\it Product Codes - PCs} in two dimensions.
The Horizontal (H) and Vertical (V) {\it Parity Code - PC} or HVPC code 
is the simplest Grid code over $k_1$ rows and $k_2$ columns:
$
H_{i, k_2} = D_{i,1} \oplus D_{i,2} \oplus \ldots  D_{i, k_2 }, \hspace{5mm}i = 1, \ldots , k_1 ,
\hspace{5mm}
V_{k_1, j} = D_{1, j} \oplus D_{2, j} \oplus \ldots D_{k_1, j},  \hspace{5mm}j =  1, \ldots , k_2 . 
$
\vspace{-2mm}
$$
P_{k_1,k_2}= 
\bigoplus_{i=1}^{k_1} H_{i,k_2} = 
\bigoplus_{j=1}^{k_2} V_{k_1,j} = 
\bigoplus_{i=1}^{k_1} \bigoplus_{j=1}^{k_2} D_{i,j} 
$$ 

If the fault-tolerance of row and column codes are given 
as $t_r$ and $t_c$ then the total number of tolerated faults is: $t_{ft} = ( t_r+1 ) ( t_c+1 ) -1.$
It is easy to see that the strips of a rectangle with $(t_r+1)(t_c+1)$ failed strips cannot be recovered,
but removing one failed strip allows recovery, hence the minus one in $t_{ft}$.  

In addition to {\it Single Parity Code - SPC} or RAID5, 
EVENODD, STAR, and X-code codes are considered in this study.
RDP which is computationally more efficient than EVENODD can replace it.
SPC requires one row or column, EVENODD, RDP, X-code require two, and STAR three. 
X-code requires a prime number of disks ($N$) 
and covers strips with $\pm 1$ slopes with respect to the bottom two rows of $N\times N$ segments.



In Fig. \ref{fig:2xcodes} we consider a $p \times p$ X-code array with P and Q parities
placed vertically (not horizontally), which is protected by SPC P' parities. 
There is data update, two parity updates for the X-code and 
three updates for the SPC code since there are updates in three columns, total of five.

\begin{figure}[h]
\begin{center}
\begin{footnotesize}
\begin{tabular}{|c|c|c|c|c||c|c|}\hline
Column\_1      &Column\_2      &Column\_3  &Column\_4         &Column\_5          &Column\_6    \\ \hline \hline
$D_{1,1}$      &$D_{1,2}$      &$\ldots$   &$D_{1,p-2}$       &${P}_{1,p-1}$      &${Q}_{1,p}$  \\ \hline
$D_{2,1}$      &$D_{2,2}$      &$\ldots$   &$D_{2,p-2}$       &${P}_{2,p-1}$      &${Q}_{2,p}$  \\ \hline
$\vdots$       &$\vdots$       &$\ddots$   &$\vdots$          &$\vdots$           &$\vdots$     \\ \hline
$D_{p,1}$      &$D_{p,2}$,     &$\ldots$   &$D_{p,p-2}$       &${P}_{p,p-1}$      &${Q}_{p,p}$  \\ \hline   \hline
${P'}_{p+1,1}$ &${P'}_{p+1,2}$ &$\ldots$   &${P'}_{p+1,p-2}$  &${P'}_{p+1,p-1}$   &${P'}_{p+1,p}$ \\ \hline
\end{tabular}
\end{footnotesize}
\end{center}
\caption{ \label{fig:2xcodes}Data in the first $p$ rows is protected 
by an X-code with P and Q parities in columns $p+1$ and $p+2$. 
(left upper corner) is inside a ($(p+2)\times(p+2)$) X-code with P and Q parities.
The $2 \times p$  space in upper right hand corner with $D'$s is only covered by the larger X-code.
Data block updates affect $P$ and $Q$ parities and the P' parities.}
\end{figure}

Data recovery in Grid files is carried alternating between row-wise and column-wise recovery
and stops when there are no further fault strips when successful.


\subsection{Hierarchical RAID - HRAID}\label{sec:HRAID} 
\vspace{2mm}

HRAID is described in Thomasian 2006a \cite{Thom06a} 
and evaluated in Thomasian et al. 2012 \cite{ThTH12}.
There are $N$ {\it Storage Nodes - SNs}, where each SN has a cached DAC serving $M$ disks.
HRAID$(k/\ell)$ provides recovery against $k$ SN failures and $\ell$ disk failures at each node,
which means $k$ (resp. $\ell$) strips out of $M$ strips per stripe at each SN 
are dedicated to interSN (resp. intraSN) check strips.
It is easy to see that its redundancy level is $(k+\ell)/M$.
Data and inter-SN check strips are protected by intra-SN check strips.
It can be said HRAID$k/\ell$ is kNFT with respect to SNs and $\ell$DFT with respect to disks at each SN.
Hierarchical RAID is shown in Fig. \ref{fig:hraid}.

\begin{figure}
\begin{scriptsize}
\begin{center}
\begin{tabular}{|c|c|c|c||c|c|c|c||c|c|c|c||c|c|c|c|}\hline
\multicolumn{4}{|c||}{ Node 1} & \multicolumn{4}{c||}{ Node 2}
& \multicolumn{4}{c||}{ Node 3} & \multicolumn{4}{c|}{ Node 4} \\ \hline \hline
$~D_{1,1}^1$ & $~D_{1,2}^1$ & $~P_{1,3}^1$ & $~Q_{1,4}^1$ &
$~D_{1,1}^2$ & $~P_{1,2}^2$ & $~Q_{1,3}^2$ & $~D_{1,4}^2$ &
$~P_{1,1}^3$ & $~Q_{1,2}^3$ & $~D_{1,3}^3$ & $~D_{1,4}^3$ &
$~Q_{1,1}^4$ & $~D_{1,2}^4$ & $~D_{1,3}^4$ & $~P_{1,4}^4$ \\ \hline
\end{tabular}
\end{center}
\end{scriptsize}
\caption{\label{fig:hraid}Hierarchical RAID HRAID1/1 with $N=4$ SNs, $M=4$ disks per SN.
P is the intraSN and Q the interSN parity. 
Data and Q parity strips are protected by P parity strips.
Only the first stripe on four SNs is shown.
Parities are rotated from row-to-row at each SN and from SN to SN.} 
\end{figure}

P parities are used due to disk failures for intraSN online recovery and rebuild due to a local disk failure.
Q parities are used due SN failures for interSN recovery and rebuild.
Q parities can be used for recovery when P parities provide insufficient protection 
Restriping as described in Rao et al. 2011 \cite{RaHG11}) is used both cases
with P and Q parity strips overwritten depending on the case.
Code to simulate HRAID reliability is given in Section \ref{sec:HRAIDsimul} in Appendix III.


\subsection{Reliable Storage at Exabyte Scale - RESAR}\label{sec:RESAR}. 
\vspace{2mm}

RESAR has parity groups weaving through a 2-D disk array Schwarz et al. 2016 \cite{SAK+16}. 
As shown in Fig. \ref{fig:RESAR} has disklets numbered such that coordinates can be converted to a single number:
\vspace{-1mm}
$$(i,j) = i \times  (k+2) +j, \hspace{3mm} 0 \leq j \leq k+1 \mbox{ with } k=4, i=2, \ldots k+1$$  
Disklet $39 = 6 \times 6 +3$ is protected by P6 horizontally and D7 diagonally.
If disklets 33, 35,  38, and 40 were placed on the same disk or disks on the same rack, 
then the failure of this disk or rack causes data loss, since the four disklets share two reliability stripes. 
Details on achieving appropriate placements are given in the paper.

\begin{figure}[h] 
\parbox{0.45\linewidth}{
\centering
\begin{footnotesize}
\begin{tabular}{|c|c|c|c|c|c|}\hline
    &2  &3 &4 &5 &$P_0$   \\ \hline
$D_0$ &8 &9 &10 &11 &$P_1$  \\ \hline
$D_1$ &14 &15 &16 &17 &$P_2$ \\ \hline
$D_2$ &20 &21 &22 &23 &$P_3$ \\ \hline 
$D_3$ & 26 & 27 &28 &29 &$P_4$ \\ \hline
$D_4$ &32 &33 &34 &35 &$P_5$  \\ \hline
D$_5$ &38 &39 &40 &41  &P$_6$ \\ \hline
D$_6$ &44 &45 &46 &47 &P$_7$ \\ \hline 
D$_7$ &50 &51 &52 &53 &P$_8$ \\ \hline
D$_8$ &56 &57 &&& P$_9$ \\ \hline
D$_9$ &&&&&             \\ \hline
\end{tabular}
\end{footnotesize}
\caption{\label{fig:RESAR}A small bipartite RESAR layout connecting 
$P_i$ and  $D_i$ parities for $0 \leq i \leq 9$ via intervening disklets.
Each data disk is protected by a P parity in its row 
and a D parity at a lower row as member of diagonal parity group.}
}
\hfill
\parbox{0.45\linewidth}{
\centering
\begin{footnotesize}
\begin{tabular}{|c|c|c|c|c|c|}\hline
       &         &$P_1$    &         &         &       \\       \hline
       &$P_2$    &         &$P_3$    &         &       \\       \hline
$P_4$  &         &         &         &$P_5$    &                \\       \hline \hline
       &         &$D_1$    &         &         &       \\       \hline
       &$D_2$    &         &$D_3$    &         &$P_6$           \\       \hline \hline  
$P_7$  &         &         &         &$P_8$    &                \\       \hline
       &         &$D_4$    &         &         &       \\       \hline
       &$D_5$    &         &$D_6$    &         &$P_9$           \\       \hline \hline 
$\vdots$  &$\vdots$ &$\vdots$ &$\vdots$ &$\vdots$  &$\vdots$      \\       \hline
\end{tabular}
\end{footnotesize}
\caption{\label{fig:3D}3D parity where each data disk is protected 
by three parities one vertically and two diagonally.}  
}
\end{figure}

RESAR-based layout with 16 data disklets per stripe has about 50 times lower probability of suffering data loss
in the presence of a fixed number of failures than a corresponding RAID6 organization.
The simulation to estimate the probability of data loss for 100,000 disks required 10,000 hours.

\begin{framed}
\subsection{Reliability of 2D Square- and 3D Cube-RAID}\label{sec:cubeRAID}
\vspace{2mm}

Reliability of 2-D size $n^2$ and 3-D size $n^3$ arrays 
are evaluated in Basak and Katz 2015 \cite{BaKa15}.
There are $n$ disks in each dimension protected by $n$ parity disks 
with an extra parity disk protecting the parities
so that the number of disks for $d$ dimensions is $n^d + d \times (n+1), d=2,3$.
The parities in each dimension are protected by an extra parity, hence the plus one. 

Four disk failures constituting a rectangle lead to data loss
and the number of cases leading to data loss is $\binom{n}{2}n$.
Given a series $k$ of such squares the total number of disks is $N=(n^2 + 2n + 2)k$.
The probability of failure with $i=3$ and $i=4$ already failed disks is 
\vspace{-1mm}
$$\alpha_3 =  \binom{N}{2}n k / \binom{N}{4}, \hspace{5mm}
\alpha_4 =  [ \binom{N}{2} n (n^2 + 2n-2 )+ n^2) ]  k / \binom{N}{5} $$

The analysis is extended to $\alpha_i = \mbox{min}(1,(i+2)\alpha_i), \hspace{2mm} i \geq 5$
and 3D cubes with $N_3=4^3$ data and 15 parity disks 
which are compared with 2D square $N_2 = 8^2 =64$ data and 18 parity disks.
The Cube-RAID provides a superior MTTDL with respect to Square-RAID
and RAID6 with the same number of data disks.

\end{framed}

\subsection{2D- and 3-Dimensional Parity Coding}\label{sec:2d}
\vspace{2mm}

2D and 3D parity protection for disk arrays is proposed  in Paris and Schwarz 2021 \cite{PaSc21}
A small example of 2D protection is given in Fig.~\ref{fig:3D}.
There are the following parity groups: 
\vspace{-1mm}
$$(P_1, D_1, D_3, D_8), (P_2, D_1, D_2, D_5, D_9), (P_3, D_3, D_4, D_7, D_{10}),
(P_4, D_3, D_4, D_6, D_9), (P_5, D_5, D_6, D_7, D_8, D_9).$$

Three disk failures can result in data loss in this case, e.g., $(P_1, P_2, D_1)$, and $(D_1,D_5,D_8)$.
In the first case there is data loss since $D_1$ loses both of its parities.
In the second case there are two failures in the three parity groups, 
which can support only one disk failure.

A small example with 3-D parities with nine  parity groups is given in Fig. \ref{fig:3D}.
$$(P_1, D_1, D_4), (P_1, D_2, D_5), (P_3, D_3, D_6),\hspace{1mm}
(P_4, D_1, D_3), (P_5,D_1, D_2), (P_6, D_2,D_3),  \hspace{1mm}
(P_7, D_4, D_6), (P_7 ,D_4, D_5), (P_9,D_5,D_6).$$ 

As few as four disk failures can result in data loss with 3D coding, 
e.g., $(P_2, P_5, D_2, P_6)$, since $D_2$ has lost the three parities protecting it.

\section{Local Redundancy Code}\label{sec:LRC}
\vspace{3mm}

The {\it Local Redundancy Code - LRC} adds a few parity disks 
to provide speedier recovery by reducing the number of disks involved in recovery.
There is also reduced data transmission cost in distributed storage with a large number of disks.
{\it Clustered RAID - CRAID} in Section \ref{sec:CRAID} differs from LRC
in that it achieves faster rebuild by parallelizing disk accesses.

The (8,2,2) Pyramid code Huang et al. 2007 \cite{HuCL07}
has two local parities each over four data disks and two global parities are over all eight data disks.
\vspace{-1mm}
$$(d_1,d_2,d_3,d_4,c_{1,1}),\hspace{3mm}(d_5,d_6,d_7,d_8,c_{1,2}),\hspace{3mm}c_2,\hspace{1mm}c_3$$ 


\begin{table}
\parbox{0.45\linewidth}{
\centering
\begin{footnotesize}
\begin{tabular}{|c|c|c|c|c|c|}\hline
\# of failed blocks &                   &1      &2     &3       &4           \\ \hline \hline
MDS code            &recovery (\%)      &100    &100   &100     &0           \\ \hline
(11,8)              &read overhead      &1.64   &2.27  &2.91    &-           \\ \hline \hline
Pyramid code        &recovery (\%)      &100   &100   &100.0    &68.89        \\ \hline
(12,8)              &read overhead      &1.25  &1.74  &2.37     &2.83         \\ \hline
\end{tabular}
\end{footnotesize}
\caption{\label{tab:pyramid}Comparison between MDS code (11,8) and Pyramid code (12,8).}
}
\hfill
\parbox{0.45\linewidth}{
\centering
\begin{footnotesize}
\begin{tabular}{|c|c|c|c|}\hline
Scheme         &Storage    &Reconstruct      &Savings       \\
               &overhead   &cost             &              \\ \hline \hline
RS(6+3)        &1.5        &6                &              \\ \hline
RS(12+4)       &1.29       &14               &14\%          \\ \hline
LRC(12,2,2)    &1.29       &7                &14\%          \\ \hline
\end{tabular}
\end{footnotesize}
\caption{\label{tab:WAS}Choice of Windows Azure Storage}
}
\end{table}

{\bf Windows Azure Storage - WAS} improves over Pyramid codes
by reading less data from more fragments Huang et al. 2012 \cite{HSX+12}.
With the (12,2,2) code 100\% of three and 86\% of all the four disk failures are recoverable.
\vspace{-1mm}
$$\left( X_0, X_1,X_2, X_3, X_4, X_5, X_6, p_X \right),
\left( Y_0, U_1, Y_2, Y_3, Y_4, Y_5, Y_6, p_Y \right), P_0, P_1 [\forall X, \forall Y ]$$
WAS design choices are justified in Table \ref{tab:WAS}:

We continue the discussion with a smaller (6,2,2) example with $N=10$ disks,

\vspace{-2mm}
$$ \left( X_0,X_1,X_2,p_X \right), \left( Y_0, Y_1, Y_2,p_Y \right), p_0,p_1 $$
This LRC can tolerate arbitrary 3 failures by using the following sets of coefficients:

\vspace{-2mm}
$$q_{x,0} = \alpha_0 x_0 + \alpha_1 x_1 + \alpha_2 x_2, \hspace{5mm}
q_{x,1} = \alpha_0^2 x_0 + \alpha_1^2 x_1 + \alpha_2^2 x_2 \hspace{5mm} q_{x,2} =  x_0 + x_1 + x_2 $$

\vspace{-2mm}
$$q_{y,0} = \beta_0   y_0 + \beta_1 y_1 + \beta_2 y_2, \hspace{5mm}
q_{y,1}   = \beta^2_0 y_0 + \beta_1^2 y_1 + \beta_2^2 y_2 \hspace{5mm} q_{y,2} =  y_0 + y_1 + y_2 $$

\vspace{-2mm}
$$p_0 = q_{x,0}+q_{y,0}, \hspace{5mm} p_1 = q_{x,1}+q_{y,1}, \hspace{5mm} p_x = q_{x,2}, \hspace{5mm} p_y = q_{y,2} $$

The values of $\alpha$s and $\beta$s are chosen such that the LRC 
can decode all information theoretically decodable four failures with probability $p_d \approx 0.86$.
The CTMC given below for $N=10$ is given as follows:
\vspace{-1mm}
$${\cal S}_{N-i}  \xrightarrow{ (N-i) \delta} {\cal S}_{N-i-1}, \hspace{3mm} 0 \leq i \leq 4
\hspace{5mm}
{\cal S}_{N-4+i} \xleftarrow{ \rho_{N-5+i}} {\cal S}_{N-5+i} \hspace{3mm} 1 \leq i \leq 4; 
\hspace{5mm}
{\cal S}_6  \xleftarrow{\lambda p_d}  {\cal S}_7, 
\hspace{5mm}
{\cal S}_7 \xrightarrow{7\delta (1-p_d)} {\cal S}_{6F}. 
$$

Each of the $N$ {\it Storage Nodes - SNs} has a capacity $S$, bandwidth $B$,
whose use for rebuild is throttled to a small fraction $\epsilon=0.1$.
The repair load is evenly distributed among $N - 1$ SNs. 
The repair rate $\rho_9 = \epsilon (N-1) B) / S C$,
where it takes three fragments to repair data fragments and local parities
and six fragments to repair global parities, so that on the average: 
$3 \times 8 + 6 \times 2 )/10=3.6.$ 
Typical parameters are: fragment size $S=16$ TB, bandwidth $B=1$ Gbps, 
and detection and triggering time $T=30$ minutes and $\rho_6=\rho_7 = \rho_8 =1/T$
The MTTDL is $2.6\times 10^{12}$, versus $3.5 \times 10^9$ for replication and $6.1 \times 10^{11}$ for (6,3) RS. 
i.e., a 10-fold increase in MTTDL at the cost of an extra disk.
The detection and triggering time $T=30$ minutes dominates repair time for more than one fragment failure,
so that $\rho_8=\rho_7=\rho_6=1/T$, $N=300$, $S=16$ TB, $B=1$ GB per second.

\subsection{A Systematic Comparison of Existing LRC Schemes}\label{sec:LRCcomparison}
\vspace{2mm}

Xorbas, Azure's LRCs, and Optimal-LRCs are compared 
using the {\it Average Degraded Read Cost -  ADRC}, 
and the {\it Normalized Repair Cost - NRC} Kolosov et al. 2018 \cite{KYL+18}.
The trade-off between these costs and the code's fault tolerance offer different choices.
Xorbas discussed in Section \ref{sec:Xorbas} is an example of a Full-LRC code.
while Pyramid and Azure-LRC are examples of so-called data-LRCs.

With information-symbol locality, only the data blocks 
can be repaired in a local fashion by $r$ blocks,
while global parities require $k$ blocks for recovery.
In codes with all-symbol locality, 
all the blocks, including the global parities, 
can be repaired locally from surviving blocks.

Ceph is an open-source {\it Software-Defined Storage - SDS} platform 
that implements object storage on a single distributed computer cluster and 
provides 3-in-1 interfaces for object-, block-, and file-level storage Maltzahn et al. 2010 \cite{MMK+10}.
The experimental evaluation on a Ceph cluster deployed on Amazon {\it Elastic Compute Cloud - EC2} 
demonstrates the different effects of realistic network and
storage bottlenecks on the benefit from various LRC approaches.
\footnote{\url{https://en.wikipedia.org/wiki/Amazon_Elastic_Compute_Cloud}}
The study shows the benefits of full-LRCs and data-LRCs depends
on the underlying storage devices, network topology, and foreground application load.


The {\it Average Repair Cost - ARC} is based on the assumption
that the probability of repair is the same for all blocks. 
In the case of (10,6,3) Azure LRC. 
$ARC = \sum_{i=1}^n \mbox{cost}(b_i) / n  = (8 \times 3) +(2 \times 6) /10 = 3.6$                      \newline
The {\it Normalized Repair Cost - NRC} amortizes the cost of repairing parity blocks over data blocks. \newline
$NRC = ARC \times ARC \times \frac{n}{k} = [ (8 \times 3) + (2 \times 6)]/6= 6$                        \newline
{\it Degraded Read Cost -DRC} is the full-node repair cost.                                            
$DRC=  \sum_{i=1}^k \mbox{cost} (b_i)  / k =   6 \times 3 / 6 =3$.

Xorbas (16,10,5) has RS-based four global parities, which can be recovered from local parities 
via an implied parity block as shown in Fig. 2 in the paper.
\vspace{-1mm}
$$( X_1, X_2, X_3, X_4, X_5, P_{1:5}), (X_6, X_7, X_8, X_9, X_{10}, P_{6:10})$$

\subsection{Combining Parity and Topology Locality in Wide Stripes}\label{sec:widestripe}
\vspace{2mm}

Wide stripes are used to minimize redundancy.
There are $n$ chunks, $k$ data chunks, and $m=n-k$, which is 3 or 4 parity chunks \cite{HCY+21}.
For the ten examples given by Table 1: $1.18 \leq \mbox{redundancy}=n/k \leq 1.5$.
The notation used in the paper is given in Table \ref{tab:notation}.

\begin{table}[h]
\parbox{0.45\linewidth}{
\centering
\begin{footnotesize}
\begin{tabular}{|c|c|}\hline
Notation            & Description                                       \\ \hline \hline
$n$                 &number of chunks of a stripe                       \\ \hline
$k$                 &number of data stripes of a chunk                  \\ \hline
$r$                 &number of retrieved chunks to repair a lost chunk  \\ \hline
$z$                 &number of racks to store a stripe                  \\ \hline
$c$                 &number of chunks of a stripe in a rack             \\ \hline
$f$                 &number of tolerable node failures of a stripe      \\ \hline
$\gamma$            &maximum allowed redundancy                         \\ \hline
\end{tabular}
\end{footnotesize}
\caption{Notation for combined locality\label{tab:notation} }
}
\hfill
\parbox{0.45\linewidth}{
\begin{footnotesize}
\begin{tabular}{|c|c|c|}\hline
System           &$(n,k,r)$                             &$(16,10,5)$     \\ \hline \hline
Azure-LRC        &$f= n-k+ \lceil k/r \rceil + 1$       &$f=5$           \\ \hline
Xorbas           &$f \leq n-k - \lceil k/r\rceil +1$    &$f=4$           \\ \hline
Optimal LRC      &$f \leq n-k - \lceil k/r\rceil +1$    &$f=4$           \\ \hline
Azure LRC+1      &$f \leq n-k - \lceil k/r\rceil$       &$f=4$           \\ \hline
\end{tabular}
\end{footnotesize}
\caption{Number of tolerable node failures $f$ for different LRCs for (n,k,r)=(16.10,5)}
}
\end{table}

$(n,k,r,z)=(26,20,5,9)$, 
where $z=9$ is the number of elements in a column as shown in Fig. \ref{fig:fig2}

Challenges for wide stripes are:
(i) Retrieving $k$ chunks is expensive,
(ii) The CPU cache may not be able to hold large number of chunks,
(iii) updating $m=n-k$ check chunks is expensive.

There is a trade-off between redundancy and repair penalty:
(i) Parity locality incurs high redundancy.
(ii) Topology locality incurs high cross-track repair bandwidth.
The Azure LRC (32,20,2) is given in Fig. \ref{fig:AzureLRC}.

\begin{figure}[h]
\begin{footnotesize}
\begin{center}
\begin{tabular}{|c|c|c|c|c|c|c|c|c|c|c|}\hline
$D_1$ & $D_3$ &$D_5$ & $D_7$ & $D_9$ & $D_{11}$ & $D_{13}$  &$D_{15}$ &$D_{17}$    &$D_19$   &$Q_1$[1:20]     \\ \hline
$D_2$ & $D_4$ &$D_6$ &  $D_8$  & $D_{10}$  & $D_{12}$ & $D_{14}$  &$D_{16}$ &$D_{18}$  &$D_{20}$ &$Q_2$[1:20] \\ \hline
$P_1$ [1,2]    & $P_2$ [3,4]   & $P_3$[5,6] & $P_4$ [7,8] & $P_5$ [9,10]  & $P_6$[11,12]
& $P_7$ [13, 14] & $P_8$ [15,16] & $P_9$[17-18] & $P_{10}$[19-20] &-                                          \\ \hline
\end{tabular}
\end{center}
\end{footnotesize}
\caption{\label{fig:AzureLRC}Azure parity locality LRC (n,k,r)=(32,20,2).}
\end{figure}


\begin{table}[h]
\parbox{0.45\linewidth}{
\centering
\begin{footnotesize}
\begin{tabular}{|c c c|}\hline
$D_1$      &$D_2$       &$D_3$          \\
$D_4$      &$D_5$       &$P_1$[1:5]     \\ \hline
$D_6$      &$D_7$       &$D_8$          \\
$D_{9}$    &$D_{10}$    &$P_2$[6:10]    \\ \hline
$D_{11}$   &$D_{12}$    &$D_{13}$       \\
$D_{14}$   &$D_{15}$    &$P_3$[11,15]   \\ \hline
$D_{16}$   &$D_{17}$    &$D_{18}$       \\
$D_{19}$   &$D_{20}$    &$P_4$[16:20]   \\ \hline
$Q_1$[1:20] & $Q_2$[1:20] & -           \\ \hline
\end{tabular}
\end{footnotesize}
\caption{\label{fig:fig2}Combined locality (25,20,5,9).}
}
\hfill
\parbox{0.45\linewidth}{
\centering
\begin{footnotesize}
\begin{tabular}{|c|c|c|c|c|c|}\hline
Scheme             &n      &k     &r    &p     &Rate             \\ \hline \hline
S1: 24-of-28       &28     &24    &2    &2     &0.857            \\
S2: 48-of-55       &55     &48    &3    &4     &0.872            \\
S3: 72-of-80       &80     &72    &4    &4     &0.9              \\
S4: 96-of-105      &105    &96    &5    &4     &0.914            \\ \hline
\end{tabular}
\end{footnotesize}
\caption{\label{sec:comp}
Wide LRC schemes used to compare different LRCs with low level of redundancy $k/n > 0.85$.
$p$ is the \# of parity groups and local parities and $r$ the \# of global parities.}
}
\end{table}

VASTData is a Flash-based SSD storage system that unifies storage, database, 
and compute into a single, scalable software platform to power AI and deep learning 
in modern data centers and clouds.
\footnote{\url{https://vastdata.com/whitepaper}}                                     
Coding is Vastdata is described here,
where recovery is attained by accessing a fraction of data chunks,
but  it is not specified how parities are computed.   
\footnote{\url{https://vastdata.com/blog/breaking-resiliency-trade-offs-with-locally-decodable-erasure-codes}} 
\footnote{\url{https://vastdata.com/blog/providing-resilience-efficiently-part-ii}}
A possible example to accomplish this is as follows.
Consider 20 chunks $D_i,1 \leq i \leq 20$, where 
$P_1=\bigoplus_{i=1}^{10}$ and $P_2 =\bigoplus_{i=1}^{20}$.
Chunk $D_{11}$ can be recovered as $D_{11} = P_1 \oplus [ \bigoplus_{12}^{20} ] \oplus P_2$

\subsection{Practical Design Considerations for Wide LRCs}\label{sec:practical}
\vspace{2mm}



The following discussion is based on Kadekodi et. al. 2023 \cite{KSCM23}.
Four different 48-of-55 Azure LRC constructions with four local parities and 3 global parities are specifiable as follows:
\hspace{-1mm}
$$c_j = \bigoplus_{12 \times (j-1) + 1}^{12 \times j}, 1 \leq j \leq 4,
\hspace{2mm} rs_1, \hspace{2mm} rs_2, \hspace{2mm} rs_3 $$

Azure-LRC+1 forms a local group of the global parities and protects them using a local parity.
\vspace{-1mm}
$$c_j =\bigoplus_{12 \times (j-1) +1}^{12 \times j}, 1 \leq j \leq 3, \hspace{2mm} c_{rs} = \bigoplus_{j=1}^3 rs_j $$

In the case of 48-of-55 optimal Cauchy LRC 
the four local parities cover, similarly to Azure parities,
but global parities cover parity blocks (as in Fig. 4 (c)).

In the case of uniform Cauchy LRC the $p$ local parities 
are uniformly distributed across the $k$ data blocks and they are all XORed with the global parity checks.
\vspace{-1mm}
$$
c_1 = \bigoplus _{i=1}^{12} d_i \hspace{5mm}
c_2 = \bigoplus _{i=13}^{25} d_i \hspace{5mm}
c_3 = \bigoplus _{i=26}^{38} d_i \hspace{5mm}
c_4 = [\bigoplus _{i=38}^{48} d_i] \oplus [ \bigoplus_{j=1}^3 rs_j ]. 
$$

The average cost of reconstructing any of the data blocks {\it Average Degraded Read Cost - ADRC} 
and two other measures are defined  as follows:
\vspace{-1mm}
$$ ADRC = \sum_{i=1}^k \mbox{cost} ({b}_i / k),
\hspace{5mm}
ARC_1 = \sum _{i=1}^n b_i /n
\hspace{5mm}
ARC_2 =\sum_{i=1,j \neq i}^n \mbox{cost} (b_{i,j}) / \binom{n}{2}.$$

Average repair or reconstruction cost takes into account local
and global parity blocks in the computation 
as ARC$_1$ and ARC$_2$ is used as the cost of reconstructing 2 blocks.
Best schemes for various metrics are given in Table \ref{sec:best}.

\begin{table}[h]
\begin{footnotesize}
\begin{center}
\begin{tabular}{|c|c|}\hline
Locality ($k$)      &Uniform Cauchy LRC - S1: 13, S2: 13, S3: 19, S4: 26             \\ \hline
ADRC                &Azure LRC -  S1: 12, S2=12, S3=18, S4=24                        \\ \hline
ARC1                &Azure LRC - S1: 12.85, S2: 12,76, S3:19, S4: 25.22              \\ \hline
ARC2                &Uniform Cauchy - LRC S1: 27.92; S2: 33.85, S3: 49.22, S4 67.69  \\ \hline
Normalized MTTDL    &S1, 1., S2, 1.01*  S3, 1.0  S4: 1.0                             \\ \hline
\end{tabular}
\end{center}
\end{footnotesize}
\caption{\label{sec:best}Best schemes based on analytic metrics used to compare different LRC constructions.
Results are normalized with respect Uniform Cauchy. *Optimal Cauchy LRC.}
\end{table}

Consider the CTMC for a 48-of-55 disk array.
Given that $i$ denotes the number failed disks a failure at state ${\cal S}_i$
leads to a transition to ${\cal S}_{i+1}$ for $i < 5$.
Some of the transition for $5 \leq i \leq 7$ lead to failure.
Transition probabilities are based on:  
which is related to MTTF as follows: 
\vspace{-1mm}
$$\mbox{AFR} = 1 - \mbox{exp}(-8766/\mbox{MTTF})\hspace{5mm}\mbox{AFR} \approx 8766 / \mbox{MTTF}.$$
The MTTDL can be determined as the mean number of visits to each state 
and the holding time: $[(55-i)\delta]^-1$ where $\delta$ is the failure rate.

\section{Reducing Rebuild Traffic in Distributed RAID}\label{sec:distr}
\vspace{3mm}

A survey of network codes for distributed storage is given in Dimakis et al. 2011 \cite{DRWS11}.
We use (4,2) MDS code as shown in Fig. \ref{fig:rashmi} to illustrate how any two disk failures.
Assuming Storage Node - SN$_1$ fails then:
\vspace{-1mm}
$$B_2 \oplus (A_2 \oplus B_2) \rightarrow A_2 , \hspace{5mm} (A_2  \oplus B_2) \oplus (A_1 \oplus A_2 \oplus B_2) $$

\begin{figure}[b]
\begin{center}
\begin{footnotesize}
\begin{tabular}{|c|c|c|c|}\hline
SN$_1$     &SN$_2$        &SN$_3$          &SN$_4$            \\ \hline \hline
$A_1$      &$B_1$         &$A_1+B_1$       &$A_2+B_1$         \\ \hline
$A_2$      &$B_2$         &$A_2+B_2$       &$A_1+A_2+B_2$     \\ \hline
\end{tabular}
\hspace{5mm}
\begin{tabular}{|c|c|c|c|}\hline
SN$_1$    &SN$_2$  &S$_3$       &SN$_4$            \\ \hline
$A_1$     &$A_2$   &$A_1+A_2$   &$A_1+2A_2$        \\
$B_1$     &$B_2$   &$B_1+B_2$   &$B_1+2B_2+A_1$    \\ \hline
\end{tabular}
\end{footnotesize}
\end{center}
\caption{An MDS (4,2) (to the left) extended with piggybacking (to the right)\label{fig:rashmi}.}
\end{figure}

Assuming SN$_4$ fails then at SN$_1$ we compute $(A_2 \oplus B_2)$ at SN$_2$:
\vspace{-1mm}
$$  (A_2 \oplus B_2) \oplus (B_1\oplus B_2) \rightarrow (A_2 \oplus B_1) , \hspace{5mm}
A_1 \oplus (A_2 \oplus B_2) \rightarrow (A_1\oplus A_2 \oplus B_2)$$

To reproduction of broken SN$_1$ and SN$_2$ proceeds as follows:        

\vspace{-2mm}
\begin{eqnarray}\nonumber
(A_2 \oplus B_2) + (A_1 \oplus A_2 \oplus B_2) \rightarrow A_1) \rightarrow A_1,   \hspace{3mm}          
A_1  \oplus (A_1 \oplus B_1) \rightarrow B_1                                      \\
(A_1 \oplus B_1)  \oplus (A_2 \oplus B_1) \rightarrow (A_1 \oplus A_2)           \hspace{3mm}
(A_1 \oplus A_2) \oplus (A_1 \oplus A_2 \oplus B_2) \rightarrow B_2              \hspace{3mm}
B_2 \oplus (A_2 \oplus B_2) \rightarrow A_2
\end{eqnarray}

Rashmi et al. 2013 \cite{RSG+13} use Piggybacking to reduce the volume of transferred data. 
Consider four nodes with two bytes of data each as shown in Figure \ref{fig:rashmi}.
	

If SN$_1$ fails its recovery would ordinarily require accessing two blocks from SN$_2$ and SN$_3$ each,
but with piggybacking the second byte from SNs 2, 3, and 4 need be accessed.
Several methods to reduce rebuild traffic in a distributed environment are discussed in this section.

\subsection{Pyramid Codes}\label{sec:pyramid}
\vspace{2mm}

Pyramid codes are not MDS, 
but provide a much lower recovery cost Huang et al. 2007 \cite{HuCL07}.
There are twelve data blocks subdivided into two groups of six blocks with local parities and three global parities.
Microsoft's Azure is a {\it Local Reconstruction Code - LRC} which tolerates all three disk failures,
but the repair of as many as four failed disks is possible Huang et al. 2012 \cite{HSX+12}.
\vspace{-1mm}
$$ 
(d_1, d_2, d_3, d_4,     d_5,     d_6,     c_{1,1}, \hspace{3mm}
(d_7, d_8, d_9, d_{10},  d_{11},  d_{12},  c_{1,2}) \hspace{3mm} 
(c_2,c_3,c_4)  $$ 


A smaller example to shorten the discussion of data recovery is as follows:
\vspace{-1mm}
$$ \left( x_0, x_1, x_2), p_x, (y_0, y_1, y_2), p_y, c_0, c_1 \right) $$
Consider the failure of $x_0$, $x_2$, and $y_1$.
Repair proceeds by repairing $y_1$ using $p_y$ and $x_0$ and $x_2$
can then be repaired using $c_0$ and $c_1$.
Recovery with four disk failures is possible in 86\% of cases, but is quite complex.

LRC can tolerate arbitrary 3 failures by choosing the following $\alpha$'s and $\beta$'s.

\begin{footnotesize}
\vspace{-1mm}
$$q_{x,0} = \alpha_0 x_0 + \alpha_1 x_1 + \alpha_2 x_2,                    
\hspace{3mm}
q_{x,1} = \alpha^2_0 x_0 + \alpha^2_1 x_1 + \alpha^2 2 x_2,        
\hspace{3mm}
q_{x,2} = x_0 + x_1 + x_2.
$$
\vspace{-1mm}
$$
q_{y,0} = \beta_0 y_0 + \beta_1 y_1 + \beta_2 y_2,               
\hspace{3mm}
q_{y,1} = \beta^2 y_0 + \beta^2_1 y_1 = \beta^2 y^2    
\hspace{3mm}
q_{y,2} = y_0 + y_1 +y_2
$$
\end{footnotesize}

The LRC equations are as follows:
\vspace{-1mm}
\begin{footnotesize}
$$p_0 = q_{x,0} + q_{y,0}, \hspace{5mm} p_1 = q_{x_1} + q_{y,1},\hspace{5mm} p_x = q_{x,2} p_y = q_{y,2}$$
\end{footnotesize}

The values of $\alpha$s and $\beta$s are selected,
so that LRC can decide all information-theoretically decodable four failures.

{\bf 1. None of four parities fails:}
The failures are equally divided between the two groups. We have the matrix.

\begin{footnotesize}
$$
G=
\begin{pmatrix}
1 & 1 & 0 & 0 \\
0 & 0 & 1 & 1 \\
\alpha_i & \alpha_j & \beta_s & \beta_t \\
\alpha^2_i & \alpha^2_j & \beta^2_s & \beta^2_t
\end{pmatrix}
\hspace{2mm}\mbox{det}(G) = (\alpha_j - \alpha_i) (\beta_t - \beta_s) (\alpha_i + \alpha_j - \beta_s - \beta_t).
$$
\end{footnotesize}

{\bf 2. Only one of $p_x$ and $p_y$ fails:}
Assume $p_y$ fails. The remaining three failures are two in group $x$ and one in group $y$.
We have three equations with coefficients given by:

\begin{footnotesize}
$$
G' =
\begin{pmatrix}
1 & 1 & 0 \\
\alpha_i & \alpha_j & \beta s \\
\alpha^2 _i & \alpha^2_j & \beta_2^2
\end{pmatrix}
det(G')= \beta_s(\alpha_j-\alpha_i)(\beta_s - \alpha_j - \alpha_i)
$$
\end{footnotesize}

{\bf 3.Both $p_x$ and $p_y$ fail}
The remaining two failures are divided between the two groups.

\begin{footnotesize}
$$
G" =
\begin{pmatrix}
\alpha_i & \beta_s \\
\alpha^2_i & \beta^2_s
\end{pmatrix}\hspace{2mm}
det(G') = \alpha_i \beta_s  (\beta_s - \alpha_i)
$$
\end{footnotesize}
To ensure all cases are decodable all three matrices should be non-singular.

The repair savings of an $(n=16,k=12,r=6)$ LRC compared to that of an $(n=16,k=12)$ RS code
in the Azure production cluster are demonstrated in Huang et al. \cite{HSX+12}.
The $(n=18,k=14,r=7)$ WAS LR code has repair degree comparable to that of an $(9,6)$ RS code,
in the Azure production cluster are demonstrated in this work.
This code has reportedly resulted in major savings for Microsoft.                  
\footnote{\url{https://www.microsoft.com/en-us/research/blog/better-way-store-data/}.}

\subsection{Hadoop Distributed File System - HDFS-Xorbas}\label{sec:Xorbas}
\vspace{2mm}

{\it Hadoop Distributed File System - HDFS} Borthakur 2007 \cite{Bort07},
which employs $n \geq 3$ replications for high data reliability is based on the following assumptions:
1. Hardware failures the norm rather than the exception.
2. Batch processing.
3. Large datasets.
4. Files once created are not changed, i.e., {\it Write-Once, Read-Many - WORM}.
5. Moving computation cheaper than moving data,'
which is the reason behind data-centric computing.
\footnote{\url{https://en.wikipedia.org/wiki/Data-centric_computing}}
6. Portability across heterogeneous platforms required.
Data-centric computing is contrasted by the following types of computing.
\footnote{\url{https://www.hitechnectar.com/blogs/learn-more-about-computing-environment-and-its-different-types/}}
Edge computing is a novel paradigm.
\footnote{\url{https://en.wikipedia.org/wiki/Edge_computing}}

HDFS Xorbas was implemented at U. Texas at Austin for Facebook (now Meta).
It is an LRC build on top of an RS code by adding extra local parities Sathiamoorthy et al. 2013 \cite{SAP+13}.
The experimental evaluation of Xorbas was carried out on an Amazon EC2 and a Facebook cluster,
in which the repair performance of $(n = 16, k = 10, r = 5)$ LR code was compared against a $[14, 10]$ RS code.
Xorbas reduced disk I/O and repair network traffic compared to RS codes.
In addition to the four RS parity blocks associated with ten file blocks, 
LRC associates local parities as shown below:
\vspace{-1mm}
$$
S_1 = c_1 X_1 + c_2 X_2 + c_3 X_3 + c_4 X_4 + c_5 X_5  \hspace{1mm}
S_2 = c_6 X_6 + c_7 X_7 + c_8 X_8 + c_9 X_9 + c_{10} X_{10} \hspace{1mm}
S_3 = {c'}_5 S_1 + {c'}_6 S_2 \mbox{ with }{c'}_5 = {c'}_6 =1
$$

With three local parities the storage overhead is 17/10,
but $S_3$ need not be stored by choosing $c_i, 1 \leq i \leq 10$ to satisfy $S_1  + S_2 + S_3 = 0$.
It is shown that this code has the largest possible distance ($d = 5$)
for this given locality ($r = 5$) and blocklength ($n = 16$).
This is an LRC $(10,6,5)$ code.



A comparison of three redundancy methods is given in Table~\ref{tab:comparison20}

\begin{table}[t]
\begin{footnotesize}
\begin{center}
\begin{tabular}{|c|c|c|c|}\hline
                   &Storage    &Repair      &MTTDL         \\ 
Scheme             &overhead   &traffic     &(days)        \\ \hline \hline 
3-replication      &2x         &1x          &2.31E+10      \\ \hline
RS(10,4)           &0.4x       &10x         &3.31E+13      \\ \hline
LRC(10,6,5)        &0.6x       &5x          &1.21E+15      \\ \hline
\end{tabular}
\end{center}
\end{footnotesize}
\caption{\label{tab:comparison20}Comparison of three redundancy schemes.}
\end{table}

\subsection{Copyset Replication for Reduced Data Loss Frequency}\label{sec:cidon} 
\vspace{2mm}

{\it Copyset Replication - CR} is a technique to significantly reduce 
the frequency of data loss events with triplicated data.
It was implemented and evaluated on three data centers 
Cidon et al. \cite{CRS+13}. 

Data replicas are ordinarily randomly scattered across several nodes for parallel data access and recovery. 
CR presents a near optimal trade-off between the number of nodes 
on which the data is scattered and the probability of data loss.
It is better to lose more data  less frequently than vice-versa:

\begin{quote}
``Even losing a single block of data incurs a high fixed cost, 
due to the overhead of locating and recovering the unavailable data. 
Therefore, given a fixed amount of unavailable data each year,
it is much better to have fewer incidents of data loss with more data each than more incidents with less data. 
We would like to optimize for minimizing the probability of incurring any data loss.
in other words fewer events are at a greater loss of data are preferred.'' 
\end{quote}

Replication characteristics of three data centers are given in Table \ref{tab:rep}.

\begin{table}
\begin{footnotesize}
\begin{center}
\begin{tabular}{|c|c|c|c|c|}\hline
System          &Chunks             &Cluster        &Scatter    &Replica        \\
                &per node           &size           &width      &scheme         \\ \hline \hline
Facebook        &10,000             &1000-5000      &10         &Small group    \\ \hline
RAMCloud        &8000               &100-10,0000    &N-1        &All nodes      \\ \hline
HDFS            &10,000             &100-10,000     &200        &Large group    \\ \hline
\end{tabular}
\end{center}
\end{footnotesize}
\caption{\label{tab:rep} Three replication schemes. 
In the case of Facebook and HDFS 2nd and 3rd replica is on the same rack.}
\end{table}

In {\it Random Replication - RR} assuming that the primary replica is placed at node $i$
the secondary replica is placed at node $i+1 \leq j \leq i+S $.
where the scatter width ($S$) is the number of nodes that may store copies of a node's data.
If $S=N-1$ the secondary replicas are drawn uniformly from all the nodes in the cluster. 

For example, for $N=9$ nodes, degree of replication $R=3$, 
$S=4$ if the primary replica is on Node 1 then the secondary replica is on Nodes (2,3,4,5).
$P_{DL}$ is the ratio \#copysets = 54 and the maximum number of sets $\binom{9}{3}$ or 0.64.
A lower scatter width decreases recovery time,
while a high scatter width increases the frequency of data loss.

Correlated failures such as cluster power outages are poorly handled by RR and (0.5\%-1\%) of nodes do not power up.
There is a high probability that all replicas of at least one chunk in the system will not be available.
According to Fig. 1 in the paper as the number nodes exceeds 300 nodes 
with degree of replication ($R=3$) the probability of data loss $P_{DL} \approx 1$.

CR splits the nodes into groups of $R$ chunks, 
so that the replicas of a single chunk can only be stored in a copyset. 
Data loss occurs when all the nodes of a copyset fail together,
e.g., with $N=9$ nodes and $R=3$ nodes per copyset: $(1,2,3),(4,5,6),(7,8,9)$,
so that data loss occurs if $(1,2,3)$, $(4,5,6)$, or $(7,8,9)$ are lost.
When a CR node fails there are $R-1$ other nodes that hold its data.

The goal of CR is to minimize the probability of data loss, 
given any $S$ by using the smallest number of copysets.
When the primary replica is on node $i$ the remaining $R-1$ replicas 
are chosen from $(i+j),j=1, \ldots S$,
With $S=4$ $(1,2,3), (4,5,6), (7,8,9), (1,4,7), (2,5,8), (3,6,9)$.
Node 5's chunks can be replicated as nodes (4,6) and nodes (2,8).
There are \#copysets=6 and if three nodes fail $P_{DL}=0.07$.
Note that the copysets overlap each other in one node and that copysets cover all nodes equally, 

Two permutations of node numbers can be used to define copysets.
The overall scatter width is $S=P(R-1)$.
The scheme creates $P N /R$ or $ S N / (R(R-1))$ copysets. 

Simulation is used to estimate $P_{DL}$ versus the number of RAMCloud nodes for $3 \leq R \leq 6$.
As shown in Fig. 2 $P_{DL} \approx =0$ for $R=6$ beyond a 1000 nodes 
and increases linearly from 0 to 50\% for 10,000 nodes.  

Fig. 3 in the paper plots $P_{DL}$ with $N=5000$, $R=3$, 
with RR and CR versus $1 \leq S \leq 500$ when 1\% of the nodes fail.
For RR $P_{DL} \approx 1 $ for $S \geq 50$, 
which is the case for Facebook HDFS and $P_{DL} > 0.0$ for $S=500$. 
 
Consider the case when the system replicates data on the following copysets:
\vspace{-1mm}
$$(1,2,3), (4,5,6),(7,8,9),(1,4,7),(2,5,8),(3,6,9).$$
That is, if the primary replica is placed on node 3, 
the two secondary replicas can only be randomly on nodes 1 and 2 or 6 and 9. 
Note that with this scheme, each node's data will be split uniformly on four other nodes.
With this scheme there are six copysets and $P_{DL}= \#copysets / 84=0.07$.

Power outage in the case of a 5000-node RAMCloud cluster with 
CR reduces the probability of data loss from 99.99\% to 0.15\%. 
For Facebook's HDFS cluster it reduces the probability from 22.8\% to 0.78\%.

\subsection{More Efficient Data Storage: Replication to Erasure Coding}\label{sec:DiskReduce}
\vspace{2mm}

Access rates to fresh data diminish with time
and data can then be stored in a more efficient manner as time progresses.
DiskReduce is a modification of HDFS enabling asynchronous compression 
of initially triplicated data down to RAID6 Fan 2014 \cite{Fan+09}.   
This increases cluster's storage capacity by a factor of three.
The compression is delayed long  enough to deliver the performance benefits of multiple data copies


If the  encoding is delayed for $t$ seconds,
we  can  obtain  the  full  performance of  having  three  copies 
from  a  block's  creation  until  it  is $t$ seconds old. 
The penalty is modeled by $r$, which stands for $100(1-r)\%$ degradation.
Different $r$s gives different expected performance bounds.
The expected performance achieved by delaying encoding can be bounded as:
$\mbox{Penalty}(t)= \Phi(t) + r \times (1 - \Phi(t)),$
where $\phi(t)$ is {\it Cumulative Distribution Function - CDF} of block access with regard to blockage, 
which is derived from the trace (Fig. 4 in papo
er). 
Plotting Penalty$(t)$ versus $t$ it can be seen in Fig. 5.  
For $r=0.8, 0.5, 0.333$ there is very little system performance penalty after an hour.

The following issues are explored based on traces collected from Yahoo!, Facebook, and Opencloud cluster:
(1) The capacity effectiveness of simple and not so simple strategies for grouping data blocks into RAID sets; 
(2) Implications of reducing the number of data copies on read performance and how to overcome the degradation; 
(3) Different  heuristics  to  mitigate  SWP - single write penalty.   
The framework has been built and submitted into the Apache Hadoop project.
\footnote{\url{https://hadoop.apache.org/}}

The {\it Quantcast File System - QFS} \cite{ORR+13} is a plugin compatible alternative to HDFS/MapReduce 
with the following efficiency improvements:                                       
\footnote{\url{https://en.wikipedia.org/wiki/Apache_Hadoop}}
(1) 50\% disk space savings through erasure coding instead of replication,        
(2) doubling of write throughput,                                                 
(3) a faster name node,                                                           
(4) support for faster sorting and logging through a concurrent append feature,   
(5) a native command line client much faster than Hadoop file system,             
(6) global feedback-directed I/O device management. 

\section{Clustered RAID5}\label{sec:CRAID}
\vspace{3mm}

{\it Clustered RAID - CRAID} reduces RAID5 load increase in degraded mode
by setting the {\it Parity Group - PG} size ($G$) to be smaller than stripe width,
which usually equals the number of disks ($N$) by Muntz and Lui 1990 \cite{MuLu90}. 
The load increase results in increased disk utilization and response times and possibly overload.

The read load increase is given by the declustering ratio $\alpha=(G-1)/(N-1) < 1$,
since each access to the failed disk results in accesses to surviving disks in the PG.
CRAID will result in a reduction of response time in degraded mode $R_n^{F/J}(\rho)$ due to a reduced $\rho$ 
and that $n = G-1 < N-1$-way F/J requests are required (refer to Section \ref{sec:degradedanal}).


The disk load increase in degraded mode RAID5 results in increased rebuild time, 
since rebuild is carried out when the disk is not processing external requests
and this load increase is reduced by CRAID. 
Another advantage of CRAID is parallelism in rebuild reading which potentially accelerates this process.
A disadvantage of CRAID is that a dedicated spare disk may become a bottleneck when disk utilizations are low.
Buffer overflow may occur since blocks are reconstructed in the DAC buffer 
at a faster rate than they can be written to the spare disk.
The utilization of the spare disk due to rebuild writes is increased due to read redirection.
The fraction of redirected reads should be decreased to allow a higher rate of rebuild writes.
Updating of materialized blocks in the spare disk (write redirection) is necessary to keep disk data up-to-date.
Distributed sparing is a possible solution.

Another advantage of CRAID is that it reduces the cost of the
{\it ReConstruct Write - RCW} method for large writes Thomasian 2005 \cite{Thom05}.
If $n > G/2$ strips are updated then it is more efficient 
to read the remaining $G - n$ strips to compute the parity strip by XORing them. 
Linux's RAID5 likewise employs a simple majority rule to select a strategy for writing,
i.e., if a majority of pages for a stripe are dirty then {\it Parity Computation - PC} 
and otherwise {\it Parity Increment - PI}  is chosen.

Six properties which are not all attainable 
for ideal layouts for CRAID in Holland et al. 1994 \cite{HoGS94} are as follows: 
(i) Single failure correcting by mapping strips in the same parity group to different disks.    
(ii) Balanced load due to parity, all disks have the same number of parity strips.                            
(iii) Balanced load in failed mode, so that reconstruction workload i evenly distributed across all disks.          
(iv) Large write optimization, each stripe should contain $N-1$ contiguous strips, where $N$ is the parity group size.  
(v) Maximal read parallelism is attained, i.e., the reading of $n \leq N $ disk blocks entails in accessing $n$ disks.  
(vi) Efficient mapping, the function that maps physical to logical addresses is easily computable.


The {\it Permutation Development Data Layout - PDDL} 
is a mapping function described in Schwarz et al. 1999 \cite{ScSB99}.
It has excellent properties and good performance 
both in light and heavy loads like the PRIME Alvarez et al. 1998 \cite{ABSC98}
and the DATUM data layout Alvarez et al. 1997 \cite{AlBC97}, respectively.
This works were conducted at UCSD under the direction of W. Burkhard.
\footnote{\url{https://cseweb.ucsd.edu/~burkhard/}}
In what follows we describe four CRAID organizations.

\subsection{Balanced Incomplete Block Design - BIBD}\label{sec:BIBD}
\vspace{2mm}

The {\it Balanced Incomplete Block Design - BIBD} Hall 1986 \cite{Hall86}
is used for balancing parity mappings at a small cost 
in Ng and Mattson 1994 \cite{NgMa94} and Holland et al. 1994 \cite{HoGS94}.
Four parameters for a declustered BIBD layout according to the latter are as follows:
$n \geq 2$ is the number of disks, $k \leq n$ is the  PG size $G$,
$b$ is the number of PGs, the number of stripes (rows) per disk is $r$.
An additional parameter $L$ which is the number of PGs common to any pair of disks. 
Only three out of five variables are free due to:
$ b k = n r \mbox{  and  } r (k-1) = L ( n-1 ).$
Theorem 5 in Alvarez et al. 1998 \cite{ABSC98} states 
that there is a placement-ideal layout with parameters $n$, $k$, $r$,
{\it if and only if - iff} $k|r$ and $(n-1)|k(r-1)$ and one of the following holds:
(1) $k=n$, (2) $k=n-1$, (3) $k=2$, (4) $k=3$ and $n=5$ (5) $k=4$ and $n=7$.

Fig. \ref{fig:BIBD} is a BIBD layout given in Ng and Mattson \cite{NgMa94} 
with $N=10$, $k=G=4$, and $L=2$, $r= 2 \times 9 /3=6$, and $b= N \times r / G= 10 \times 6 / 4 = 15$.
Each set of six consecutive stripes (rows) constitute a segment.

\begin{figure}[t]
\begin{footnotesize}
\begin{center}
\begin{tabular}{|c|*{10}{c}|} \hline
Disk \#     &1   &2   &3   &4   &5   &6   &7   &8    &9     &10  \tabularnewline \hline\hline
            & 1  &1   &2   &3   &1   &2   &3   &1    &2     & 3  \tabularnewline
            & 2  &4   &4   &4   &5   &6   &7   &5    &6     & 7  \tabularnewline
Parity      & 3  &6   &5   &5   &8   &8   &8   &9    &10    &11  \tabularnewline
Groups      & 4  &7   &7   &6   &0   &9   &9   &12   &12    &12  \tabularnewline
            & 8  &9   &10  &11  &11  &11  &10  &14   &13    &13  \tabularnewline
            &12  &13  &14  &15  &13  &14  &15  &15   &15    &14  \tabularnewline \hline
\end{tabular}
\end{center}
\end{footnotesize}
\caption{\label{fig:BIBD}BIBD data layout of a segment with $N=10$ disks and PG size $G=k=4$.}
\end{figure}

The following disk accesses are required to reconstruct the six strips
on failed Disk\#3 specified as strip\#(disk\#1.disk\#2,disk\#3):            
2(1,6,9), 4(1,2,4), 5(4,5,8), 7(2,7,10), 10(5,7,9), 14(6,8,10).  
Two strips are read from each one of surviving disks denoted by D\#i.                       
D\#1(2,4), D\#2(4,7), D\#4(4,5), D\#5(5,10), D\#6(2,14), D\#7(7,10), D\#8(5,14), D\#9(2,10), D\#10(7,14).
The on-demand reconstruction load is balanced and there is a 3-fold increase in rebuild bandwidth.

One of the strips, say the first), in all  PGs, may be set consistently to be the parity.
Given $b=15$ PGs and $N=10$ disks there are 1.5 parities per disk on the average.
Given that five disks have one parity and another five two parities per segment,
ten rotations of segments can be balance parity loads with 15 parities
in each column or an average of 1.5 parities per segment.

BIBD parity placement for $N=5$ and $G=4$ is given in Holland et al. 1994 \cite{HoGS94}.
BIBD designs are not available for all $N$ and $G$ values,
e.g., layouts for $N=33$ with $G=12$ do not exist,
but designs with $G=11$ and $G=13$ can be used instead Hall 1986 \cite{Hall86}.

\subsection{Thorp Shuffle CRAID Implementation of CRAID}\label{sec:Thorp}
\vspace{2mm}

The Thorp shuffle is used to implement CRAID in Merchant and Yu \cite{MeYu96}. 
Given examples are restricted to $N/G=2^i, 0 \leq i \leq 3$ with $G=8$.
The Thorp shuffle cuts a deck of cards into two equal half decks A and B and
then the following is repeated until no cards are left:
``Pick deck A or B randomly and drop its bottom card and
then drop the bottom card from the other half deck atop it'' Thorp 1973 \cite{Thor73}.

A shuffle-exchange network where the propagation through the network
is determined by random variables was used for this purpose.
With $N$ disks and $B$ strips per disk the overall time complexity 
of computing block addresses is O($\mbox{log}(B) + \mbox{log}(N)$) according to Holland et al. \cite{HoGS94},
i.e., the Thorp shuffle method has the disadvantage of incurring 
high computational cost per access: 14 $\mu$s on an SGI computer. 
\footnote{\url{https://en.wikipedia.org/wiki/Silicon_Graphics}}
but disk capacities which affect $B$ have increased by several orders of magnitude in the last quarter century, 
which would make the Thorp shuffle more costly to implement,
In addition it seems sufficient load balancing can be attained by the following simpler methods.

\subsection{Nearly Random Permutation - NRP}\label{sec:NRP}
\vspace{2mm}

The {\it Nearly Random Permutation - NRP} in Fu et al. 2004 \cite{FTHN04}
incurs less mapping cost by permuting individual stripes (disk rows).
To build the mapping consecutive PGs of size $G < N$
are placed in the matrix with $N$ columns in as many rows as disk capacity allows,
so that PG$_i$ occupies strips $iG:iG+(G-1)$.
In the case of RAID5 (resp. RAID6) the $P$ (resp. $P$ and $Q$) parities
are consistently placed as the first or last two strips in each PG.

P parities appear on half of the disks as shown in the first two lines as ``Initial allocation'' in Fig. \ref{fig:NRP}.
This imbalance is alleviated by randomizing strip placement,
so that approximately the same number of parity blocks are allocated per disk.
Even parity placement can be attained simply by rotating pairs of rows by one,
which is the approach pursued in Section \ref{sec:shifted}.
Algorithm 235 in Durstenfeld 1964 \cite{Durs64} is used for random permuting.

\begin{quote}
Consider an array $A$ with $N$ elements. Set $n=N$. \newline
L: Pick a random number $k$ between $1$ and $n$.    \newline
If $k \neq n$ then $A_n \leftrightarrow A_k$.       \newline
Set $n=n-1$ and go back to L if $n <2 $.
\end{quote}

A pseudo-{\it Random Number Generator - RNG} yields a random permutation of $\{ 0,1,\dots,N-1 \}$ as:
$\underline{P_I} = \{P_0, P_1, \ldots, P_{N-1} \}$, e.g. for $N=10$, 
${\bf P}= \{ 0, 9, 7, 6, 2, 1, 5, 3, 4, 8 \}$

If $N\;mod(G)=0$ then the permutation is applied once
and otherwise it is repeated for $K = \mbox{GCD}(N,G)/N$ rows.
A permutation obtained using a polynomial function $f(I)$ with $I = j K, j \geq 1$ is used as a seed.
The same permutation is applied to stripes $j K, j K + 1, \ldots, jK +(K-1)$.

For $N=10$ and $G=4$ with $K=\mbox{GCD}(10,4)=2$ the same permutation
should be applied to the two rows in Fig.~\ref{fig:NRP}.
This assures that a PG straddling two rows is mapped onto different disks,
since initially all strips were on different disks
(this is the case for $D_6$, $D_7$, $D_8$, $P_{6:8}$).
Data strips in a PG are assigned to different disks to ensure recovery, 
but this also allows parallel access.

\begin{figure}
\begin{footnotesize}
\begin{center}
\begin{tabular}{|c|*{10}{c}|} \hline
Disk \#      &1       &2         &3        &4          &5        &6          &7         &8          &9        &10 \\\hline \hline
Initial      &$D_1$   &$D_2$     &$D_3$    &$P_{1:3}$  &$D_4$    &$D_5$      &$D_6$     &$P_{4:6}$  &$D_7$    &$D_8$ \\ \hline
allocation   &$D_9$   &$P_{7:9}$ &$D_{10}$ &$D_{11}$   &$D_{12}$ &$P_{10:12}$ &$D_{13}$ &$D_{14}$   &$D_{15}$ &$P_{13:15}$
\\ \hline \hline
Final        &  $D_1$  & $D_5$  &  $D_4$  &  $P_{4:6}$ &  $D_7$  & $D_6$  & $P_{1:3}$ &  $D_3$ &  $D_8$   &  $D_2$ \\
allocation   &  $D_9$  & $P_{10:12}$ &  $D_{12}$ & $D_{14}$ &   $D_{15}$ & $D_{13}$ & $D_{11}$  & $D_{10}$ & $P_{13:15}$ & $P_{7:9}$
\\ \hline
\end{tabular}
\end{center}
\end{footnotesize}
\caption{\label{fig:NRP}The two rows show the preliminary allocation of strips
before they are permuted using NRP ($N=10, G=4, K=2$).
Permuted data blocks are shown in the bottom two lines.}
\end{figure}

\subsection{Shifted Parity Group Placement}\label{sec:shifted}
\vspace{2mm}

When $N$ is divisible by $G$, i.e., $N\mbox{mod}G=0$ the PGs in 
every $G^{th}$ row need be shifted $G-1$ times in the following $G-1$ rows.
If $G$ is coprime with $N$, i.e., the {\it Greatest Common Divisor - GCD} 
$m=\mbox{GCD}(N,G)=1$ then $i G \mbox{mod} N, i \geq 1$ 
generates all values $0:N-1$ in a periodic manner.
The number of entries equals the {\it Least Common Multiplier - LCM}
$L=(N \times G)/\mbox{GCD}(N,G)$, number of rows $K=L/N$, number of parity groups $p=L/G$. 
For $N=10$, $G=3$, $m=1$, $L=30$, $K=3$, $p=10$  
For $N=11$, $G=4$, $m=1$, $L=44$, $K=4$, $p=11$.
For $N=10$, $G=4$, $m=2$, $L=20$, $K=2$  $p=5$,  
For $N=15$, $G=6$, $m=3$, $L=30$, $K=3$  $p=5$   
and $K-1=2$ two shifts are required as shown in Fig. \ref{fig:tab7}.


\begin{figure}[htb]
\begin{scriptsize}
\begin{center}
\begin{tabular}{|c|*{15}{c}|} \hline
Disk \#     &1  &  2  &  3  &  4  &  5 &  6   &   7   &  8   &  9  & 10 & 11 & 12 & 13 & 14 & 15
\tabularnewline \hline\hline
Groups & $D_1$ & $D_2$ & $D_3$ & $D_4$ & $D_5$ & $P_{1:5}$ &
$D_6$ & $D_7$ & $D_8$ & $D_9$ & $D_{10}$ & $P_{6:10}$  & $D_{11}$ & $D_{12}$ & $D_{13}$
\tabularnewline
 0 shift    & $D_{14}$ & $D_{15}$ & $P_{11:15}$  &
$D_{16}$  & $D_{17}$ & $D_{18}$ & $D_{19}$ & $D_{20}$ & $P_{16:20}$ &
$D_{21}$ & $D_{22}$  & $D_{23}$ & $D_{24}$ &  $D_{25}$ & $P_{21:25}$
\tabularnewline \hline
Groups & $D_{27}$ & $D_{28}$ & $D_{29}$ & $D_{30}$ & $P_{26:30}$ & $D_{31}$ &
$D_{32}$ & $D_{33}$ & $D_{34}$ & $D_{35}$ & $P_{31:35}$  & $D_{36}$ &
$D_{37}$ & $D_{38}$ & $D_{39}$
\tabularnewline
 1 shift  &  $D_{40}$ & $P_{36:40}$ & $D_{41}$  & $D_{42}$  & $D_{43}$ & $D_{44}$ & $D_{45}$ &
$P_{41:45}$ & $D_{46}$ & $D_{47}$ & $P_{48}$ & $D_{49}$ & $D_{50}$  & $P_{46:50}$ & $D_{26}$
\tabularnewline \hline
Groups & $D_{53}$ & $D_{54}$ & $D_{55}$ & $P_{51:55}$ & $D_{56}$ & $D_{57}$ &
         $D_{58}$ & $D_{59}$ & $D_{60}$ & $P_{56:60}$ & $D_{61}$ & $D_{62}$ &
         $d_{63}$ & $d_{64}$ & $d_{65}$
\tabularnewline
 2 shifts   & $P_{61:65}$  & $D_{66}$ & $D_{67}$ & $D_{68}$ & $D_{69}$ & $D_{70}$ &  $P_{66:70}$ &
$D_{71}$ & $D_{72}$  & $D_{73}$ & $D_{74}$ &  $D_{75}$ & $P_{71:75}$ & $D_{51}$ & $D_{52}$
\tabularnewline \hline
\end{tabular}
\end{center}
\end{scriptsize}
\caption{\label{fig:tab7}$N=15$ and $G=6$, $K=GCM(15,6)=3$, so that two shifts are required.}
\end{figure}

\section{Flash Solid State Drives - SSDs}\label{sec:flash}
\vspace{3mm}

NAND Flash SSDs introduced by Toshiba in 1984 provides 
low latency, high bandwidth, are nonvolatile, and consume less power than HDDs.
Due to higher bandwidth Flash SSDs sustain the same workload with fewer drives. 
Flash SSDs are replacing HDDs in some areas such as PCs as their prices are dropping.
NAND flash memory technology is covered in Micheloni et al. 2010 \cite{MiCM10}.
\footnote{\url{https://en.wikipedia.org/wiki/Flash_memory}}
Flash storage is discussed in Section 2.3.20 in Thomasian 2021 \cite{Thom21}.


Flash SSDs are classified according to cell type or number of bits per cell: 
SLC (single), MLC (multiple-2), TLC (triple), QLC (quad), PLC (penta). 
There is a rapid drop in {\it Program/Erase - P/E} cycles 
\footnote{\url{https://www.techtarget.com/searchstorage/definition/P-E-cycle}}
as the cell size is decreased from 5 to 1 nanometer, 
see Table \ref{tab:penta} summarizing Fig. 1 in Jaffer et al. 2022 \cite{JaMS22}). 
\footnote{\url{https://blocksandfiles.com/2019/08/07/penta-level-cell-flash/}}

\begin{table}[h]
\centering
\begin{footnotesize}
\begin{tabular}{|c|c|c|c|c|c|}\hline
Cell type   &SLC(1b)  &MLC(2b)  &TLC(3b)  & QLC(4b) &PLC(5b) \\ \hline \hline 
5Xnm        &11,00    &10,000   &2,500    & 800     & 400         \\ \hline
3Xnm        &10,000   &5,000    &1,250    & 350     & 175         \\ \hline
2Xnm        & 7,500   &3,000    & 750     & 150     &  75         \\ \hline
1Xnm        &5,000    &1,500    & 500     & 70      &  35         \\ \hline
\end{tabular}
\end{footnotesize}
\caption{P/E cycles for varying number of bits-per-cell as cell size is reduced.
PLC numbers are estimated.\label{tab:penta}}
\end{table}
\begin{table}
\centering
\begin{footnotesize}
\begin{tabular}{|c|c|c|c|} \hline
RAID             & Capacity      & Seq'l Write        & Rebuild time      \\
                 & TB            & MB/sec             & Minutes           \\  \hline \hline
Disk-1/2/3       & 0.72/1/4      & 80/115/460         & 15/145/580        \\ \hline
FlashMax III     & 2.2           &1,400               & 26                \\ \hline
Intel D3600      & 2             &1,500               & 22                \\ \hline
Micron 9100      & 3.2           &1,500               & 27                \\ \hline
Intel DC 3608    & 4             &3,000               & 22                \\ \hline
\end{tabular}
\end{footnotesize}
\caption{\label{tab:SSDrebuild}Minimum rebuild times for disks and flash SSDs.
Capacity in TB times 1000 divided by Seq'l write in MB/s yields rebuild time in seconds.}
\end{table}


Performance-wise SSDs outperform HDDs.
Huawei OceanStor  Dorado 18000 V6, set a peak for SPC-1 Benchmark performance in Oct. 2020, 
achieving 21,002,561 IOPS according to {\it Storage Performance Council - SPC}: 
\footnote{\url{https://e.huawei.com/en/products/storage/all-flash-storage/dorado-8000-18000-v6}}
HDDs led in price-performance with the SPC-2 benchmark in March 2017.  
\footnote{\url{https//www.spcresults.org}}
One hundred 1, 2, 4 TB SSDs from the viewpoint of Random IOPS, Seq MB/s, Copy MB/s, and Avg power are compared here.
\footnote{\url{https://www.tomshardware.com/features/ssd-benchmarks-hierarchy}}

Rebuild times for HDDs and SSDs are given in Table \ref{tab:SSDrebuild}.
\footnote{\url{https://www.theregister.com/2016/05/13/disak_versus_ssd_raid_rebuild_times/}}
The one and four TB HDDs differ in the number of platters, 
Seagate 18 TB 554 MB/s takes about nine hours to read.      
which is a consequence of the fact that disk transfer rates have not kept up with disk capacities.
It can be seen that rebuild time is much less of a problem with SSDs. 

In spite of ``flash wear'' SSDs are much more reliable than HDDs.
Counting the AFRs - Annual Failure Rates  of Backblaze's drives;      
HDDs had a 10.56\% AFR, while SSDs AFR was just 0.58\%.   
\footnote{\url{https://www.backblaze.com/blog/backblaze-hard-drive-stats-q1-2021/} }
This makes higher levels of redundancy less of a consideration for SSDs.  
Feature-based SSD failure prediction is a proactive fault-tolerance technique, 
which requires no redundancy given that SSDs are more reliable than HDDs Zhang et al. 2023\cite{ZHN+23}. 
SSDs are inherently and physically much more resilient than traditional hard disks. 
the fact that SSDs lack HDD moving and mechanical components makes them insensitive to shocks and vibrations.

Techniques for making effective use of flash for database applications 
in Athanassoulis et al 2010 \cite{AAC+10} are as follows: 
(i) as a log device for transaction processing, 
(ii) as the main data store for transaction processing,  
(iii) as an update cache for HDD-resident data warehouses.

Indexing structures support on flash SSDs is discussed in Fevgas et al. 2020 \cite{Fev+20}.
Not taking into account idiosyncrasies, like erase-before-write, 
wear-out and asymmetric read/write may lead to poor performance. 
Reinvention of indexing techniques designed primarily for HDDs is required.
MySQL DB performance has been compared on SSDs and HDDs.
\footnote{\url{https://download.semiconductor.samsung.com/resources/white-paper/best-practices-for-mysql-with-ssds.pdf}}

HDDs still provide less costly storage per GB,
but NAND Flash technology is progressing rapidly yielding higher capacity chips at lower cost. 
\footnote{\url{https://www.intel.com/content/www/us/en/products/docs/memory-storage/solid-state-drives/ssd-vs-hdd.html}}   
The disk/flash crossover is discussed here.                        
\footnote{\url{https://blocksandfiles.com/2023/02/15/pure-the-disk-flash-crossover-event-is-closer-than-you-think/}} 
PureStorage projected in 2023 that no disk drives will be sold after 2028.   
\footnote{\url{https://blocksandfiles.com/2023/05/09/pure-no-more-hard-drives-2028/}}

Data updates in SSDs are carried out by writing a new copy, as in LFS, 
rather than overwriting old data, marking prior copies of data invalidated.
Writes are performed in units of pages, similarly to disks, even if data to be written is smaller.
Space is reclaimed in units of multipages by erasing,
which necessitates copying of any remaining valid pages in the block before reclamation.

The efficiency of the cleaning process greatly affects performance under random workloads;
{\it Write Amplification factor - WAF} defined as the ratio of 
the amount of data an SSD controller writes in relation to the amount of data 
that the host's flash controller writes reduces application throughput.

Log-Structured File System for Infinite Partition is a scheme
eliminates garbage collection in flash SSDs Kim et al 2022 \cite{Kim+22}.
The scheme separates {\it Logical Partition Size - LPS} from the physical storage size 
and the LPS is large enough so that there is no lack of free segments during SSD's lifespan, 
allowing the filesystem to write the updates in append-only fashion without reclaiming the invalid filesystem blocks. 
The metadata structure of the baseline filesystem can efficiently handle the storage partition with $2^{64}$ sectors. 
Interval mapping minimizes the memory requirement for the {\it Logical Block Address - LBA} to
{\it Physical Block Address -PBA} translation in {\it Flash Translation Layer - FTL}.
FTL's role is to emulate a block-type peripheral such as HDDs.
and is a hardware and software layer is part of flash memory controllers.   
\footnote{\url{https://en.wikipedia.org/wiki/Flash_memory_controller}}

The effect of Flash memories on transaction performance 
has been investigated in simulation  and measurement studies.
NoFTL allows for native Flash  access  and integrates parts  of  the FTL  functionality
into the DBMS yielding significant performance increase 
and simplified I/O stack Hardock et al. 2013 \cite{Har+13}.
A Flash emulator integrated with a DBMS (Shore-MT) demonstrate 
a 3.7-fold performance improvement under various TPC workloads.

Performance evaluation of the write operation in flash-based SSDs 
based on greedy garbage collection is reported in Bux and Iliadis 2010 \cite{BuIl10}.
Scheduling and performance modeling and optimization is discussed by Bux et al. 2012 \cite{BHIH12}.
Write performance in flash SSDs is analyzed in Desnoyers 2014 \cite{Desn14}
where a nearly-exact closed-form solution for write amplification under greedy cleaning 
for uniformly-distributed random traffic is validated using simulation.

Flash SSD storage is available from PureStorage, Dell/EMC, NetApp, Fungible, Oracle (Exadata), etc.
Flash SSDs outperform HDDs in power consumption 3-fold (2-5 watts versus 6-15 watts).
From a TCO viewpoint SSDs will be preferred when they are less than 5-fold expensive than HDDs per GB. 
\footnote{\url{https://www.intelice.com/ssd-or-hdd-hard-drives/}}

{\it Distributed SSD - DSSD} Flash coding is based on Grid files in Section \ref{sec:gridfiles} Bonwick 2014 \cite{Bonw16}.
An obvious problem with DDSD is that check strips chips 
will be updated more often than data chips and susceptible to earlier failures.
DSSD was acquired by EMC in 2014, but its product stopped shipping after EMC joined Dell in 2017.
\footnote{\url{https://www.datacenterknowledge.com/archives/2014/05/07/emc-acquires-flash-storage-startup-dssd}}

The {\it Multi-View and Multi-Task Random Forest - MVTRF} scheme 
predicts SSD failures based on features extracted from 
both long and short-term monitoring of SSD data Zhang et al. 2023 \cite{ZHN+23}. 
MVTRF can simultaneously predict the type of failure and when it will occur.
At Tencent data centers it showed high failure prediction accuracy: 
improved precision by 46.1\% and recall by 57.4\% on average compared with the existing schemes. 
\footnote{\url{https://en.wikipedia.org/wiki/Precision_and_recall}}

\subsection{Hadoop Adaptively-Coded Distributed File System - HACDFS}\label{sec:HACDFS}
\vspace{2mm}

HACDFS is based on {\it Product Code - PC}
which was used by Grid files described in Section \ref{sec:gridfiles}
It uses two codes: a fast code with low recovery cost 
and a compact code with low storage overhead Xia et al. \cite{XSBP15}.
It exploits the data access skew observed in Hadoop workloads 
to decide on the initial encoding of data blocks.
HACDFS uses the fast code to encode a small fraction of the frequently accessed data
and provides overall low recovery cost for the system.
The compact code encodes the majority of less frequently accessed data blocks with a low storage overhead.

After an initial encoding, e.g., with the compact code,
HACDFS dynamically adapts to workload changes by using upcoding and downcoding 
to convert data blocks between the fast and compact codes.
Blocks initially encoded with fast code are upcoded into compact code 
enabling HACDFS system to reduce the storage overhead. 
Downcoding data blocks from compact code to fast code representation lowers 
the overall recovery cost of the HACDFS system. 
The upcode and downcode operations are efficient and only update the associated parity blocks.
HACDFS exploits the data access skew observed in Hadoop workloads.    
\footnote{\url{http://blog.cloudera.com/blog/2012/09/what-do-real-life-hadoop-workloads-look-like/}}

\begin{figure}[h]
\begin{footnotesize}
\begin{center}
\begin{tabular}{|c|c|c|c|c|c|}  \hline
D   &D  &D  &D  &D &P   \\ \hline
D   &D  &D  &D  &D &P   \\ \hline
P   &P  &P  &P  &P &P   \\ \hline
\end{tabular}
\hspace{5mm}
\begin{tabular}{|c|c|c|c|c|c|} \hline
D   &D  &D  &D &D &P   \\ \hline
D   &D  &D  &D &D &P   \\ \hline
P   &P  &P  &P &P &P   \\ \hline
\end{tabular}
\hspace{5mm}
\begin{tabular}{|c|c|c|c|c|c|}  \hline
D   &D  &D  &D &D &P   \\ \hline
D   &D  &D  &D &D &P   \\ \hline
P   &P  &P  &P &P &P   \\ \hline
\end{tabular}
\end{center}
\begin{center}
\begin{tabular}{|c|c|c|c|c|c|}    \hline
D   &D  &D  &D &D &P          \\ \hline
D   &D  &D  &D &D &P          \\ \hline
D   &D  &D  &D &D &P          \\ \hline
D   &D  &D  &D &D &P          \\ \hline
D   &D  &D  &D &D &P          \\ \hline
D   &D  &D  &D &D &P          \\ \hline
P   &P  &P  &P &P &P          \\ \hline
\end{tabular}
\end{center}
\end{footnotesize}
\caption{\label{fig:xia15}Parity encodings to attain reliability in Hadoop.
Upcoding (2x5) arrays to (6x5), results in a reduced redundancy from 18/10=1.8 to 42/30=1.4.
Horizontal parity codes are simply copied,
while the vertical parity bits are the XOR of the parities in the high density code.}
\end{figure}

\begin{figure}[h]
\begin{center}
\begin{footnotesize}
\begin{tabular}{|c|c|c|c|c|c|c|}  \hline
D   &D  &D  &D  &D &D   &G   \\ \hline
D   &D  &D  &D  &D &D   &G   \\ \hline
L   &L  &L  &L  &L &L   &-  \\ \hline
\end{tabular}
\hspace{5mm}
\begin{tabular}{|c|c|c|c|c|c|c|} \hline
D   &D  &D  &D &D &D  &G  \\ \hline
D   &D  &D  &D &D &D  &G  \\ \hline
-   &-  &L  &- &- &L  &-  \\ \hline
\end{tabular}
\end{footnotesize}
\caption{\label{fig:LRCode}LRC code Upcode and Downcode operations.
LRC(12,6.2) on the left, the local L codes for LRC(12,2,2) are XORs of three L codes.}
\end{center}
\end{figure}

\begin{table}
\begin{center}
\begin{footnotesize}
\begin{tabular}{|c||c|c||c|c|}\hline
-     &PC$_{fast}$          &PC$_{compact}$         &LRC$_{fast}$          &LRC$_{compact}$       \\ \hline \hline
DRC  &2                     &5                      &2                     &6                     \\ \hline
RC   &2                     &5                      &3.25                  &6.75                  \\ \hline      
SO   &1.8x                  &1.4x                   &1.66x                 &1.33x                 \\ \hline    
MTTF &$1.4 \times 10^{12}$  &$2.1 \times 10^{11}$   &$6.1 \times 10^{11}$  &$8.9 \times 10^{10}$  \\ \hline
\end{tabular}
\end{footnotesize}
\caption{\label{tab:PCvsLRC}Fast and Compact PC and LRC codes 
(DRC: Degraded Read Cost, RC: Reconstruction Cost, SO: Storage Overhead).}
\end{center}
\end{table}

Fast and compact codes with PC and LRC shown in Fig. \ref{fig:LRCode}
are compared in Table \ref{tab:PCvsLRC} and PC is shown to be superior. 

\subsection{Fast Array of Wimpy Nodes - FAWN}\label{sec:FAWN}
\vspace{2mm}

FAWN was a project at CMU's {\it Parallel Data Laboratory -PDL} 
combined low-power nodes with flash cluster storage providing fast
and energy-efficient processing of random queries in a {\it Key Value - KV} store Andersen et al. 2022 \cite{AFK+11}.
\footnote{\url{https://www.kdnuggets.com/2021/04/nosql-explained-understanding-key-value-databases.html}} 
FAWN-KV begins with a log-structured per-node datastore to serialize writes and make them fast on flash.
KV-stores store retrieves and manages associative arrays using a dictionary or hash table.

FAWN uses replication between cluster nodes to provide reliability and strong consistency.
The FAWN prototype with 1000 queries/Joule (watts $\times$ second)                
demonstrated a significant potential for I/O-intensive workloads
In 2011 4-year-old FAWN nodes delivered over an order of magnitude more queries per Joule than disks.

\subsection{Differential RAID for SSDs}\label{sec:diffRAID}
\vspace{2mm}

Diff-RAID design takes into account the fact that flash memories
have very different failure characteristics from HDDs,
i.e., the {\it Bit Error Rate - BER} of an SSD increases with more writes Balakrishnan et al. 2010 \cite{BKPM10}.

By balancing writes evenly across the array, RAID schemes wear together at similar rates, 
making all devices susceptible to data loss at the same time.
Diff-RAID reshuffles the parity distribution on each drive replacement
to maintain an age differential when old devices are replaced by new ones.

Diff-RAID distributes parity blocks unevenly masking the BER higher on aging SSDs to:   
(1) Retain the low overhead of RAID5.                               
(2) Extend the lifetime of commodity SSDs.   
(3) Alleviate the need for expensive error correction hardware.


For a  workload consisting  only  of  random  writes,  
the relative aging rates of devices for a given parity assignment. 
Let $a_{i,j}$ represent the ratio of the aging rate of $i^{th}$ device to that of $j^{th}$ device, 
and $p_i$ and $p_j$ the percentages of parity allotted to the respective devices, then:

\vspace{-2mm}
\begin{eqnarray} 
a_{i,j}= \frac{p_i (n-1) + (100-p_i)}{p_j(n-1) + (100-p_j)}.
\end{eqnarray}

Consider $n=4$ SSDs and 50\% of the parity at the first device
and the rest evenly distributed over three disks, so that $(70,10,10,10)$ 
Using the above formula the aging rate of the first device is twice that of the others.
\vspace{-1mm}
$$(70 \times  3 + 100-70) / (10 \times 3 + 100 - 10) = 240/120=2   $$                                    
After numerous replacements the ages of remaining devices 
at replacement time converge to $(5750, 4312.5, 2875, 1437.5)$.
The implementation of device replacement can be accomplished via rebuild processing,
while maintaining the same parity layout.

A simulator to evaluate Diff-RAID's reliability by using BERs from 12 flash chips
showed that it is more reliable than RAID5 by orders of magnitude.

\subsection{Distributed DRAM-based Storage - RAMCloud}\label{sec:RAMCloud}
\vspace{2mm}

The crossover point at which the more costly DRAM will replace disks is explored in Gibson 1992 \cite{Gibs92}.
Plotted in 1992 Graph 2.1(a) and 2.1(b) give 2010 and 2020 as optimistic and pessimistic crossover points.
Neither projection has materialized.

RAMCloud is a distributed storage system keeping all data in DRAM 
offering exceptionally low latency for remote accesses,
e.g., small reads completed in less than 10 $\mu$ second in a 100,000 node datacenter Ousterhout et al. 2015 \cite{Oust15}.
Memory prices per GB have been falling over the years.
\footnote{\url{https://aiimpacts.org/trends-in-dram-price-per-gigabyte/}}
RAMCloud's 1 PB or more storage supports Web applications, via a single coherent key-value store.
an associative array where each key is associated with one and only one value in a collection.
RAMCloud maintains backup copies of data on HDDs.

\subsection{Write-Once Memory - WOM Codes to Enhance SSD Lifetimes}\label{sec:WOM}
\vspace{2mm}

Increased storage density is achieved with higher bits per cell, 
but this is accompanied by order of magnitude lower P/E cycles,
which decreases the number of times the SSD can be rewritten and hence their lifetime. 

WOM codes are one way of improving drive lifetime.  
i.e., rewrite on top of pre-existing data without erasing previous data. 
WOM codes alter the logical data before it is physically written, 
thus allowing the reuse of cells for multiple writes. 
On every consecutive write, zeroes may be overwritten with ones, 
but not vice-versa Yaakobi et al. 2012 \cite{Yaa+12}.
A WOM(x,y) code encodes a code word of x bits into a code word of y bits

WOM increases the total logical data 
that can be written on the physical medium before requiring an erase operation.  
Traditional WOM codes are not scalable and only offer up to 50\% increase 
in total writable logical data between any two erase operations. 
There are simple and highly efficient family of generic WOM codes that could be applied to any N-Level cell drive. 
The focus of Jaffer et al. \cite{JaMS22} is QLC drives.
The application of coding at various levels is given in Table \ref{tab:WOM},
where D=data, V=voltage level, G-generation, shared states 2 generations.

Trace-driven simulation shows that WOM-v codes can reduce erase cycles 4.4-11.1-fold with minimal performance overheads.
The increase in the total logical writable data before an erase is 50-375\%. 
The total logical writable data between two erase operations may be increased by up to 500\% by the choice of internal ECC.

\begin{table}[t]
\parbox{0.25\linewidth}{
\centering
\begin{tiny}
\begin{tabular}{|c|c|c|}                                 \hline
v(3,4)                  &v(2,4)            &v(1,4)      \\  \hline \hline
D111,V15,G2             &D11,V15,G5        &D1,V15,-    \\ \hline
D110,V14,G2             &D10,V14,G5        &D0,V14,G14  \\ \hline
D101,V13,G2             &D01,V13,G5        &D1,V13,G13  \\ \hline          
D100,V12,G2             &D00,V12,G5,G4     &D0,V12,G12  \\ \hline
D011,V11,G2             &D11,V11,G4        &D1,V11,G11  \\ \hline
D010,V10,G2             &D10,V10,G4        &D0,V10,G10  \\ \hline
D001,V09,G2             &D01,V09,G4,G3     &D1,V09.G09  \\ \hline
D000,V08,G2             &D00,V08,G3        &D0,V08,G08  \\ \hline
D111,V07,G1,G2          &D11.V07,G3        &D1,V07,G07  \\ \hline 
D110,V06,G1             &D10,V06,G3,G2     &D0,V06,G06  \\ \hline
D101,V05,G1             &D01,V05,G2        &D1,V05,G05  \\ \hline
D100,V04,G1             &D00,V04,G2        &D0,V04,G4  \\ \hline
D011,V03,G1             &D11,V03,G2,G1     &D1,V03,G3  \\ \hline
D010,V02,G1             &D10,V02,G1        &D0,V03,G2  \\ \hline
D001,V01,G1             &D01,V01,G1        &D1,V02,G1   \\ \hline
D000,V00,G0             &D00,V00,G0        &D0,V01.-     \\ \hline
\end{tabular}
\end{tiny}
\caption{Voltage-based codes for encoding 3,2,1 bits into 16 levels.
\label{tab:WOM}}
}
\hspace{8mm}
\parbox{0.475\linewidth}{
\centering
\begin{footnotesize}
\begin{center}
\begin{tabular}{|c|c|c|}\hline
Nanosecond events       &Microsecond events                    &Millisecond events    \\ \hline \hline             
register file: 1-5 ns   & datacenter network O(1 $\mu$s)      &disk O(10) ms \\ \hline
cache accesses:4-30 ns  &new NVM memory O(1 $\mu$s)            &low-end-flash O(1) ms   \\ \hline
memory access 100 ns    &high end flash O(10 $\mu$s)           &wide area networking O(10) ms \\ 
                        &GPU/accelerator O(10 $\mu$s)          &                              \\ \hline
\end{tabular}
\end{center}
\end{footnotesize}
\caption{\label{tab:times}Events and their latencies.} 
}
\end{table}

Intel Optane SSD provides 1.2 GB/s with a latency per read-I/O of about 10 $\mu$second,
while 3D Xpoint based on {\it Phase-Change Memory - PCM} latency is one $\mu$second,
\footnote{\url{https://en.wikipedia.org/wiki/Phase-change_memory}}
The access time for {\it Dynamic RAM - DRAM} used in main memories is 60-100 ns,
\footnote{\url{https://en.wikipedia.org/wiki/Dynamic_random-access_memory}}
while the access time for {\it Static RAM - SRAM} used by CPU caches is 10 ns.
\footnote{\url{https://en.wikipedia.org/wiki/Static_random-access_memory}}

\subsection{Redundant Array of Independent Zones - RAIZN}\label{sec:RAIZN}
\vspace{2mm}

Zoned NameSpace (ZNS) SSDs is a host-managed flash storage, 
enabling improved performance than traditional block interface SSDs referred to as mdraid.
RAIZN provides more stable throughput and lower tail latencies than mdraid arrays at a lower cost-per-byte. 
\footnote{\url{https://linux.die.net/man/8/mdadm}}
RAIZN logical volume manager has a ZNS interface 
and stripes data and parity across ZNS SSDs Kim et al. 2021 \cite{KJA+23}.
RAIZN achieves superior performance because device-level garbage collection slows down conventional SSDs. 
RAIZN benefits translate to higher layers in the case of F2FS file system, RocksDB key-value store, and MySQL database.
Miscellaneous database systems are reviewed in Chapter 9 in Thomasian 2021 \cite{Thom21}.

RAIZN achieves 14-fold higher throughput and lower tail latency.
In RAIZN {\it Time to Repair - TTR} scales with the amount of data rebuild, 
and is bottlenecked by the SSD write throughput.
mdraid always rebuilds the entire address space,
resulting in the same TTR regardless of the amount of valid data present on the array

\subsection{NVMe-SSDs and Predictable Microsecond Level Support for Flash}\label{sec:predictable}
\vspace{2mm}

Barroso et al. 2017 \cite{BMPR17} discuss the issue of 
lack of support for microsecond ($\mu$) scale events. 
It is argued that disk accesses are in milliseconds (ms) and the CPU accesses its registers, cache 
and even main memory in nanoseconds (ns) (refer to Table~\ref{tab:times}).
Flash memory access times which are in $\mu$s are not adequately supported by the interface.
Number of instructions between I/O events is as follows
with data in cases 2 and 3 is based on trace analysis for unavailable new memories. 
(1) Flash: 225K instruction=O(100$\mu$s;                  
(2) Fast Flash: 20K instruction=O(10$\mu)$s;             
(3) New NVM memory: 2K instructions =O(1$\mu)$;         
(4) DRAM: 500 instructions = O(100ns-1$\mu$s).         

SSD performance is inherently non-deterministic due to the internal management activities 
such as the garbage collection, wear leveling, and internal buffer flush.
\footnote{\url{https://en.wikipedia.org/wiki/Wear_leveling}}
I/O Determinism interface is a host/SSD co-designed flash array with predictable latency, 
which does not sacrifice aggregate bandwidth Li et al. 2023 \cite{LPS+23}.
Minimal changes to the NVMe interface and flash firmware was required to achieve near ideal latencies. 
\footnote{\begin{scriptsize}\url{https://en.wikipedia.org/wiki/NVM_Express}\end{scriptsize} }

\subsection{Computational Storage Drives - CSD}\label{sec:CSD}
\vspace{2mm}

Active Disks projects at the turn of the century took advantage of processing power 
on individual disk drives to run database, data mining, and multimedia 
applications as discussed in Section 8.14 in Thomasian 2021 \cite{Thom21}.
Active disks dramatically reduce data traffic and take advantage of the storage parallelism, 
since there are many disks Riedel et al. 1998 \cite{RiGF98}. 
A scheme supporting a Data Mining workload on an OLTP system almost for free
is possible via so-called freeblocks scheduling Lumb et al. 2000 \cite{LSG+00} with a small impact on the throughput 
and response time of the existing workload Riedel et al. 2000 \cite{RFGN00}.

Modern CSDs can efficiently deals with high volumes of data for AI applications.
{\it Field Programmable Gate Arrays - FPGAs} are used in some cases in preference to CPUs.
Table 5 in Do et al. 2020 \cite{DFB+20} lists five research and four commercial products of 
which Newport developed at \url{NGD.com} is the only one with a Linux OS and a file system.
Newport hardware is given in Fig. 1 and Fig. 5 shows its connection to a host running Linux.
Major applications of Newport are {\it Nearest Neighbor -NN} search and tracking objects in videos.
Results for {\it Batch Similarity Search - BSS} in Fig. 12 specify 
{\it queries per second - qps} versus energy consumption in {\it kiloJoules - kJ}.
Experiments with four configurations yield:
(1) host-only (1 SSD) 120 kJ, 2.00 qps.
(2) host only (8 SSDs) 110 kJ, 2.25 qps.
(3) hybrid (8 SSDs) 80 kJ, 5.52 qps.
(4) hybrid (24 SSDs), 45 KJ, 9.64 qps.  

{\it You Only Look Once - YOLO} is a real-time object detection software Redmon 2016 \cite{RDGF16} 
versions of which were compared with others running  a GPU., CPU, and CSD 
are given in Table \ref{tab:tab2}, whch is based in Table 2 in the paper.
{\it Intersection over Union - IoU} is the key metric for quality of object detection.
\footnote{\url{https://www.superannotate.com/blog/intersection-over-union-for-object-detection}}

\begin{table}[t]
\begin{center}
\begin{footnotesize}
\begin{tabular}{|c|c|c|c|c|}\hline
Computing          &Object          &Overall       &Avg-per video   &Energy usage      \\    
resource           &tracker         &throughput    &throughput      &Joules/frame      \\ \hline \hline
GPU                &YOLOv3          &34.79         &34.66             &6.03            \\ \hline
CPU                &YOLO-lite       &2.72          &2.76              &14.26            \\ \hline
CSD                &YOLO-lite       &2.36          &0.23              &2.00             \\ \hline
\end{tabular}
\end{footnotesize}
\caption{Throughput and energy consumption of versions of YOLO object tracker running on different platforms
FPS=Frames per Second \label{tab:tab2}}
\end{center}
\end{table}

Energy consumption (Joules/frame and accuracy of executing object trackers (IoU)
on different computing resources is given in Fig. 13.
The results for YOLO on GPU, CPU and CSD are as follows:
(15.26 J/frame, 0.13), (6.03 J/frame, 0.35), (2.00 J/frame 0.13), respectively.

{\it Storage Networking Industry Association - SNIA} maintains a library on computational storage.
\footnote{\url{https://www.snia.org/education/what-is-computational-storage}}

\section{Interconnection Networks}\label{sec:IN}
\vspace{2mm}

{\it Direct Access Storage - DAS} via I/O channels was an early connectivity paradigm in dominant IBM mainframes.
\footnote{\url{\https://en.wikipedia.org/wiki/Channel_I/O}}
A simplified analysis of delays incurred by circuit-switched I/O channels 
and related studies appears in Chapter 10 of Lazowska et al. 1984 \cite{LZGS84}.
{\it Rotational Position Sensing - RPS} made available with IBM 3330 drives
disconnects channel and controller during disk latency phase for improved performance.
An RPS miss is failure to reconnect possibly using a different path 
result in extra rotation Rafii 1976 \cite{Rafi76}

{\it Storage Area Network - SAN} typically use Fibre Channel connectivity, 
\footnote{\url{https://en.wikipedia.org/wiki/Fibre_Channel}}
while {\it Network Attached Storage - NAS} use standard Ethernet connection. 
\footnote{\begin{small}\url{https://en.wikipedia.org/wiki/Ethernet}\end{small} }
SANs store data at the block level, while NAS at file level. 
\footnote{\url{https://en.wikipedia.org/wiki/Network-attached_storage}}   

Fibre Channel connection - FICON provide high-speed data transfer between systems and storage devices. 
Fibre Channel networks consist of servers, storage controllers, and storage devices as end nodes, 
which are interconnected by Fibre Channel switches, directors, and hubs.
FICON is available from IBM, Dell/EMC and Hitachi Schulz 2004 \cite{Schu04}.
\footnote{\url{https://www.computerweekly.com/feature/Mainframe-storage-Three-players-in-a-market-thats-here-to-stay}}
The I/O organization of latest (z16) mainframes which are based in Telum processor are described in Chapter 4.
\footnote{\url{https://www.redbooks.ibm.com/redbooks/pdfs/sg248951.pdf}}
The PCIe Gen. 3 uses 128b/130b encoding for data transmission reducing encoding overhead 
to approximately 1.54\% versus the PCIe Gen. 2 overhead of 20\% that uses 8b/10b encoding.
\footnote{\url{https://en.wikipedia.org/wiki/64b/66b_encoding}}

Disaggregated storage allows storage devices to function as a storage pool,
where devices can be allocated to any server on the network.
\footnote{\url{https://en.wikipedia.org/wiki/Disaggregated_storage}}
Fungible Inc. founded in 2015 and acquired by Microsoft in 2023 
\footnote{\url{https://en.wikipedia.org/wiki/Fungible_Inc.}} 
developed {\it Data Processing Units - DPUs} enabling composable disaggregated infrastructure. 
\footnote{\url{https://blogs.nvidia.com/blog/whats-a-dpu-data-processing-unit/}}

There is a significant increase in bandwidths used at hyperscaler data storage centers, e.g., 100 Gb/s. 
\footnote{\url{https://www.nextplatform.com/2019/08/18/the-future-of-networks-depends-on-hyperscalers-and-big-clouds/}}
Interconnection network reliability and performance are discussed in 
Goyal and Rajkumar 2020 \cite{GoRa20} and Rojas-Cessa 2017 \cite{Roja17} 
(Part III - Data Center Networks - DCNs), respectively.
Jupiter DCN may scale to more than 30K servers with 40 Gb/s per-server, 
supporting more than 1 Pb/sec of aggregate bandwidth.
\footnote{\url{https://cloud.google.com/blog/topics/systems/the-evolution-of-googles-jupiter-data-center-network}}

{\it Non-Volatile Memory Express - NVMe} is an optimized high performance,
software standard for accessing NVM over {\it Peripheral Component Interconnect Express - PCIe}. 
\footnote{\url{https://nvmexpress.org}}
NVMe-SSDs which are PCIe-based are superior to SATA and SAS SSDs in terms of bandwidth and latencies
sustaining up to one million IOPS and a 90 $\mu$second latency.
The {\it NVMe-over-Fabrics - NVMe-oF} remote storage protocol reduces 
the remote access overhead to less than 10$\mu$second Guz et al. 2018 \cite{GLSB18}. 

{\it Remote Direct Memory Access - RDMA} 
supports zero-copy networking by enabling the network adapter to transfer data 
from the wire directly to application memory or vice-versa, 
eliminating the need to copy data between application memory and the data buffers in the OS. 
Such transfers require no work to be done by CPUs, etc.
\footnote{\url{https://en.wikipedia.org/wiki/Remote_direct_memory_access}}
Two versions of {\it RDMA over Converged Ethernet - RoCE} are: 
v1 communication between any two hosts,
v2: am internet layer protocol which allows packets to be routed. 
\footnote{\url{https://en.wikipedia.org/wiki/RDMA_over_Converged_Ethernet}}

InfiniBand is a computer networking communications standard used 
in high-performance computing with high throughput and very low latency. 
It is used for data interconnect both among and within computers.
\footnote{\url{https://en.wikipedia.org/wiki/InfiniBand}\\
\url{https://network.nvidia.com/pdf/whitepapers/IB_Intro_WP_190.pdf} }

\section{Conclusions: Cloud Storage}\label{sec:conclusions}
\vspace{3mm}

Cloud Storage allows user to maintain their data in the cloud. 
There are the following considerations: 
{\bf Advantages:}
1. Accessibility via the internet or dedicated lines. 
2. Scalability. 
3. Cost-effective, pay-as you-go.
4. Redundancy ensuring high availability.
5. Security. 
{\bf Disadvantages:}
1. Dependence on connectivity.
2. Privacy concerns (handled by encryption).
3. Limited control. 
4. Risk of data loss.
5. Cost structure.

Teradata Vantage can run on premise and three different public clouds simultaneously.
\footnote{\url{https://assets.teradata.com/resourceCenter/downloads/Datasheets/Teradata-Vantage-on-Google-Cloud-MD006022.pdf}}
The user can create an Ubuntu VM with 4 CPU's and 8GB of RAM, a 70GB balanced disk.
Dedicated private connections or Internet are two connectivity choices. 
Vantage relies on block storage with fixed sized blocks rather than object and file storage.
\footnote{\url{https://quickstarts.teradata.com/vantage.express.gcp.html}}
AWS's {\it Relational Data Services - RDS} allows Oracle's MySQL and PostgreSQL.
\footnote{\url{https://www.theregister.com/2023/11/29/aws_launch_ibms_db2_database/}}
Options provided by Google's storage are listed here.
\footnote{\url{https://cloud.google.com/compute/docs/disks}}

Amazon {\it Elastic Block Store - EBS} supports both SSDs and HDDs provides block-level storage volumes 
for users of Amazon {\it Elastic Compute Cloud - EC2} instances.
\footnote{\url{https://aws.amazon.com/pm/ec2/}}
EBS supports RAID1, RAID1, RAID5 , RAID6 and RAID10.
\footnote{\url{https://aws.amazon.com/blogs/aws/new-amazon-s3-storage-class-glacier-deep-archive/}}
AWS {\it Simple Storage Service - S3} Glacier storage classes deliver cost-optimized archive storage. 
Amazon EBS charges users by storage capacity (GBs provisioned per month). 
\footnote{\url{https://aws.amazon.com/ebs/pricing/}}
Users also pay for IOPS and throughput.
Amazon EBS does not recommend RAID 5 or RAID 6 because of extra IOPS (20-30\%) with respect to RAID0,
which may be considered a shift in RAID paradigm..

Cloud storage is unacceptable solution for real time applications 
such as those associated with {\it Internet of Things - IoT}.
\footnote{\url{https://en.wikipedia.org/wiki/Internet_of_things}} 
Industrial IoT for smart manufacturing has real time requirements.
\footnote{\url{https://www.automate.org/editorials/the-relationship-between-real-time-computing-and-the-internet-of-things-iot}}

Readers may follow rapid advances in the field by referring 
to the comprehensive list of trade and academic publications 
provided in the Appendix in Thomasian 2021 \cite{Thom21}.

 


\clearpage
\section{Appendix I: RAID Performance Evaluation}\label{sec:perfeval} 
\vspace{3mm}

\begin{small}
\ref{sec:factors} Application performance as affected by storage. 
\ref{sec:RAID5perfanal}$^*$ RAID performance analysis with M/G/1 queueing model.        \newline 
\ref{sec:diskservice}$^*$ Components of disk service time.                              \newline 
\ref{sec:seekdistance}$^*$ Seek distance distribution without and with ZBR.             \newline
\ref{sec:normalanal}$^*$. RAID5 performance analysis in normal mode.                    \newline
\ref{sec:degradedanal}$^*$ Fork/Join response time analysis for degraded mode.          \newline  
\ref{sec:rebuildanal}$^*$ Analysis of rebuild processing with {\t Vacationing Server Model - VSM}. \newline 
\ref{sec:VSM}$^*$ M/G/1 VSM analysis with multiple vacations of two types.              \newline 
\ref{sec:LST}$^*$ The Laplace Stieltjes Transform for rebuild seek time.                \newline
\ref{sec:alternative}$^*$ An alternative method to estimate rebuild time.               \newline 
\ref{sec:optimal} Optimizing storage allocation.                                        \newline
\ref{sec:tapelibrary} Performance analysis of a tape library system.                    \newline 
\end{small}

\subsection{Application Performance as Affected by Storage}\label{sec:factors}
\vspace{2mm}

With the advent of faster CPUs, the performance of applications with stringent response time requirements, 
such as OLTP, rely heavily on storage performance, which is especially a problem when the data is held by disks. 
Transition to Flash SSDs is expected to alleviate this performance bottleneck
leading to improved response times at a higher equipment cost.

Potential storage accesses are obviated by caching, 
e.g., higher levels of a B+ tree index,
see e.g., Section 9.4 in Ramakrishnan and Gehrke 2002 \cite{RaGe02}, 
may reside permanently in database caches.
Generally database performance can be improved by obviating disk accesses 
by utilizing larger buffers and improved buffer management policies, 
as discussed in Sections 8.1 and 9.4 of the Db2 Handbook.
\footnote{\url{https://docs.broadcom.com/doc/db2-for-zos-performance-handbook}}

The CPU checks if a referenced database or file block is held in the main memory buffer. 
\footnote{\url{https://en.wikipedia.org/wiki/Database_caching}}
If not an I/O request is issued which is intercepted by the DAC,
which checks if it caches the specified block and id so returns the block.
DAC caches with {\it Battery Backed Unit - BBU} are mainly used for writing.
\footnote{\url{https://www.thomas-krenn.com/en/wiki/RAID_Controller_and_Hard_Disk_Cache_Settings}}
If the block is not cached disk's onboard cache is checked 
before accessing the disk taking into account faulty sectors.
\footnote{\url{https://en.wikipedia.org/wiki/Disk_buffer}}

Disk transfer rate or bandwidth is an important metric given as:                                 
\vspace{-1mm}
$$\mbox{Sectors\_Per\_Track} \times \mbox{sector\_size} \times \mbox{RPM}/60$$              
The data transfer rate for {\it Constant Angular Velocity - CAV} disks 
with ZBR from outer tracks is higher than inner disks.  
Sustained disk data rates take into account head switching times across tracks and cylinders.
There is a transfer rate from disk to the onboard buffer and then onward to the main memory. 


{\it Least Recently Used - LRU} was an early replacement policy, but there are many others.
\footnote{\url{https://en.wikipedia.org/wiki/Page_replacement_algorithm}} 
LRU-K algorithm is self-tuning, and does not rely on external hints 
and adapts in real time to changing patterns of access O'Neil 1993 \cite{ONOW93}.
A recent study advocates FIFO in preference to LRU Yang et al. 2023 \cite{YQZ+23}. 
Other page replacement algorithms are discussed in Section 1.15.1 in Thomasian 2021 \cite{Thom21}.
Large capacity of caches requires very long traces making trace-driven simulation difficult if not impossible. 
Online measurement of page miss ratios versus memory size 
using hardware and software is explored in Zhou et al. 2004 \cite{ZPS+04}.

Part of the DAC cache is NVRAM and duplexed NVRAM is as reliable as disk storage 
for logging according to Menon and Cortney 1993 \cite{MeCo93},
This allows a fast-write capability and deferred destaging of dirty blocks.
Dirty blocks are overwritten several times before destaging thus reducing disk load.
Batch destaging allows disk scheduling to minimize disk utilization.

While prefetching is known to improve performance,
excessive prefetching results in a reduction the cache hit rate by overwriting viable data.
A scheme that maximizes the sum of hit rates due to prefetching and caching 
is described in Baek and Park 2008 \cite{BaPa08}.
Five prefetching categories are:
(1) history-based, 
(2) application hint-based, 
(3) offline optimal, 
(4) sequential prefetching, 
(5) combined with caching.
Aligning sequential prefetch in strip or stripe boundaries 
shows a performance improvement 3.2x for 128 clients and 2.4x for a  single sequential read.
on I/O performance with respect to sequential prefetching schemes in Linux
Prefetching methods are reviewed in Section 1.31 in Thomasian 2021 \cite{Thom21}.

Results from I/O trace analysis by Treiber and Menon 1995 \cite{TrMe95} 
were incorporated into the performance analysis in Thomasian and Menon 1997 \cite{ThMe97},
e.g., occurrences of two blocks a short distance apart on a track, treated as one disk access,
which is an instance of disk access coalescing or proximal I/O  Schindler et al. 2011 \cite{ScSS11}.

Disk cache miss ratio analysis and design considerations are discussed in Smith 1985 \cite{Smit85}.
The fractal structure of cache references model by a statistically self-similar underlying process, 
which is transient in nature is discussed in McNutt 2002 \cite{McNu02}.
Self-similarity in nature was observed by B. Mandelbrot and is discussed in Peitgen et al. 2004 \cite{PeJS04}.
\footnote{\url{https://en.wikipedia.org/wiki/Self-similarity}}           


IBM z16 mainframe supports up to 40 TB=$10^{12}$ bytes) main memory.
\footnote{\url{https://www.ibm.com/downloads/cas/6NW3RPQV}}
Rather than just caching the whole database may be held in main memory, 
e.g., IBM's {\it Information Management System - IMS} FastPath 
\footnote{\url{https://en.wikipedia.org/wiki/IBM_Information_Management_System}}
Oracle's TimesTen Lahiri et al. 2013 \cite{LaNF13}, and SAP's HANA.
\footnote{\url{https://www.sap.com/products/technology-platform/hana/what-is-sap-hana.html}}
The sizes of largest databases in 2015 were as follows.   
\footnote{\url{https://www.comparebusinessproducts.com/fyi/10-largest-databases-in-the-world}}
Top ten largest databases using different criteria are given here.
\footnote{url{https://www.wheelhouse.com/resources/the-top-10-largest-databases-in-the-world-past-and-present-a7843}}


\subsection{Analysis of RAID5 Performance$^*$}\label{sec:RAID5perfanal} 
\vspace{2mm}

An M/G/1 queueing model is adopted in this section with Poisson arrivals 
with exponential interarrival times with mean $\bar{t}=1/\lambda$ ({\bf M}) 
and generally distributed disk service times ({\bf G}) with $\overline{x^i}$ as the $i^{th}$ moment
Kleinrock 1975 \cite{Klei75}. 

\subsubsection{Components of Disk Service Time}\label{sec:diskservice}
\vspace{2mm}

Disk service time is the sum of seek (s), latency ($\ell$), and transfer time (t).
Given the mean and variance of service time components 
we have the following mean, variance and second moment for disk service time:

\vspace{-1mm}
$$\bar{x}_{disk} = \bar{x}_s + \bar{x}_\ell + \bar{x}_t, \hspace{5mm}
\sigma^2_{disk} = \sigma^2_s + \sigma^2_\ell + \sigma^2_t, \hspace{5mm}
\overline{x^2}_{disk} = \sigma^2_{disk} + (\bar{x}_{disk})^2.$$ 

The latter equality requires the independence of random variables,
especially latency and transfer time.
This is true for accesses to small blocks.
The second moment of disk service time is: 

Seagate Cheetah 15k.5 model ST3146855FC parameters extracted in 2007 at CMU's PDL are as follows.
\footnote{\url{https://www.pdl.cmu.edu/DiskSim/diskspecs.shtml}}                                       
146 GB, maximum logical block number b=286749487 (number of 512 B sectors), 
c=72,170 cylinders, number of surfaces 4, RPM=15015, $h=4$ heads,
The disk rotation time is $T_R \approx 60,000/\mbox{RPM}=4$ ms.
The mean number of sectors per track is: $b/(c \times h) \approx 993$.

\begin{framed}
\subsection{Seek Distance Distribution without and with ZBR$^*$}\label{sec:seekdistance}
\vspace{2mm}

Most analytic studies make the false assumption that disk are fully loaded, which is not true in practice.
It was observed in a 1970s that disk arms rarely move,
which may be due to the batch workloads, e.g., a job reading successive blocks of a disk file. 
Given that the probability of not moving the disk arm is $0 \leq p \leq 1 $
and the remaining cylinders are accessed uniformly then according to Lavenberg 1983 \cite{Lave83} 
the distribution of seek distance is given in Eq. \ref{eq:dist} 

\begin{eqnarray}\label{eq:dist}
P_D [d] = 
\begin{cases}
p, \mbox{  for  }d=0, \\ 
(1-p)\frac{2(C-d)}{C(C-1)}, \mbox{  for  } 1 \leq d \leq C-1.
\end{cases}
\end{eqnarray}
There are $2(C-d)$ ways to move $1 \leq d \leq C-1$ cylinders, 
which are normalized  by $\sum_{d=1}^{C-1} 2 (C-d) = C(C-1)$.
For uniform distribution across all cylinders:

\vspace{-2mm}
\begin{eqnarray}\label{eq:dist2}
P_D[0]=p=\frac{1}{C} \mbox{   and   }P_D[d]=\frac{2(C-d)}{C^2}, 1 \leq d \leq C-1.
\end{eqnarray} 

The mean seek distance in this case is:
\vspace{-2mm}
\begin{eqnarray}\label{eq:meanseek}
\bar{d} = \sum_{d=1}^{C-1} d P_D (d) = \sum_{d=1}^{C-1} d \frac{2(C-d)}{C(C-1)} \approx \frac{C}{3}.
\end{eqnarray}

In a continuous domain with $f_x(x)=1/L, x \in [0,L]$ the mean distance $Y=|x_1-x_2|$ 
can be obtained by noting that $f(x_1,x_2) = f(x_1)f(x_2)=1/L^2)$.
In what follows we have taken advantage of symmetry by multiplying by two.
\begin{eqnarray}\label{meanseek2}
E[Y] = \int_0^L \int_0^L (x_1-x_2)f(x_1,x_2) dx_1 dx_2 =
\frac{2}{L^2} \int_0^L \int_0^L (x_1-x_2) dx_1 dx_2 = \frac{2L^3}{6L^2}=\frac{L}{3}.
\end{eqnarray}  

The seek time versus seek distance in cylinders or tracks ($d$): 
$t_{seek}(d),1 \leq d \leq C-1$ is called the seek time characteristic. 
Curve-fitting to experimental results yields equations of the following forms, 
see e.g., Hennessey and Patterson 2017 \cite{HePa17}: 
\vspace{-1mm}
$$t_{seek} (d) = a + b \sqrt{d-1}, \mbox{ or  } t_{seek} (d) = a + b \sqrt{d-1} + c (d-1) \hspace{5mm} 1 \leq d \leq C-1$$  

The average seek time can be expressed as $t_{seek} (\bar{d})$,
where the mean seek distance $\bar{d}$ is given by Eq. (\ref{eq:meanseek}) 
but a more accurate estimate is obtainable as follows:
\vspace{-1mm}
$$\overline{x^i_s} = \sum_{d=1}^{C-1} P_D (d) t^i_{seek} (d),\hspace{5mm}\sigma^2_s = \overline{x^2_s} - (\bar{x_s})^2. $$

The analysis in Merchant and Yu 1996 \cite{MeYu96} postulates a seek-time formula of the form:
$t_{seek} (f) = a + b \sqrt{f}$, where $0 \leq f \leq 1 $ is the fraction of uniformly traversed tracks. 
This implies a mean normalized seek distance $\bar{f}=1/2$. 
The same mistake is repeated in Varki et al. 2004 \cite{VMXQ04}. 

The analysis of seek times for disks with ZBR taking into account 
the variability of track capacities is given in Thomasian et al. 2007 \cite{ThFH07}.
The number of sectors on cylinder $c$ is $s_c, 1 \leq c \leq C$. 
With the assumption of fully loaded disks and uniformly accessed sectors 
the probability of accessing cylinder $c$ is proportional to the number of its sectors: $P_c= s_c / C_d$, 
where the disk capacity is $C_d = \sum_{c=1}^C s_c$.
Starting with the probability of seek distance $d$ starting at cylinder $c$,
requires $c+d < C$ and $c>d$. Unconditioning on $c$:
\vspace{-1mm}
$$P_D (d|c) = P_c (c+d) + P_c (c-d), \hspace{5mm}   P_D(d)= \sum_{c=1}^{C} P_D(d|c) P_c $$

A mapping of files onto disk cylinders and their access rates can be used 
to obtain more accurate estimates of seek time.

\end{framed}

Rebuild analysis requires the LST of seek time and transfer time.
Provided that the {\it Rebuild Unit - RU} is a track,
there is almost no latency due to {\it Zero Latency Access - ZLA} 
which allows track reading to start at the first encountered sector,
see Section 21.1.3 Jacob et al. 2008 \cite{JaNW08},
Given that track transfer time equals disk rotation time ($T_R$), 
the LST is: ${\cal L}^*_t (s) = T_R/s$.

Latencies to access small blocks are uniformly distributed over track rotation time: $(0,T_R)$.
The LST and $i^{th}$ moment of latency is given as: 
\vspace{-1mm}
$${\cal L}^*_{\ell} (s) = (1- e^{- sT_R ) / (s T_R}), \hspace{5mm} \overline{x^i_\ell} \approx T^i_R /(i+1).$$  

Given that the transfer time of $j=8$ sectors from cylinder $c$ is $x_t (j,c) = j T_R / s_c$, 
it follows that the $i^{th}$ moment of transfer time is:
\vspace{-1mm}
$$\overline{x_t^i} (j) = \sum_{c=1}^C x_t^i (j,c) P_c , \hspace{5mm} \sigma^2_t = \overline{x^2_t} - (\bar{x_t})^2 .$$
The transfer time of fixed size blocks in nonZBR disks is a constant $c=x_t$, 
so that $X^*_t (s)|c_{constant} = c/s$
With ZBR the variability of track transfer times should be taken into account
by uncondioning on the variation of transfer time,
which increases almost uniformly from innermost to outmost cylinder.

The LST of response time $R^*(s) = W^*(s)B^*(s)$
where the LST of service time is $B^*(s) = X^*_s (s) X^*_\ell (s) X^*_t (s) $,
where according to Chapter 5 in Kleinrock 1975 \cite{Klei75}.
\vspace{-1mm} 
$$W^*(s) = \frac{\bar{x}(1-\rho)}{1-\lambda (1- B^*(s)}.$$ 
The $i^{th}$ moment of disk response time is: 
\vspace{-1mm}
$$\overline{R^i} = (-1)^i \frac {d R^*(s) } {ds^i} |_0 $$

\begin{framed}
\subsection*{Shingled Magnetic Recording - SMR}
\vspace{2mm}

SMR writes partially overlaps written tracks increasing recording density radially 
while ZBR increases disk capacity linearly.
\footnote{\url{https://en.wikipedia.org/wiki/Shingled_magnetic_recording}}
The emergence of SMR may be attributed to slowdown in disk capacity growth after 2010, 
while there was rapid growth in the period 2000-2010 Gray and Shenoy 2000 \cite{GrSh00}.

SMR design issues are presented in Amer et al. 2010 \cite{ALM+10}.
Theoretical justifications for SMR are given in Sanchez 2007 \cite{Sanc07},
which in Section 1.2 introduces the trilemma: (1) thermal stability, 
(2) writeability, and (3) media {\it Signal to Noise Ratio - SNR}.
A technical description of SMR drives is given in Feldman and Gibson 2013 \cite{FeGi13}.

The performance of SMR drive is unpredictable in that clean drives can be written quickly, 
but if the drive has too many writes queued, 
or has insufficient idle time to reorganize or discard overwritten data, 
then write speeds can be significantly lower than the usual disk bandwidth, e.g., 1030 Mbits/s for 7200 RPM disks.
\footnote{\url{https://en.wikipedia.org/wiki/Hard_disk_drive_performance_characteristics}}
Because of their access time variability SMR disks should be excluded from RAID5 arrays. 
\footnote{url{https://www.servethehome.com/surreptitiously-swapping-smr-into-hard-drives-must-end/}}

\end{framed}

\subsection{RAID5 Performance in Normal Mode$^*$}\label{sec:normalanal}
\vspace{2mm}

RAID5 utilizes left symmetric parity layout over successive $(N+1)\times (N+1)$ segments
The parities are placed at $D(N+1-i,N+1-i), 0 \leq i \leq N$
he parities can be used to recover from disk failures 
and to cope with unreadable sectors as long as they do not belong to the same parity group.
There is extra processing to update parities as corresponding data blocks are updated.
Given a modified disk block $d_{new}$, if not cashed its previous version $d_{old}$ has to be read from disk 
to compute $d_{diff} = d_{old} \oplus d_{new}$.
Similarly $p_{old}$ is accessed if not cached to compute $p_{new} = p_{old} \oplus p_{diff}$.
The fact that a single write requires four disk accesses 
is called the {\it Small Write Penalty - SWP} Chen et al. 1994  \cite{Che+94}.
RAID(4+k), $\ell \geq 1$ requires $2(k+1)$ reads and writes to update $k$ data and $k$ check blocks.

The writing of $d_{new}$ and $p_{new}$ to disk be deferred 
by first writing to {\it NonVolatile Storage - NVS} e.g., {\it NonVolatile RAM - NVRAM}.
Dirty blocks in NVRAM can be overwritten several times 
before being destaged or written to disk thus obviating unnecessary writes.  
Destaging batches of blocks allows permuting the order 
in which blocks are written to reduce positioning times.  
Other methods to deal with SWP are presented in Chen et al. 1994 \cite{Che+94}.

Let $f_r$ and $f_w=1-f_r$ denote the fraction of logical read and write requests.
Each read request results in a {\it Single Read - SR} access to disk if the data is not cached 
and each logical write may require two SR and two {\it Single Write - SW} disk accesses.
The fraction of SRs and SWs are as follows: 
\vspace{-1mm}
$$f_{SR} = (f_r + 2 f_w)/(3 f_r+ 2 f_w),  \hspace{5mm}f_{SW}=1-f_{SR}.$$
The moments of disk service time are given as follows:

\vspace{-2mm}
\begin{eqnarray}
\overline{x^i}_{disk} = f_{SR} \overline{x^i_{SR}} + f_{SW} \overline{x^i_{SW}}.
\end{eqnarray}

Given that the arrival rate to the RAID5 disk array is $\Lambda$            .
then RAID striping results in uniform accesses to $N+1$ disks: $\lambda = \Lambda / (N+1)$,
so that the disk utilization factor is: 
\vspace{-1mm}
$$\rho=\lambda \bar{x}_{SR} + 2 f_w (\bar{x}_{SR} + \bar{x}_{SW}).$$

Given the mean and variance of disk service time with FCFS scheduling
the  coefficient of variation of disk service time is: $c^2_X=\sigma^2_{disk}/ (\bar{x}_{disk})^2.$
The mean, second moment, and variance of waiting time in M/G/1 queues is given 
by the Pollaczek-Khinchine formula Kleinrock 1975 \cite{Klei75}:

\vspace{-2mm}
\begin{eqnarray}\label{eq:MG1}
W = \frac{ \lambda \overline{x^2}_{disk} }{2(1-\rho)} = 
\frac{\rho \bar{x}(1+c^2_X)}{2(1-\rho)},\hspace{5mm}
\overline{W^2} = 2 W^2 + \frac{ \lambda \overline{x^3}_{disk} } {3(1-\rho)}, \hspace{5mm}
\sigma^2_W = \overline{W^2} - W^2.
\end{eqnarray}

The prioritized processing of SR requests on behalf of logical reads 
is only affected by disk utilization due to other logical read requests.
The fraction of such requests is  ${f'}_{SR} = f_r / (f_r + 4 f_w)$
and moments of disk service time ($x_{disk}$) should be recomputed with ${f'}_{SR}$. 
The waiting time of SR requests with nonpreemptive priorities is given in Kleinrock 1976 \cite{Klei76}. 

\vspace{-2mm}
\begin{eqnarray}\label{eq:priority}
W_{pr} = \frac{\lambda \overline{x^2}_{disk} }{2(1-\rho_{SR})}
\mbox{  with  }\rho_{SR} = {f'}_r \lambda \bar{x}_{SR}.
\end{eqnarray}

The read response without and with priorities is:
$R_{SR} = \bar{x}_{SR} + W\mbox{  and  }R_{SR} = \bar{x}_{SR} + W_{pr}.$

A write is considered completed when both the data and parity blocks are written to disk,
which is an instance of F/J processing discussed in Section \ref{sec:degraded}. 
Write response times are of no interest since they do not affect 
application response time due to the fast write capability onto NVRAM. 

No-Force transaction commit does not require dirty blocks 
in database buffer to be written to disk or NVS,
because ''after images'' of such blocks are written to disk as part of transaction commits. 
Logging for database recovery is discussed in Chapter 18 
on Crash Recovery in Ramakrishnan and Gehrke 2002 \cite{RaGe02}.

Since for the FCFS policy the waiting time is independent from service time then: 
$E[\tilde{W}_{SR} \tilde{x}_{SR}] =  \bar{W}_{SR} \times \bar{x}_{SR}$ and the second moment of SR requests is:

\vspace{-2mm}
\begin{eqnarray}
\overline{R^2}_{SR} = \overline{W^2}_{SR}  +\overline{x^2}_{SR} +2 W \bar{x}_{SR} \mbox{ and }
\sigma^2_{SR} = \overline{R^2}_{SR} - (R_{SR})^2.
\end{eqnarray}

\subsection{RAID5 Operation in Degraded Mode}\label{sec:degraded}
\vspace{2mm}

A block $d_1$ on failed Disk$_1$ can be reconstructed as: 
$d_1 = d_2 \oplus d_3 \oplus \ldots \oplus p_{1:N+1}.$

The read load  on surviving disks is doubled,
since given balanced read loads due to striping each read access
to the failed disk will results in an access to $N$ surviving disks to reconstruct blocks on demand
In the case of a write to Disk$_1$ $N-2$ disk accesses are required.
$p_{1:N+1}^{new} = d_1^{new} \oplus d_2 \oplus d_3 \oplus \ldots \oplus d_N.$
The load increase for a fraction $f_R$ for read and $f_W$ 
for write accesses according to Ng and Mattson 1994 \cite{NgMa94} is:

\vspace{-3mm}
\begin{eqnarray}\label{eq:degraded}
\mbox{LoadIncr}(f_R) =
\frac{ U_{faulty} }{ U_{faultfree} } = \frac{N}{N-1} + \frac{ (N-2)f_R + (N-8) f_W}{(N-1)(f_R + 4 f_W)},
\mbox{  e.g., LoadIncr} (0) = 1.333.              
\end{eqnarray}

\begin{figure}[t]
\begin{center}
\includegraphics[scale=0.650,angle=00]{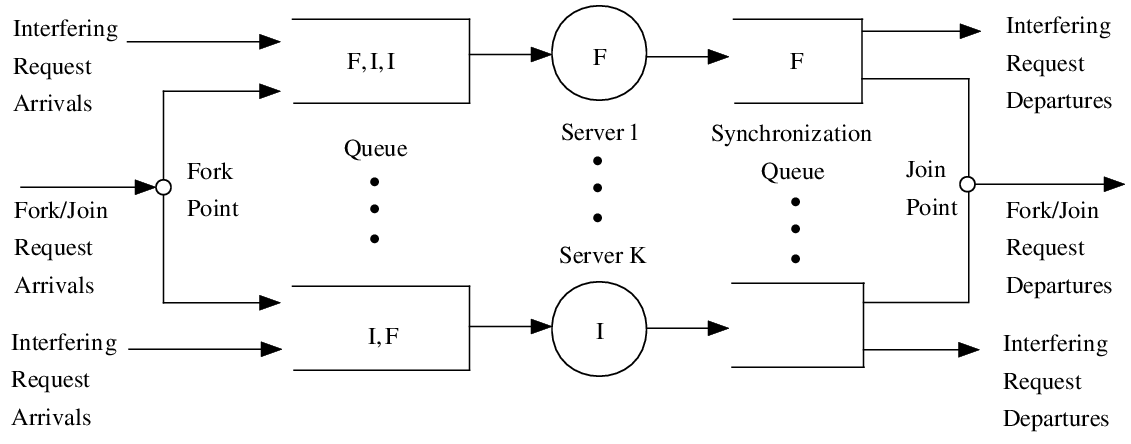}
\caption{Fork/join processing with interfering requests processing \label{fig:FJ}}
\end{center}
\end{figure}

In the case of a RAID5 with $N+1$ disks an $N$-way (resp. $N-1$-way) F/J request 
is required to process read (resp. write) requests involving failed disks in degraded mode of operation. 
In addition to F/J requests disk process their own requests termed interfering in Fig. \ref{fig:FJ}.

Before rebuild processing starts F/J requests on behalf of read requests to a failed disk
constitute one-half of read requests processed by surviving disks, 
but this fraction gets smaller as rebuild progresses due to ``read redirection'', 
i.e., reading reconstructed data from the spare disk Muntz and Lui 1990 \cite{MuLu90}.
This term is a misnomer since redirection is also used by writes 
to compute the parity block when $d_{old}$ is already reconstructed on a spare.
When $d_{old}$ or $p_{old}$ are not reconstructed yet on the spare 
there is no need to compute and write $d_{new}$ or $p_{new}$,
since this is done more efficiently later as part of rebuild.

Piggybacking proposed by Muntz and Lui 1990 \cite{MuLu90} writes blocks reconstructed on demand on the spare disk. 
\footnote{This piggybacking is different from that in Section \ref{sec:distr}.}
Piggybacking small writes for rebuild tends to be inefficient Holland et al. 1994 \cite{HoGS94}. 
Piggypbacking at the level of tracks on which reconstructed data blocks reside is considered in Fu et al. \cite{FTHN04a}. 
Instead of half a disk rotation to read a block a full rotation will be required.

The load in clustered RAID5 and RAID6 is obtained using decision trees in Thomasian 2005b \cite{Thom05b}.
\footnote{\url{https://en.wikipedia.org/wiki/Decision_tree}} 
Let $\bar{x}_{\cdot}$ denotes read, write, and {\it Read-Modify-Write - RMW} mean disk service times.
RMW is a read followed by a write after a disk rotation and 
can be substituted by independent single reads and writes.
With arrival rate $\Lambda$ to a RAID5 with $N$ disks 
disk utilizations factors due to read and write requests with one failed disk are:
\footnote{Surviving disk loads should be increased by a factor $N/(N-1)$, 
see e.g. Thomasian et al. 2007 \cite{ThFH07}.}

\vspace{-5mm}
\begin{eqnarray}\nonumber 
\rho^{RAID5/F1}_{read} =
\frac{ \Lambda f_{read} }{N-1}
\left[ \frac{N+G-2}{N} \bar{x}_{read} \right],                         
\hspace{5mm}
\rho^{RAID5/F1}_{write} =
\frac{ \Lambda f_{write}} {N-1}
\left[ \frac{G-2}{N} \bar{x}_{read} +\frac{2}{N} \bar{x}_{write} + \frac{2(N-2) }{N} \bar{w}_{RMW} \right]
\end{eqnarray}



\subsection{Methods to Obtain Response Times for Fork/Join Requests$^*$}\label{sec:degradedanal}
\vspace{2mm}

The only exact result for F/J queueing systems is a 2-way F/J with two parallel M/M/1 queues. 
With arrival rate $\lambda$ and mean service time $\bar{x}=1/\mu$ server utilization $\rho = \lambda \bar{x} $.
The response time distribution in M/M/1 queues is exponentially distributed according to Eq. \ref{eq:Rdistr}, 
i.e., $R(t) = 1- e^{-t / \rho}$, where $R(\rho)=\bar{x}/ (1-\rho)$ is the mean response time.
The analysis in Flatto and Hahn 1984 \cite{FlHa84} is used 
to derive $R_2^{F/J} (\rho)$ in Nelson and Tantawi 1988 \cite{NeTa88}.

\vspace{-2mm}
\begin{eqnarray}\label{eq:FlHa84}
R_2^{F/J} (\rho) = \left[ H_2 - \frac{\rho}{8} \right] R(\rho)  = \frac{12-\rho}{8} R(\rho),
\mbox{   where the Harmonic sum is: } H_K \stackrel{\text{def}}{=}  \sum_{k=1}^K \frac{1}{k}.
\end{eqnarray}

Note that $R_2^{FJ} (0) = R_2^{max} (0) = \frac{H_2}{\mu}$ 
The following approximate equation for $2 \leq n \leq 32$-way M/M/1 F/J 
queueing systems is derived in Nelson and Tantawi 1988 \cite{NeTa88}.

\vspace{-2mm}
\begin{eqnarray}\label{eq:NeTa88}
R_n^{F/J} (\rho) =  \left[ \frac{H_n}{H_2} + (1- \frac{H_n}{H_2} ) \alpha (\rho) \right] R_2^{F/J} (\rho),
\end{eqnarray}
Curve-fitting to simulation results yielded $\alpha (\rho ) \approx (4/11) \rho $.

This equation is used in Menon 1994 \cite{Meno94} where disk service times are assumed to be exponentially distributed 
to estimate F/J response times in accessing the failed disk. 
This is an underestimate because disks process interfering requests.

Simulation results for the mean F/J response time versus server utilization by varying the fraction of F/J requests,
while the total $\rho$ is kept fixed for different service time distributions 
are given in Thomasian and Tantawi 1994 \cite{ThTa94}.
As the fraction of F/J requests decreases with respect to interfering requests,
their mean response times increases and tends to the expected value of the maximum in Eq. (\ref{eq:FJappr}).

\vspace{-2mm}
\begin{eqnarray}\label{eq:FJappr}
R_n^{F/J} (\rho) \approx R_n^{max} (\rho).
\end{eqnarray} 

We next consider schemes to estimate the expected value of the maximum of $n$ i.i.d. random variables.
Given the mean $\mu_X$ and standard deviation $\sigma_X$ of the components 
of an F/J request according to David and Nagaraja 2003 \cite{DaNa03}:

\vspace{-2mm}
\begin{eqnarray}\label{eq:Xmax}
\overline{X}_n^{max} \approx \mu_X + \sigma_X G(n). 
\end{eqnarray}

For the exponential distribution $G(n) = H_n - 1$ and since $\sigma_X = \mu_X$ 
then $\overline{X}_n^{max}= \mu_X H_n$, which is also given by Eq. (\ref{eq:FJ}), 
which is exact as shown below.

The {\it Probability Distribution Function - PDF} of the maximum of $n$ random variables 
with PDF $F_Y(y)$ is also the $n^{th}$ order statistic Trivedi 2001 \cite{Triv01}: 
\vspace{-2mm}
\begin{eqnarray}
F_{Y_n^{max}} (y)  = [F_Y (y)]^n.
\end{eqnarray} 

$$\mbox{Given }F_y(y) = 1- e^{-\lambda y}\mbox{ for }n=2 \hspace{5mm}\bar{Y}_2^{max} 
= \frac{2}{\lambda} - \frac{1}{2\lambda}= \frac{H_2}{\lambda}.$$
Induction can be used to obtain $Y_n^{max}$ or using the method  
in Section 4.6.2 in Trivedi 2001 \cite{Triv01} or:
\footnote{\url{https://mikespivey.wordpress.com/2013/05/13/expectedmaxexponential/}}
 
\vspace{-2mm}
\begin{eqnarray}\label{eq:FJ}
\bar{Y}_n^{max} = H_n \bar{Y}.
\end{eqnarray}

Eq. (\ref{eq:ThTa94}) in Thomasian and Tantawi 1994 \cite{ThTa94} 
for the expected value of the maximum of $n$ response times is motivated by Eq. (\ref{eq:Xmax}):
\vspace{-2mm}
\begin{eqnarray}\label{eq:ThTa94}
R_n^{F/J} (\rho) = R (\rho) + F_n \sigma_R (\rho) \alpha_n (\rho),
\end{eqnarray}
where $R(\rho)$ and $\sigma_{R} (\rho) $ are the mean 
and standard deviation of response time at utilization factor $\rho$.
Surface-fitting to simulation results was used in this study
to obtain $\alpha_n (\rho)$ for several distributions.

It is shown in Nelson and Tantawi 1988 \cite{NeTa88} that $R_n^{max} (rho)$ is an upper bound 
to $R_n^{F/J}(\rho)$ for the exponential distribution.
This effect is observed via simulation in Thomasian and Tantawi 1994 \cite{ThTa94} for other distributions
and that $R_n^{max}(\rho)$ is  a good approximation to $R_n^{F/J}(\rho)$  
when F/J requests constitute a small fraction of processed requests.
This is the case for RAID5 in degraded mode as rebuild progresses 
and most requests to the failed disk are satisfied by redirection and the fraction of F/J requests drops.

A simple method to obtain the maximum of $n$ i.i.d. random variables 
is to use the two-parameter {\it Extreme Value Distribution - EVD} approximation
Johnson et al. 1995 \cite{JoKB95}, Kotz and Nadarajah 2000 \cite{KoNa00}. 

\vspace{-1mm}
\begin{eqnarray} 
F_Y (y) =  P(Y < y) = exp (- e^{- \frac{y-a}{b} } ), \hspace{3mm}
\overline{Y} = a + \gamma b, \hspace{3mm} \sigma^2_Y = \frac{\pi^2 b^2}{6}, 
\end{eqnarray}
where $\gamma = 0.577721 $ is the Euler constant.
Matching the first two moments: $\bar{Y} =R(\rho)$ and $\sigma_Y=\sigma_R$ leads to:
\vspace{-1mm}
$$b=  \sqrt{\frac{6}{\pi}} \sigma_Y \mbox{   and   } a=\bar{Y} - \gamma b.$$  
The maximum of $n$ random variables with EVD has a simple form:

\vspace{-2mm}
\begin{eqnarray}\label{eq:meanEVD}
\overline{Y_n^max} = ( a + \gamma b) + b \mbox{ln} (n)
= \overline{Y} +  \frac{ \sqrt{6} } { \pi}  \sigma_Y.
\end{eqnarray}


It follows from simulation results in Thomasian et al. 2007 \cite{ThFH07}  
that EVD overestimates $\overline{y^n}_{max}$ and a better approximation is  obtained 
if the second summation term is divided by 1.27.
Three types of EVDs are and the above discussion is based on Gumbel distribution:
\footnote{\url{https://core.ac.uk/download/pdf/15566772.pdf}}

The coefficient of variation of response times of M/G/1 queues 
for accesses to small randomly placed disk blocks 
for disk characteristics in Thomasian and Menon 1997 \cite{ThMe97} is:

\vspace{-2mm}
\begin{eqnarray}
c^2_R (\rho) = \frac{\sigma^2_R (\rho)}{R^2 (\rho)}=
\frac{c^2_X + \frac{\rho s_X }{3(1-\rho)} +\frac{\rho(1+c2_X)}{ 4(1-\rho)^2} }
{1 +\frac{\rho (1+c^2_X }{1-\rho}+\frac{\rho(1+c2_X)}{ 4(1-\rho)^2} },
\end{eqnarray}
where $s_X= \overline{x^3}_{SR}/(\bar{X}_{SR})^3$. 
Where $c_R = c_X$ for $\rho=0$ and that as $\rho \rightarrow 1$ $c_R \rightarrow 1$, 
i.e., an exponential distribution, hence $c_R < 1$ for $C_X <1$,
which was the case the study.

Approximating the response time distribution with a distribution, 
which is an exponential in form has the advantage 
that its expected value of its maximum can be computed easily by integration.
In the case of the Erlang distribution whose $c^2 < 1$ 
the number of its stages is: $k=\lceil 1/c^2 \rceil$ 
and the mean delay per stage is $1/\mu = R(\rho) / k$ 
$R_n^{max} (\rho) $ for an $n$-way  $k$-stage Erlang distribution is then:

\vspace{-2mm}
\begin{eqnarray}\label{eq:balanced}
R^{max}_n (\rho) = 
\int_0^\infty \left[ 1 -  \prod_{i=1}^n 
\left( 1 - e^{ - \mu_i t }
\sum_{j=0}^{k_i -1}\frac{ (\mu_i t )^j }{ j!}
\right)
\right]  dt  .
\end{eqnarray}

Provided disks have an XOR capability updating parities results in a 2-way F/J request with unequal service times:
$d_{new}$ and the block address is sent to the disk
and after $d_{old}$ is read $d_{diff}$ is computed and $d_new$ 
overwrites $d_{old}$ after a disk rotation.
$d_{diff}$ and the address of $P_{old}$ are then sent via the DAC to the parity disk 
where it is XORed with $p_{old}$ to compute $P_{new}$,
which overwrites $p_{old}$ after one disk rotation.
Write response time is the maximum of data and parity writes.

\vspace{-2mm}
\begin{eqnarray}\label{eq:asymmetric}
R^{max}_{2} (\rho) = \sum_{i=1}^2 R_i (\rho)  -
\sum_{m=0}^{k_1 - 1}
\sum_{n=0}^{k_2 - 1}
\binom{m+n}{m}
\frac{\mu_1^m \mu_2^n }
{(\mu_1 + \mu_2 )^{m+n+1} } ,
\mbox{  where  }\mu_1= k_1 / R_1 (\rho)\mbox{ and }\mu_2 = k_2 / R_2 (\rho).
\end{eqnarray}

The more flexible Coxian distribution used by Chen and Towsley 1993 \cite{ChTo93}.
can be applied to any response time distribution regardless of the value of $c_R$. 

Two moment approximations for maxima are given in Crow et  al. 2007 \cite{CrGW07}.
Formulas for special cases when the coefficient of variation $c < 1$  are provided.
When $c>1$ a 3-moment approximation to the Hyperexponential-2 distribution given by Eq. (4.4) is suggested.

\subsection{Vacationing Server Model for Analyzing Rebuild Processing$^*$}\label{sec:rebuildanal}
\vspace{2mm}

Rebuild is the systematic reconstruction of the contents of the failed disk on a spare disk.
Rebuild is started after an appropriate delay to allow for fail-slow events in Section \ref{sec:failslow}. 
Strip $D_1$ on failed Disk$_1$ is reconstructed in the DAC's buffer 
by reading and XORing $N-1$ corresponding strips,
which are written to strip ${D'}_1$ on a spare disk which replaces failed Disk$_1$.
${D'}_1 = D_2 \oplus D_3 \oplus \ldots P_{1:N}.$

The rebuilding of a failed disk is carried out concurrently with 
the processing of external requests to maintain high availability 
due to the high cost of downtime in Section \ref{sec:intro}. 
Rebuild time is lengthened due to the interference of external disk requests
and conversely disk response times is increased by rebuild reads.

Distributed sparing in RAID5 provides sufficient storage on $N-1$ surviving disks 
to hold the data strips of the failed disk.
Spare strips are allocated in repeating diagonals similarly 
to parity strips as shown in Fig. 1 in Thomasian and Menon 1997 \cite{ThMe97}.   
The advantage of distributed sparing over dedicated sparing is
that the bandwidth of the spare disk is not wasted.
Rebuild reading in distributed sparing takes longer than dedicated sparing,
since rebuild reading from surviving disks is slowed down by rebuild writes.
In fact distributed sparing is advantageous for CRAID, 
since a single spare disk may become a bottleneck in this case.
A disadvantage of distributed and restriping rebuild discussed below 
is that an additional copyback step to copy the contents of spare strips onto a spare disk is not required.

Restriping which overwrites check strips 
is preferable to distributed sparing in that it attains higher reliability Rao et al. \cite{RaHG11}. 
Starting with RAID7 there are the following transitions:                                    \newline
RAID7 $\rightarrow$ RAID6 $\rightarrow$ RAID5 $\rightarrow$  RAID0 $\rightarrow$ Data Loss. \newline
In parity sparing proposed in Reddy et al. \cite{ReCB93} two RAID5 arrays are merged and 
parity blocks on one RAID5 are used as spare areas to hold missing strips. 

Rebuild time in multi-TB HDDs takes hours making RAID5 vulnerable to a second disk failure,
especially if disk failures are correlated, which was an initial reason for RAID6.
Most rebuild failures are due to LSEs - Latent Sector Errors, 
which stop rebuild from progressing as discussed in Section \ref{sec:relanal}. 
A queueing model to estimate rebuild time in RAID5 with HDDs is discussed in Section \ref{sec:rebuildanal}

Before the introduction of ZBR fixed sized RUs were set to a single or multiples disk tracks.
Fixed size RUs can be adopted for ZBRs with RU sizes large enough to make positioning time negligible.


RAID5 rebuild time is investigated via simulation in Holland et al. 1994 \cite{HoGS94},
where it is concluded that {\it Disk-Oriented Rebuild - DOR} is superior to {\it Stripe-Oriented Rebuild - SOR},
which proceeds by reading one stripe at a time,
since this introduces  unnecessary synchronization delays but reduces buffer requirements with respect to DOR.
A fixed size {\it Rebuild Unit - RU} equaling one track was postulated.

DOR which reads disks when they become idle is analyzed in Thomasian and Menon \cite{ThMe94,ThMe97}
using the {\it Vacationing Server Model - VSM} with multiple vacations 
of two types in Doshi 1985 \cite{Dosh85} and cited in Takagi 1991 \cite{Taka91}. 

The analysis yields the increased response time of external disk requests 
due to interference caused by rebuild processing
and the effect of external disk requests expanding rebuild processing time.
Rebuild reading with VSM is restarted when the disk becomes idle.
The accuracy of the analyses was verified using a random-number driven simulation in both studies,
since the system under consideration with codename Hager in Chen et al. 1994 \cite{Che+94} 
was under development, while NCR's  6298 RAID5 discussed in Section 5.3 of the tutorial was released in 1991
\footnote{\url{https://techmonitor.ai/technology/ncr_offers_redundant_array_scsi_chips_controller}}

Rebuild reading is halted with the arrival of external requests 
and the accumulated requests are served after rebuild read completes and the cycle repeats.
This is implicitly a priority queueing model with higher priority assigned to external requests.
The application of VSM to analyzing rebuild processing is also discussed in Thomasian 2018 \cite{Thom18}.

\begin{figure}[h]
\begin{center}
\includegraphics[scale=0.55,angle=00]{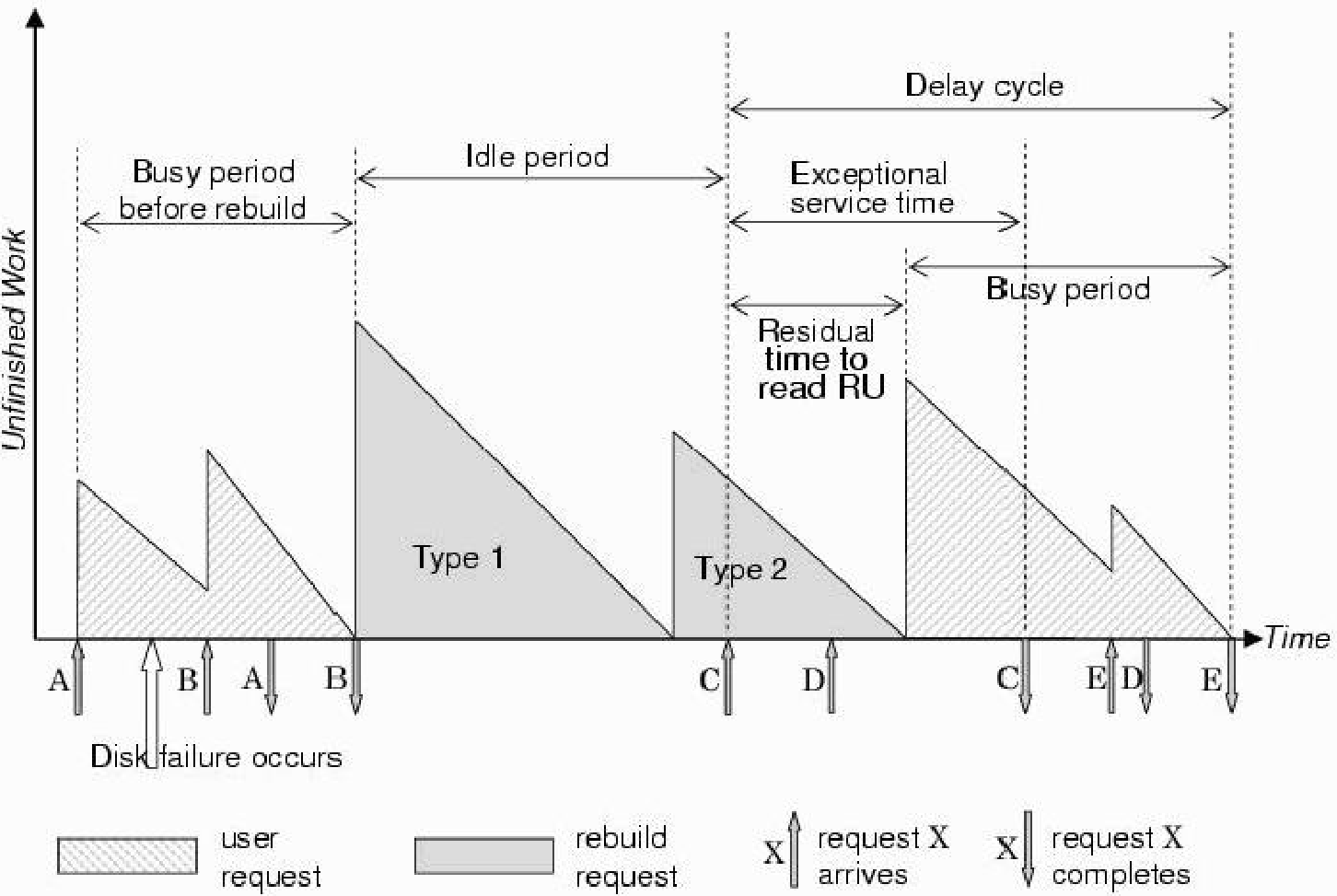}
\caption{Key VSM parameters associated with rebuild processing.}
\label{fig:rebuild}
\end{center}
\end{figure}

The initial RU or track read takes longer, 
since it involves a seek and latency as shown in Fig. \ref{fig:rebuild}).
The reading of succeeding RUs take less time, since a seek is not required,
e.g., a disk rotation if the RU size is a track.

The delay encountered by an external request requests with VSM follows 
from the fact that {\it Poisson Arrivals See Time Averages - PASTA}.  
The mean waiting time for external disk requests is the sum of three delays:              \newline
(1) the mean waiting time in the queue,
which is a product of mean queue length ($\bar{N}_q$) and mean service time per request ($\bar{x}$).    \newline
(2) if the server is busy with probability $\rho$ there is an additional delay given 
by the mean residual service time ($\bar{x}_r=\bar{x^2}/(2 \bar{x})$ Kleinrock 1975 \cite{Klei75}).  \newline
(3) the system is idle with probability $1-\rho$,
in which case the delay is the mean residual vacation (RU reading) time: 
$\bar{v}_r=\overline{v^2}^2 / (2 \bar{v})$),
where the $\overline{v^i}$ is the $i^{th}$ moment of vacation or rebuild read time. 

\vspace{-1mm}
$$W = \bar{N}_q \bar{x} + \rho \bar{x}_r + (1-\rho) \bar{v}_r \mbox{ due to Little's result:}
\bar{N}_q = \lambda W,  \hspace{2mm}\rho = \lambda \bar{x}$$

Simplifying the above equation yields:
\vspace{-2mm}
\begin{eqnarray}\label{eq:VSM}
W_{VSM} = W_{M/G/1} + \bar{v}_r = \frac{ \lambda \bar{x^2} } { 2 (1- \rho) } + \bar{v}_r.
\end{eqnarray}

An analysis of VSM with multiple vacations of two types is given in Section \ref{sec:VSM}.


The M/G/1 queue alternates between busy periods with mean $\bar{g}$ 
during which external disk requests are served 
and idle periods given with mean interarrival time ($1/\lambda$).
This is when RUs are reads although the processing of 
the first external request is delayed until a rebuild read is completed, 
unless preemption is allowed as discussed in Thomasian 2005b \cite{Thom05b}.
The utilization factor $\rho$ which is the fraction of time the server is busy can be expressed as:

\vspace{-2mm}
\begin{eqnarray}\label{eq:meanbusy}
\rho = \frac{\overline{g}}{\overline{g}+1/\lambda} \Longrightarrow \overline{g} = \frac{\bar{x}_{disk}}{1-\rho}.
\end{eqnarray}

The delay cycle is different from a busy period in that it starts with a special request
whose mean is the sum of mean residual rebuild time ($\bar{v}_r$) and 
mean disk service time ($\bar{x}_{disk}$) as shown in Fig. \ref{fig:rebuild}.
The mean duration of the delay cycle using the definition of $\rho$ 
according to Thomasian 2018 \cite{Thom18} is:

\begin{eqnarray}\label{eq:dc}
\rho =
\frac{
\bar{T}_{dc} (\rho) - \bar{v}_r}
{ 
\bar{T}_{dc} (\rho) + 1/\lambda
}
\hspace{2mm}
\Longrightarrow
\hspace{2mm}
\bar{T}_{dc} (\rho) = \frac{\bar{x}_{disk} + \bar{v}_r }{ 1- \rho}.
\end{eqnarray}

The cycle time is defined as the time between 
the start of successive busy periods and is given by Eq. (\ref{eq:cycletime}): 

\vspace{-2mm}
\begin{eqnarray}\label{eq:cycletime}
\bar{T}_{ct} (\rho) = \bar{T}_{dc} (\rho) + \frac{1}{\lambda}.
\end{eqnarray}

The moments of busy period and delay cycle can be obtained 
from their LSTs Kleinrock 1975/76 \cite{Klei75,Klei76}. 
The variances obtained in this manner can be used to obtain 
a better estimate of rebuild time as given by Eq. (\ref{eq:Xmax}),
but this is not necessary since simulation results show that 
there is little variation in the completion time of disk rebuild reading times 
(see e.g., Fig. 4 in Thomasian and Menon 1997 \cite{ThMe97}).

Assuming that the RU is a track, the number of disk tracks: $N_{track}$
and the mean number of tracks read per cycle: $\bar{n}_{track}$, 
the rebuild read time is:  

\vspace{-2mm}
\begin{eqnarray}\label{eq:rebuildtime}
T_{rebuild} (\rho) = \frac{N_{track}}{\bar{n}_{track}} \times T_{cycletime}.
\end{eqnarray}


Rebuild reading is overlapped with rebuild writing,
$i^{th}$ RU is written as RUs to reconstruct the $(i-1)^{st}$ RU are being being read  
so that rebuild time can be approximated with RU read time. 

The spare disk is not a bottleneck unless we have a {\it Clustered RAID - CRAID} in Section \ref{sec:CRAID}.
CRAID provides parallelism for rebuild reading, 
which would result in buffer overflow with VSM rebuild to a dedicated spare.
Rebuild reads can be throttled taking into account the degree of parallelism due to CRAID.
A simpler scheme is to monitor the number of RUs to be written to the spare disk in buffer
and stopping rebuild reading when their number exceeds a certain threshold.

\subsection{M/G/1 VSM Analysis of Multiple Vacations with Two Types$^*$}\label{sec:VSM}
\vspace{2mm}

This section is based on Thomasian 1994/1997 \cite{ThMe94,ThMe97}
which utilize VSM in Doshi 1985 \cite{Dosh85} and summarized in Takagi 1991 \cite{Taka91}. 
Both analyses assume that the RU size is a track.

The first RU/track read at the start of an idle period requires a seek followed by the reading of a track,
while successive rebuild reads just take a disk rotation, as shown in Fig. \ref{fig:rebuild}.
Due to ZLA reading of a track starts with the first encountered sector, 
i.e., the rotational latency (time to access a sector) is negligible. 
Track and cylinders skew in reading successive tracks is negligible Jacob et al. 2008 \cite{JaNW08}.

\begin{framed}
\subsubsection*{The Laplace Stieltjes Transform for Rebuild Seek Time$^*$}\label{sec:LST}
\vspace{1mm}

LSTs are discussed in Appendix I in Kleinrock 1975 \cite{Klei75}  and Appendix D in Trivedi 2001 \cite{Triv01}. 

To obtain the LST of seek time required for type 1 vacations
we approximate the distribution in Eq. (\ref{eq:dist}) by a continuous function,
whose {\it probability density function - pdf} is: 

\vspace{-2mm}
\begin{eqnarray}\label{eq:4p}
f_X (x) = p \delta (x) + (1-p) \frac {2 (C-x) } { C(C-2)}\mbox { note that }\int_{x=1}^{C-1} f_X(x) =1,
\end{eqnarray}
where $\delta (x)$ is the unit impulse at $x=0$.

Motivated by Abate et al. 1968 \cite{AbDW68} we derive the LST of seek time 
by using a piecewise linear approximation to the seek time characteristic:
($T_k,C_k$), where $C_1=1$ and $C_{K+1}=C-1$. 
\vspace{-1mm}
$$t(x)=\alpha_k x - \beta_k  \hspace{5mm} C_k \leq x \leq C_{k+1}.  $$
\vspace{-1mm}
$$ \alpha_k = \frac{C_{k+1} - C_k} {T_{k+1}- T_k} \mbox{   and  }\beta_k= \alpha_k T_k - C_k.$$
Given that $x$ denotes the number of moved tracks: 

\vspace{-2mm}
\begin{eqnarray}\label{eq:5}
x= \sum_{k=1}^K (\alpha_k t- \beta_k ) [ u(t - C_k)- u(t- C_{k+1} ],
\end{eqnarray}
where $u(x)$ is the unit step function Kleinrock 1975 \cite{Klei75}.
\vspace{-1mm}
$$ 
u(x)= 
\begin{cases}
1 \mbox{ for }x \geq 0 \\
0\mbox{  for  } x < 0. 
\end{cases}
$$

The LST of seek time using Eq. (\ref{eq:4p}) and Eq. (\ref{eq:5}) is, but $p=1/C$ was used in the study: 
\begin{eqnarray}
{\cal L}^*_s (s) = \frac{p}{s} + \frac{2(1-p)}{C(C-2)}
\sum_{k=1}^K  \alpha_k (C+\beta_k ) \frac{ e^{-s T_k} - e^{-sT_{k+1}} }{s} \\
\nonumber
-\alpha_k^2  \left( \frac {  T_k e^{-s T_k} - T_{k+1} e^{-s T_{k+1}} } {s}  
-     \frac{ e^{-s T_k} -         e^{-s T_{k+1}} } {s^2}  
\right)  
\end{eqnarray}

The LST of disk service time according to Eq. (ref{eq:diskservice}) is the product of the LSTs of its components:
\vspace{-1mm}
$${\cal L}^*_{disk} (s) =  {\cal L}^*_s (s) \times {\cal L}^*_{\ell} (s) \times {\cal L}_t (s) .$$  

The moments of disk service time can be obtained 
by taking the derivatives of ${\cal L}^*_{disk}$, such as Eq. (\ref{eq:vacmoments}).


\end{framed}

Reading of RUs is modeled as two vacation types as noted earlier: 
Type 1 vacations start once the disk is idle and require a seek to access the next RU to be read.
Type 2 vacations are reads of consecutive RUs until an external request arrives.

The PDFs of two vacation types are denoted by $V_i (t),i=1,2$
In the case of Poisson arrivals the probability of no arrival (resp. an arrival) 
is $e^{-\lambda t}$ (resp. $1-e^{-\lambda t}$) in time interval $(0,t)$.
Let $p_i$ denote the probability that a request arrives during the $i^{th}$ vacation.
Unconditioning on $V_i (t)$:

\vspace{-2mm}
\begin{eqnarray}
p_i= \left[ 1- \int_0^\infty e^{-\lambda t}dV_i (t) \right] 
\prod_{j=1}^{i-1} \int_0^\infty e^{-\lambda T} dV_j(t)
=[1-  V_i^* (t)] \prod_{j=1}^{i-1} V_j^* (\lambda).
\end{eqnarray}
where $V^*_j(\cdot)$ is the LST of the $j^{th}$ vacation. 
The mean number of tracks read or vacations taken per idle period is:

\vspace{-2mm}
\begin{eqnarray}\label{eq:ntrack} 
\bar{n}_{track} = \bar{J} = \sum_{j=1}^\infty j p_j = 
1 + \frac { V^*_1 \lambda) }{1 - V^*_2 ( \lambda) }
\end{eqnarray}

The split seek-option skips track reading following a seek
if there is an arrival during the seek Thomasian and Menon 1994 \cite{ThMe94}.
Preemption during the latency and even track transfer phases 
are studied via simulation in Thomasian 1995 \cite{Thom95}.

While preemption results is an improvement in disk response times, 
rebuild time is elongated since more requests are processed 
and this results in an increased number of requests during rebuild processing (with increased response times).
The cumulative sum of response times is an appropriate metric in comparing rebuild options. 

The probability of an arrival during the first and $i^{th}$ vacation is:

\vspace{-2mm}
\begin{eqnarray}
p_1= 1 - V^*_1 (\lambda) \hspace{5mm}
p_i = [1 - V^*_2 (\lambda) ] V_1^* (\lambda) ] V^*_1 (\lambda) [V^*_2 (\lambda)]^{i-2}
\end{eqnarray}

The probability that the $i^{th}$ vacation occurs is $q_i = p_i / \bar{J}$,
where $\bar{J}$ was given by Eq. (\ref{eq:ntrack}).
The probability $q_i, i=1,2$ that an external arrival occurs during a type one or two vacation is:
\vspace{-1mm}
$$q_1 = \frac{1}{ 1 + V^*_1 (\lambda) / (1-V^*_2\lambda) }, \hspace{5mm}q_2= 1 - q_1.$$
It follows that overall LST for vacations is:

\vspace{-2mm}
\begin{eqnarray}\label{eq:LTvac}
V^* (\lambda) = \frac{ (1- V^*_2 (\lambda) ) V_1^*(s) + V^*_1 (\lambda) V_2^* (s) }
{ 1- V_2^* (\lambda) + V_1^* (\lambda) }.
\end{eqnarray}   

Derivatives of the LST Eq. (\ref{eq:LTvac} set to zero yield 
the $j^{th}$ moment of vacation/rebuild read time:

\vspace{-2mm}
\begin{eqnarray}\label{eq:vacmoments}
\overline{v^j} = 
(-1)^j \frac{dV^j(\lambda)}{d\lambda} |_{\lambda=0} = 
\sum_{i=1}^\infty q_j \overline{v_i^j}  =
\frac{ (1 - V^*_2 (\lambda) ) \overline{v}_{1^j} + V^*_1 (\lambda) \overline{v}_{2^j} }
{ 1 - V_2^* (\lambda) + V_1^* (\lambda) }.
\end{eqnarray}

The mean residual vacation/rebuild read time as noted earlier is: 
$\overline{v_r} =  \overline{v^2} / ( 2 \overline{v} )$.

The variation in disk utilization due to redirection in computing $T_{rebuild} (\rho_i)$
is taken into account by computing rebuild processing time over $k$ intervals with small load variation.
A read only workload to simplify the discussion results in the doubling of disk loads,
but otherwise the load decreases as rebuild progresses are given by Eq. (\ref{eq:degraded}).
The rate of rebuild processing accelerates as disk utilizations drop due to redirection.

\vspace{-2mm}
\begin{eqnarray}
T_{rebuild} (\rho) = \sum_{i=1}^k T_{rebuild} (\rho_i), \hspace{2mm}
\rho_i = [ 2 - \frac{i}{k}] \rho.
\end{eqnarray}

RAID5 rebuild time with readonly workload at an initial disk utilization $\rho$ 
can be approximated by Eq. (ref{eq:beta}). 
which results in highest load increase in degraded mode.
IBM 18ES 9 GByte, 7200 RPM HDDs were simulated in Fu et al. 2004 \cite{FTHN04a}.

\vspace{-2mm}
\begin{eqnarray}\label{eq:beta}
T_{rebuild} (\rho) = \frac{T_{rebuild} (0) }{1- \beta \rho},\mbox{   where  }\beta \approx 1.75.  
\end{eqnarray}

\subsection{Distributed Sparing in RAID5}\label{sec:distsparing}
\vspace{2mm}

Distributed sparing in RAID5 allocates sufficient spare areas 
on surviving disks to accommodate a single disk failure, 
i.e., to hold the data blocks of a failed disk as shown in Fig. 1 in Thomasian and Menon 1997 \cite{ThMe97}.
Empty strips are placed in right to left diagonals in parallel with parity strips at each disk. 

The main advantage of distributed sparing with respect to dedicated sparing 
is that all disks contribute to the processing of the workload in normal mode.
Rebuild reading time is slowed down because surviving disks 
in addition to being read from are written onto by reconstructed RUs. 
An iterative solution is used in \cite{ThMe97} to equalize the rate at which RUs are being read and written,
This is to prevent the overflow of DAC's buffer,
which more of a problem with CRAID where rebuild read bandwidth 
may exceed exceed that of the spare disk if the external disk load is low.

The analysis for distributed sparing for RAID5 is applicable to restriping analysis in RAID(4+K).
Once a spare disk becomes available the contents of spare areas are copied onto it in copyback mode
to return the system to its original state, which is a disadvantage of both methods.   

\subsection{Rebuild Processing Using the Permanent Customer Model}\label{sec:PCM}
\vspace{2mm}

The {\it Permanent Customer Model - PCM} proposed and analyzed in Boxma and Cohen 1991 \cite{BoCo91}
postulates a circulating permanent customer in an M/G/1 queue. 
The queue which serves external customers processes one permanent customer, 
who once completed rejoins a FCFS queue from which all customers are served. 
PCM is adopted in Merchant and Yu 1996 \cite{MeYu96} to model rebuild processing in RAID5.
FCFS scheduling is not compatible with modern disk scheduling policies 
as discussed in Section \ref{sec:sched}.

VSM for rebuild processing outperforms PCM by attaining a lower response times for external disk requests, 
since it processes rebuild read requests at a lower priority than external requests.

In PCM rebuild reads spend more time in the FCFS queue 
and this reduces the probability of consecutive rebuild reads with respect to VSM and hence increased rebuild time.
The probability that rebuild requests are intercepted by user requests arriving 
with rate $\lambda$ with VSM and PCM rebuild are given in Fu et al. 2004a \cite{FTHN04a}:

\vspace{-1mm}
$$
P_{VSM} = 1 - exp( - \lambda \bar{x}_{RU} ), \hspace{2mm}
P_{PCM} = 1 - exp( - \lambda ( \bar{x}_{RU} + W_{RU} )),
$$
$\bar{x}_{RU}$ is the time to read an RU and $W_{RU}$ is the mean waiting time in the queue for rebuild reads in PCM.
$P_{VSM} < P_{PCM}$ since $\bar{x}_{RU} + W_{RU}  > \bar{x}_{RU}$. 

Both effects are verified via simulation in Fu et al. 2004a \cite{FTHN04a},
where it is shown that VSM provides a shorter than rebuild time with PCM
fewer external requests are processed while rebuild processing is in progress.

\subsection{Rebuild in Heterogeneous Disk Array}\label{sec:HDA}
\vspace{2mm}

\begin{figure}[htb]
\centerline{
\begin{tabular}{c}
\includegraphics[scale=0.75]{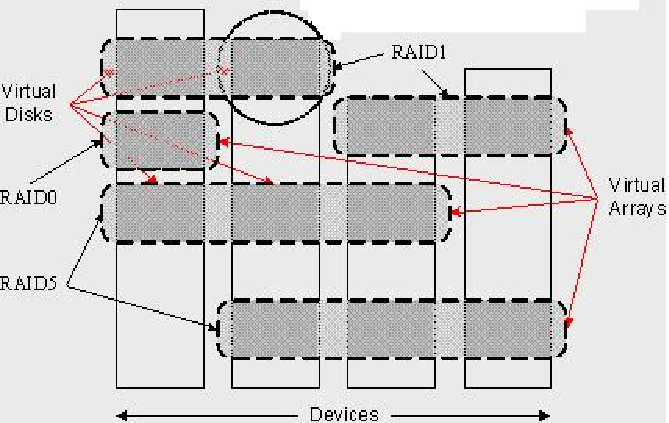}
\end{tabular}}
\caption{\label{fig:HDA} HDA with four disks
and five VAs with different RAID levels:
two RAID1 VAs with 2 disks each,
two RAID5 VAs with 3 disks each,
and a single disk referred to as RAID0,
since it has no redundancy.}
\end{figure}

{\it Heterogeneous Disk Array - HDA} postulates a DAC, 
which can manage multiple RAID levels Thomasian and Xu 2011b \cite{ThXu11b} as shown in Fig. \ref{fig:HDA}
The choice of RAID1 and RAID5 for {\it Virtual Arrays - VAs} is based on their suitability for given applications,
e.g., RAID1 for OLTP and RAID5 for data mining requiring table scans. 
The VA width $W$ is selected to limit the physical disk load per {\it Virtual Disk - VD}
below a certain percentage of disk bandwidth expressed in IOPS.

VAs with higher dependability requirements are allocated as RAID$(4+k),k \geq 1$.
Variable redundancy levels makes it unnecessary to adopt the highest redundancy level 
required by a few VAs to all VAs.
In the case of cloud storage where users are charged for IOs
allowing diversity in RAID levels reduces storage costs.

Given that disks are bandwidth rather than capacity bound 
the number of allocations can be increased using the clustered RAID paradigm in Section \ref{sec:CRAID}. 

The study is concerned with alternative data allocation methods 
for {\it Virtual Disks - VDs} constituting VAs.
Allocations are carried out in degraded mode, as if one of the VDs of a VA has failed to avoid overload if a disk fails,
but in the case of RAID(4+k) allocations with up to $i \leq k $ VDs can be allowed.

Two dimensions: disk bandwidth and capacity per VD are considered in this study,
based on an improved best-fit allocation algorithm for the HDA to meet this goal.
Capacities are less relevant for HDDs, but would be so if SSDs are included, 
and vice versa especially for accesses to small blocks.
Note that even in 2012 a bleak feature was predicted for Flash SSDs and hence SSDs were not included in this early study.
\footnote{\url{https://www.usenix.org/conference/fast12/bleak-future-nand-flash-memory}}

As far as rebuild processing is concerned given that the disk layout 
is known there is no need to read unallocated disk areas.
since they can be assumed to be set to zero and need not participate in rebuild processing,
e.g., if 4th disk fails then the blocks of RAID5 can be reconstructed on Disk 1 
by XORing corresponding blocks on 2nd and 3rd disks in Fig. \ref{fig:HDA}.
It is also possible to prioritize the rebuild processing of VAs used by more critical applications, 
e.g., OLTP for e-commerce generates more revenue than data mining. 

When RAID(4+k) arrays share space on a broken disk rebuilding of arrays 
with smaller $k$ is prioritized to prevent data loss.
More specifically the rebuild of RAID(4+k) with smallest $(k-i)$, 
where $i$ is the number of its failed VDs, should be prioritized.  
Reliability and performance of two HDA configurations are compared in Section \ref{sec:HDArel}.


A two level allocation scheme was considered in Thomasian et al. 2005 \cite{ThBH05,Thom05}.
Heterogeneous disks but not Flash SSDs were not considered in this study.
The characteristics of disks used in this study are given in Table \ref{tab:HDA}.

\begin{table}
\begin{center}
\begin{footnotesize}
\begin{tabular}{|c|c|c|c|}\hline
Model         &Seagate    &Seagate          &Maxtor      \\
              Barracuda   &Atlas &10K       &15K         \\ \hline   
Year          &1997       &1999             &2003        \\ \hline   
Capacity      &2 GB       &9.1 GB           &73 GB       \\ \hline
IOPS (4KB)    &74.5       &12.84            &1882        \\ \hline
IOPS/Capacity &37,25      &12.84            &2.58        \\ \hline
\end{tabular}
\end{footnotesize}
\caption{HDDs used in this study and bandwidth/capacity ratio\label{tab:HDA}.}
\end{center}
\end{table}

URSA minor cluster based storage in Abd-el-Malek et at. 2005 \cite{Abd+05}  
and IBM's {\it RAID Engine and Optimizer - REO} project
Kenchamana-Hosekote et al. 2007 \cite{KeHH07} have similarities to HDA.

\subsection{Optimal Single Disk Rebuild in RAID6 Arrays}\label{sec:singlediskrebuild}
\vspace{2mm}

The rebuild method proposed in the context of RDP is optimal in the number of accessed pages 
and balancing disk loads Xiang et al. 2010 \cite{XXLC10}.
It is erroneously assumed that symbols are as large as strips.
The method was extended to the EVENODD layout in Xiang et al. 2011 \cite{XXL+11},
which uses small symbol sizes like RDP as discussed in Section \ref{sec:RDP}.  
 
The method however is applicable to X-code arrays Xu and Bruck 1999 \cite{XuBr99}, 
since it applies coding at strip level Xiang et al. 2014 \cite{XLL+14}.
A similar method to minimize page accesses in X-code was proposed in Thomasian and Xu 2011 \cite{ThXu11}.

Given $N$ disks, which is a prime number, 
X-code consists of repetitive segments of $N \times N$ strips.
Rows in a segment are indexed as $0:N-1$, rows $0:N-3$ hold data strips,
row $N-2$ holds P diagonal parities with positive slope, 
and row $N-1$ Q parities with negative slope, thus the redundancy level is $2/N$
and each data block is protected by two parity groups, which are diagonals with $\pm 1$ slopes.
hence we have an MDS array since it is shown in Xu and Bruck \cite{XuBr99}, 
where it is shown that recovery is possible for all two disk failures.

Fig.~\ref{fig:slope--1} show the data layouts for $N=7$ columns.
$B_{i,j}$ denotes a data block (strip) at the $i^{th}$ row (stripe) and the $j^{th}$ column (disk).
PG memberships are specified by the number for the corresponding P and Q parities.
The two parities are independent and given that ${\langle{X}\rangle}_N \equiv X \ \mbox{mod} \ N $
then $p(i)$ and $q(i)$ for $0 \le i \le N-1$ are computed as follows:
\vspace{-1mm}
$$p(i) = B_{N-2,i} = \sum_{k=0}^{N-3} B_{k, {\langle{i-k-2}\rangle}_N},  \hspace{5mm}
q(i) = B_{N-1,i} = \sum_{k=0}^{N-3} B_{k, {\langle{i+k+2}\rangle}_N}.$$

The parities to be updated are determined by the {\it Parity Group - PG} to which the block belongs. 
The PG size is $N-1=6$ for both parities and either PG can be accessed to reconstruct a block,
so that the load increase in degraded mode is the same as RAID6.
Data recovery with two disk failures in X-code is investigated in Thomasian and Xu 2011 \cite{ThXu11},
where it is shown that the load increase is affected by the distance of two failed disks.
A balanced load increase can be attained by rotating successive $N \times N$ segments.

\begin{figure}[t]
\begin{footnotesize}
\begin{center}
\begin{tabular}{|c||c|c|c|c|c|c|c|}\hline
    Row\# & $D_0$ & $D_1$ & $D_2$ & $D_3$ & $D_4$ & $D_5$ & $D_6$ \\ \hline\hline
    0     & $B_{0,0}^2$ & $B_{0,1}^3$ & $B_{0,2}^4$ & $B_{0,3}^5$ & $B_{0,4}^6$ & $B_{0,5}^0$ & $B_{0,6}^1$ \\ \hline
    1     & $B_{1,0}^3$ & $B_{1,1}^4$ & $B_{1,2}^5$ & $B_{1,3}^6$ & $B_{1,4}^0$ & $B_{1,5}^1$ & $B_{1,6}^2$ \\ \hline
    2     & $B_{2,0}^4$ & $B_{2,1}^5$ & $B_{2,2}^6$ & $B_{2,3}^0$ & $B_{2,4}^1$ & $B_{2,5}^2$ & $B_{2,6}^3$ \\ \hline
    3     & $B_{3,0}^5$ & $B_{3,1}^6$ & $B_{3,2}^0$ & $B_{3,3}^1$ & $B_{3,4}^2$ & $B_{3,5}^3$ & $B_{3,6}^4$ \\ \hline
    4     & $B_{4,0}^6$ & $B_{4,1}^0$ & $B_{4,2}^1$ & $B_{4,3}^2$ & $B_{4,4}^3$ & $B_{4,5}^4$ & $B_{4,6}^5$ \\ \hline
    5     & $B_{5,0}^0$ & $B_{5,1}^1$ & $B_{5,2}^2$ & $B_{5,3}^3$ & $B_{5,4}^4$ & $B_{5,5}^5$ & $B_{5.6}^6$ \\ \hline
    6     & & & & & & &\\ \hline
\end{tabular}
\hspace{5mm}
\begin{tabular}{|c||c|c|c|c|c|c|c|}\hline
\hline
      Row\# & $D_0$ & $D_1$ & $D_2$ & $D_3$ & $D_4$ & $D_5$ & $D_6$ \\ \hline\hline
    0 & $B_{0,0}^5$ & $B_{0,1}^6$ & $B_{0,2}^0$ & $B_{0,3}^1$ & $B_{0,4}^2$ & $B_{0,5}^3$ & $B_{0,6}^4$ \\ \hline
    1 & $B_{1,0}^4$ & $B_{1,1}^5$ & $B_{1,2}^6$ & $B_{1,3}^0$ & $B_{1,4}^1$ & $B_{1,5}^2$ & $B_{1,6}^3$ \\ \hline
    2 & $B_{2,0}^3$ & $B_{2,1}^4$ & $B_{2,2}^5$ & $B_{2,3}^6$ & $B_{2,4}^0$ & $B_{2,5}^1$ & $B_{2,6}^2$ \\ \hline
    3 & $B_{3,0}^2$ & $B_{3,1}^3$ & $B_{3,2}^4$ & $B_{3,3}^5$ & $B_{3,4}^6$ & $B_{3,5}^0$ & $B_{3,6}^1$ \\ \hline
    4 & $B_{4,0}^1$ & $B_{4,1}^2$ & $B_{4,2}^3$ & $B_{4,3}^4$ & $B_{4,4}^5$ & $B_{4,5}^6$ & $B_{4.6}^0$ \\ \hline
    5   &  & & &  &  &  & \\ \hline
    6 & $B_{6,0}^0$ & $B_{6,1}^1$ & $B_{6,2}^2$ & $B_{6,3}^3$ & $B_{6,4}^4$ & $B_{6,5}^6$ & $B_{6,6}^6$\\ \hline
\end{tabular}
\end{center}
\end{footnotesize}
\caption{ \label{fig:slope--1}
X-code array with $N=7$ disks with parity group for $p(i)=B(5,i), 0 \leq i \leq 6$ with slope=+1 
and parity group for $q(i) = B(6,j), 0 \leq i \leq 6$ at row 6 with slope=-1.}
\end{figure}


\subsubsection{Optimizing Storage Allocation}\label{sec:optimal}
\vspace{1mm}

File migration between disks and archival storage is studied in A. J. Smith 1981 \cite{Smit81}
and it is shown that algorithms based on both file size and time since last use work best.
Miss ratios are compared with page replacement algorithms such as P. J. Denning's Working Set. 
\url{https://denninginstitute.com/pjd/PUBS/Starting.html}
Data migration policies are reviewed here.
\footnote{\url{\https://en.wikipedia.org/wiki/Data_migration}}

ExaPlan uses a queueing model for a given budget and workload to determine both the data-to-tier assignment and 
the number of devices in each tier that minimize the system's mean response time Iliadis et al. 2017 \cite{Ili+17}.
IBM's {\it Hierarchical Storage Management - HSM} moves data both  horizontally across disks to balance loads,
but also vertically from slower to faster devices and vice-versa.
\footnote{\url{https://en.wikipedia.org/wiki/Hierarchical_storage_management}  }
       
Heuristic algorithms to improve overall storage performance and 
throughput do not work well for all conditions and workloads. 
There are numerous tunable parameters for users with continually optimizing their own storage systems and applications. 
Storage systems are usually responsible for most latency in I/O dominated applications, 
and the effect of small I/O latency improvement can be significant. 

{\it Machine learning - ML} techniques learn patterns and generalize from them, 
and enable optimal solutions that adapt to changing workloads. 
KML is applied to optimal readahead and read size values in {\it Network File System - NFS} Akgun et al. 2023 \cite{AAB+23} 
\footnote{\url{https://en.wikipedia.org/wiki/Network_File_System}} 
KML consumes little OS resources, adds negligible latency, 
and yet can learn patterns that can improve I/O throughput by as much as 2.3x or 15x in two cases.
KML is ti be extended with reinforcement learning Kaelbling et al. 1996 \cite{KaLM96}.

{\it Adaptive Intelligent Tiering - AIT} described in Pang et al. 2023 \cite{PAR+23}
dynamically adapts to the changes of storage accesses 
by the running applications using a deep learning model to generate a set of candidate movements 
and employs a reinforcement learning mechanism to further refine the candidates. 
Simulations of a 3-tier system show that the proposed scheme
enhances workload performance up to 85\% on storage traces with a wide range of characteristics.

\subsection{An Alternate Method to Estimate Rebuild Time$^*$}\label{sec:alternative}
\vspace{2mm}

Rebuild time is estimated as the ratio of utilized disk capacity ($U$) and the mean rebuild rate ($b_d$).
Memory bandwidth is not expected to affect RAID5 rebuild time,
as shown in Section 5.2 of Dholakia et al. 2008 \cite{DEH+08}. 

Zoning subdivides disk tracks into $I$ bands with the same number of sectors,
so that for small variation of track circumference the number of sectors is not varied track to track.
There are $n_i$ tracks with $c_i$ sectors per track in the $i^{th}$ band.
The data portion of the sector ($s_{sector}$) was increased from 512B to 4096 B circa 2010. 
The mean disk transfer rate is:                        
\vspace{-1mm}
$$t_d \approx \sum_{i=1}^I  n_i  \times c_i \times s_{sector}  / T_R .$$              
 
Given that the RU size is $s_{RU}$ the transfer time per RU is $s_{RU}/t_d$,
but access to the first RU according to VSM requires a seek ($\bar{x}_s$) 
plus rotational latency estimated as follows.
Let $f$ denote the size of RU as a fraction of tracks size,
then if the R/W head lands inside the RU then we need a full disk rotation to transfer an RU ($f \times T_R$) 
and otherwise the cost is $(1-f)^2 T_R /2$, hence:
\vspace{-1mm}
$$\bar{x}_{\ell} =  f \times T_R + (1-f)^2 T_R/2 =  (1+f^2)T_R/2$$
Track or cylinder skew should be added if track or cylinder boundaries are traversed.
For large RU sizes track and cylinder skew Jacob et al. 2008 \cite{JaNW08} have a negligible effect. 

The number of bytes transferred per cycle is the product 
of the number of RUs transferred ($\bar{n}_{RU}$) and RU size $(s_{RU})$.
The cycle time is the sum of interarrival time ($1/\lambda$),
plus the delay cycle for processing external disk requests given by Eq. (\ref{eq:dc}):
\vspace{-1mm}
$$\bar{T}_{dc} = \frac{ \bar{x}_{disk}+\bar{v}_r}{1-\rho}$$ 

\vspace{-2mm}
\begin{eqnarray}\label{eq:bd}
b_d = \frac { \bar{n}_{RU} \times s_{RU} }
{ \bar{T}_{dc} + \bar{x}_{seek} + \bar{x}_{\ell} + \bar{n}_{RU} \times \bar{x}_{RU} }
\end{eqnarray}

where $\bar{x}_{RU}= \bar{s}_{RU}/t_d$. 
An upper bound to rebuild time can be obtained by setting $\bar{n}_{RU}=1$,
which is also the assumption made in Dholakia et al. 2008 \cite{DEH+08}.
The number of consecutively transferred RUs with no intervening seeks can be derived as follows. 
\vspace{-1mm}
$$p_n = (1- e^{-\lambda x_{RU}}) (e^{-\lambda x_{RU}})^{n-1} , n \geq 1
\mbox{  hence  }\bar{n}_{RU} = (1 - e^{-\lambda s_{RU}})^{-1}$$

In the case of ZBR outer disk tracks have higher capacities
and a small fraction of disk tracks are utilized for critical data, 
since this allows short strocking.
With a fraction $U$ of disk capacity: $C_d$ utilized rebuild time is:

\vspace{-3mm}
\begin{eqnarray}
T_{rebuild} = U C_d / b_d.
\end{eqnarray} 

Dholakia et al. 2008 \cite{DEH+08} estimates $b_d$ in Eq. (13) as follows:  
\vspace{-1mm}
$$b_d = s_{req} / ( 1/r_{io} +s_{req} / t_d ) ,$$
where $r_{io}$ is the request rate and $s_{req}$ is the request size. 

\subsection{Performance Studies of Log-structured File Systems - LFS}\label{sec:LFS}
\vspace{2mm}

A performance study of LFS in McNutt 1994 \cite{McNu94}
postulates a directory for the most recent copy of objects written more than once.
A mathematical model of garbage collection is used to show how load
relates to the utilization of storage and the amount of update locality present in the pattern of updates.
A realistic statistical model of updates, based upon trace data analysis is applied
and alternative policies are examined for determining which areas to target.

The following relationship for LSA is quoted in Menon 1995 \cite{Meno95}
\vspace{-1mm}
$$\mbox{ASO}=(1-\mbox{BSO}) / \mbox{ln}(1- \mbox{BSO}), \hspace{5mm}\mbox{e.g., if ASO=0.6 then BSO=0.324}.$$
where {\it Average Segment Occupancy - ASO} and lowest or {\it Best Segment Occupancy - BSO}.
Two types of write hits are considered: 
(1) to dirty blocks in NVRAM caches, (2) to clean pages.
Given a request rate $K$ the fraction of request to clean pages is $C$ 
and the fraction of writes to dirty ages is $D$.

The reading of segments is attributable to garbage collection
and only a fraction BSO of segment writing is attributable to garbage collection and the rest to destage.
The read miss rate is $K \times B$, the miss rate due to write hits in clean pages is $K \times C$
and the destage rate due to write misses $K \times D$.

Given that $X$ blocks are destaged together and $Z$ is the segment size in tracks
the rate at which dirty blocks are created in the NVS caches is $K(C+D)/X$.
It is then argued that segments are written and read at rate $(KC+KD/(X (1-BSO) Z)$
The effect of data compression and seek affinity is investigated in this study. 
It is observed that LSA outperforms RAID5 in throughput, but offers higher response times.

A new algorithm for choosing segments for garbage collection in LFS and LSA
proposed in Menon and  Stockmeyer 1998 \cite{MeSt98}
is compared against a greedy and cost-benefit algorithms via simulation.
Segments which have been recently filled by writes 
are not to be considered for garbage collection until they achieve a certain age-threshold, 
because as time progresses such segments are less likely to be updated
and this makes old segments eligible for garbage collection.
The algorithm chooses segments that yield the most free space.

\subsection{Performance Analysis of a Tape Library System}\label{sec:tapelibrary}
\vspace{2mm}

An analytical model to evaluate the performance of a tape library system 
is presented in Iliadis et al. 2016 \cite{IKSV16}. 
Earlier studies of this topic is are reviewed.

The analysis takes into account the number of cartridges and tape drives 
as well as different mount/unmount policies to determine the mean waiting time for tapes. 
The accuracy of the model is confirmed by validation against measurements. 

A tape library consists of tape drives, robot arms,
a storage rack for the tape cartridges, and a cartridge control unit.
To serve a request, a robot arm fetches the appropriate tape cartridge
from the storage rack and delivers it to a free tape drive.
The tape drive control unit mounts the tape, 
positions the head to the desired file and then transfers its data.
To free a tape drive, a robot arm unmounts the tape cartridge and returns it to the storage rack.
Tape read/write requests contain the cartridge id, 
the position of the data block in the cartridge, and its size.

Requests submitted for cartridges are queued and served according to a scheduling policy.
A hierarchical scheduling algorithm is used to ensure fairness and avoiding starvation as follows:
{\bf Upper level:} Cyclic or round-robin scheduling among the queues (mounting cartridges).
{\bf Lower level:} A FCFS policy for serving requests within a queue (reading from a mounted cartridge).


When all requests of a cartridge are served (exhaustive service),
and there are still pending requests at some other, non-mounted cartridge,
an unused cartridge is unmounted and another cartridge with pending requests is mounted.
If, however, there are no other pending requests to any other non-mounted cartridge,
the cartridge can either remain mounted in anticipation of future requests
or be unmounted so as to save time when future requests arrive for other non-mounted cartridges.
Two mount/unmount policies deployed in this context are:
{\bf 1. Always-Unmount - AU:} A tape cartridge is immediately unmounted upon completion of all pending requests 
for it in anticipation of the next request arriving for another non-mounted cartridge.
{\bf 2. Not-Unmount - NU:}
A tape cartridge remains mounted upon completion of all pending requests for it in anticipation 
of the next request arriving for this same currently mounted, but idle cartridge.

The performance analysis of the resulting polling system 
relies on state-dependent queues Takagi 1986 \cite{Taka86}.  
Parameters considered in this study are based on Table II in Iliadis et al. 2021 
of the IBM TS4500 tape library in Table \ref{tab:tapelibrary}
\footnote{\url{https://www.gotomojo.com/wp-content/uploads/2019/08/IBM-TS4500-Tape-Library-Data-Sheet.pdf}}

\begin{table}
\begin{footnotesize}
\begin{center}
\begin{tabular}{|c|c|c|}    \hline
Parameter        &Values        & Definition            \\ \hline \hline
c               &3200           & number of cartridges  \\ 
d               &32             & number of tape drives \\ 
a               &1,2            & number of arms        \\ 
R               &5 s (fixed)    & robot transfer time   \\ 
$t_L$           &24 s (fixed)   & load ready time       \\
$\bar{t}_R$     &59 s           & mean rewind time      \\  
$\bar{t}_U$     &24 s           & unload ready time     \\ 
$s_{max}$       &118 s          & maximum seek time     \\ 
$\bar{Q}$       &843 MB         & mean request size     \\ 
$\overline{Q^2}$ &2.8 GB        & standard deviation    \\ 
$b_w$           &360 MB/s       & bandwidth             \\ \hline
$n$             &100            & \# cartridges per tape drive \\ 
$M$             & 20 s (fixed)  & mount time $M=R+t_L$  \\ 
$\bar{U}$       & 88 s          & mean unmount time ($\bar{U}=\bar{t}_R+t+U +R$) \\
$\bar{t}_T$     &2.34           & mean transfer time ($t_T = \bar{Q}/b_w$   \\ \hline
\end{tabular}
\end{center}
\end{footnotesize}
\caption{\label{tab:tapelibrary}
Parameter values according to Table II in the paper.}
\end{table}

Predicting the performance of a tape library system is key to efficiently dimensioning it.
The effect of two mount/unmount policies was analytically assessed.
It was observed that for light loads, the AU - Always-Unmount policy yields a mean delay
that is lower than that of the NU - Not-Unmount policy,
with the difference becoming negligible as the load increases.
The effect of the number of robot arms was assessed by means of simulation.
It is shown that given fast robot arms multiple arms do not provide a significant performance improvement,
but multiple arms may be necessary to attain higher availability.
The analysis has been extended in Iliadis et al. 2021 \cite{IJLS21}

\clearpage
\section{Appendix II: RAID Reliability Analysis}\label{sec:reliability}
\vspace{3mm}

\begin{small}
\ref{sec:RAIDrel} RAID5 reliability analysis.
\ref{sec:cubeRAID} Reliability of 2D square- and 3D cube-RAID.                           
\ref{sec:metrics}. System reliability metrics.                                           \newline
\ref{sec:placement}. Clustered versus declustered data placements.                       \newline
\ref{sec:relanal}$^*$. RAID5/6 reliability analysis with unrecoverable sector errors.    \newline
\ref{sec:DEH}$^*$. Disk scrubbing and Intra-Disk Redundancy to deal with Latent Sector Errors - LSEs.    \newline
\ref{sec:IDRschemes}. Schemes to implement IDR.                                          \newline
\ref{sec:IDRperf}. IDR's effect on disk performance.                                     \newline
\ref{sec:comparison}. Comparison of disk scrubbing and IDR with IPC.                     \newline
\ref{sec:combined}. Combining scrubbing with SMART and Machine Learning.                 \newline
\ref{sec:failslow} Fail-slow faults in drives and their detection.                       \newline 
\end{small}

\subsection{RAID5 Reliability Analysis}\label{sec:RAIDrel}
\vspace{2mm}

RAID$(4+k)$ disk arrays with $N$ data and $k$ checks disks 
continue their operation after a single disk failure without data loss 
by on-demand recovery of requested data blocks but with degraded performance.
Denoting disk reliability as $R(t)=1-F(t)$, where $F(t)$ is the PDF of disk lifetime 
reliability of RAID$(4+k), k \geq 1$ arrays with no repair is given as follows:

\vspace{-2mm}
\begin{eqnarray}\label{eq:RAID(4+k)}
R_{RAID(4+k)} (t) = \sum_{i=0}^k \binom{N+k}{i} R^{N+k-i}(t)  (1-R(t))^i.
\end{eqnarray}

The {\it Mean Time To Failure - MTTF} of PC disks was crudely estimated 
based on customer returns of broken disks to manufacturers Gibson 1992 \cite{Gibs92} 
using statistical methods for small samples in Lawless 2003 \cite{Lawl03}.
Notable studies of storage failure are: 
Bairavasundaram et al. 2007 \cite{BGPS07}:
(1) there is a high degree of temporal locality between successive LSEs.
(2) the vast majority of disks developed relatively few errors during first three years,
but these few errors cause significant data loss if not detected proactively by disk scrubbing.
Schroeder and Gibson 2007 \cite{ScGi07}, 
Pinheiro et al. 2007 \cite{PiWB07},
Jiang et al. \cite{JHZK08}, 
DRAM faults Schroeder et al. 2009 \cite{ScPW09},
{\it Latent Sector Errors - LSE} Schroeder et al. 1010 \cite{ScDG10},
NAND flash-based SSDs in Schroeder et al. 2017 \cite{ScML17}, 
SSDs Maneas et al. 2021 \cite{MMES21}.

While the two parameter Weiball distribution Trivedi 2001 \cite{Triv01} 
as utilized by Elerath and Schindler 2014 \cite{ElSc14}, is a better fit for disk lifetimes. 
Due to its mathematical tractability disk reliability  was approximated 
a single parameter negative exponential distribution by Gibson 1992 \cite{Gibs92}: 
$r(t) = e^{-\delta t}$, where $\delta$ is the disk failure rate.
\footnote{We are using $\delta$, since we use $\lambda$ to denote the arrival rate in queueing systems.}
The HDD {\it Mean Time to Failure - MTTF}$=\int_0^\infty R(t)dt = 1/\delta$ 
and the RAID5 {\it Mean Time to Repair - MTTR}$=1/\mu$ are two key input parameters for RAID analysis.

RAID5 rebuild processing may be started after a brief delay to allow for fail-slow faults \ref{sec:failslow}
without human intervention provided a hot spare is available. 
Data loss occurs if a second disk fails before rebuild is completed,
but a much more common cause of rebuild failures are {\it Latent Sector Errors - LSEs} 
as discussed in Section \ref{sec:DEH}.

RAID6 provides an extra check disk based Q whose blocks 
are computed using Reed-Solomon coding Fujiwara 2006 \cite{Fiji06}
Note that RAID6 remains operational in degraded mode with two failed disks.

RAID5 reliability analysis is given in Section 5.4.1 in Gibson 1992 \cite{Gibs92},
which is repeated below..
\footnote{Out of print and also not available as an EECS Berkeley dissertation.} 
A similar analysis in the context of mirrored disks is given by Example 8.34 in Trivedi 2001 \cite{Triv01}. 


\begin{framed}
\subsection*{Transient Analysis of CTMC for RAID5 Rebuild}
\vspace{2mm}

The evolution of the system is modeled by a CTMC with three states ${\cal S}_i$, 
where $i=0,1,2$ is the number of disks failed disks. We have the following transitions:

\vspace{-1mm}
$$
{\cal S}_0 \xrightarrow{(N+1) \delta} {\cal S}_1 , \hspace{5mm} 
{\cal S}_0 \xleftarrow{\mu} {\cal S}_1 , \hspace{5mm}
{\cal S}_1 \xrightarrow{N \delta} {\cal S}_2.
$$

The system evolution is described by the following set of linear {\it Ordinary Differential Equations - ODEs}: 

\vspace{-1mm}
$$\frac{dP_0(t)}{dt} = - (N+1) \delta P_0(t) + \mu P_1(t), \hspace{2mm}
\frac{dP_1(t)}{dt} = - (\delta + \mu ) P_1(t) (N+1) \delta P_0(t), \hspace{2mm}
\frac{dP_2(t)}{dt} = N \delta P_1 (t). $$ 


The LST $L^* (s) = \int_0^\infty P(t) e^{-st} dt$ for the ODEs are given by Eq. (\ref{eq:LT})  

\vspace{-2mm}
\begin{align}\label{eq:LT}
s {\cal L}^*_0 (s) - P_0(0) &= -(N+1) \delta {\cal L}^*_0 (s) + \mu {\cal L}^*_1 (s),\hspace{5mm}P_0(0)=1. \\
\nonumber
s {\cal L}^*_1 (s) - P_1)0) &= -(\delta + \mu ) {\cal L}^*_1 (s) + (N+1) {\cal L}^*_0 (s)\hspace{5mm}P_1(0)=0. \\
\nonumber
s {\cal L}^*_2 (s) - P_2(0) &= N \delta {\cal L}^*_1 {s}\hspace{5mm}P_2(0)=0. 
\end{align}


Solving for ${\cal L}^*_2(s)$ leads to  

\vspace{-2mm}
\begin{eqnarray}\nonumber
{\cal L}^*_Y (s) = s {\cal L}^*_2 (s) = \frac{2 \delta ^2} {s^2 + ((2N+1)\delta + \mu)s +2N \delta^2}  
= \frac{2 \delta^2} {\zeta-\eta} \left( \frac{1}{s+\zeta} - \frac{1}{s+\eta} \right)         
\end{eqnarray}

The reliability is the sum of probabilities that the system is functional. 

\vspace{-2mm}
$$ R (t) = P_0 (t) + P_1 (t) = 1- P_2 (t)\mbox{ so that } f_Y(t) = - \frac{dR(t)}{dt} = \frac{dP_2(t)}{dt}. $$

\vspace{-3mm}
\begin{eqnarray}\label{eq:RAID5t}
R_{RAID5} (t) = \frac{\zeta e^{\eta t} - \eta e^{\zeta t}}{\zeta - \eta}
\mbox{ where: }
\zeta,\eta =  \frac{1}{2}[-(2N+1)\delta+\mu \pm \sqrt{\delta^2+ \mu^2+2(2N+1)\delta \mu}]. 
\end{eqnarray}


Given $R(t)$ the MTTDL can be obtained where we have
substituted $\delta^{-1} = \mbox{MTTF}$ and $\mu^{-1}=\mbox{MTTR}$ leads to:


\vspace{-2mm}
\begin{eqnarray}\label{eq:R5MTTDL}
\mbox{MTTDL}_{\mbox{RAID5}} = 
\int_{t=0}^\infty R_{RAID5} (t) dt =
\frac{ (2N+1) \delta + \mu }{ N(N+1) \delta^2 }\approx \frac{\mu}{N(N+1)\delta^2} = 
\frac{MTTF^2}{N(N+1)\mbox{MTTR}}.
\end{eqnarray}

\end{framed}

	

A simple method to derive Eq. (\ref{eq:R5MTTDL}) is to multiply 
the number of times the normal state: ${\cal S}_0$ and the degraded state: ${\cal S}_1$ are visited, 
before there is a transition to the failed state: ${\cal S}_2$.
Given that the probability of an unsuccessful rebuild is $p_f = N \delta / (N \delta + \mu)$,
then the probability of $k$ successful rebuilds and the mean number of successful is: 

\vspace{-1mm}
$$p_k = p_f (1-p_f)^{k-1}, k \geq 1 \mbox{  with  mean  }
\bar{k}= 1/p_f=  1 +\mu/(N \delta) \approx \frac{MTTDL}{N \times MTTR}.$$

$\mbox{MTTDL}_{RAID5}$ is obtained by multiplying $\bar{k}$ by the 
sum of holding times in ${\cal S}_0$ and ${\cal S}_1$.

Chen et al. 1994 \cite{Che+94} provide the MTTDL for RAID5 and RAID6 
with $N$ disks in total and parity group size $G$:

\vspace{-2mm}
\begin{eqnarray} 
\mbox{MTTDL}_{\mbox{RAID5}} \approx \frac{\mbox{MTTF}^2}{N(G-1)\mbox{MTTR}}, \hspace{5mm}
\mbox{MTTDL}_{\mbox{RAID6}} \approx \frac {\mbox{MTTF}^3 }{N(G-1)(G-2)\mbox{MTTR}^2}.
\end{eqnarray}

Generalizing the formula and with one parity group ($N=G$) we have:

\vspace{-2mm}
\begin{eqnarray}\label{eq:MTTDL}
\mbox{MTTDL}_{\mbox{RAID}(4+k)} \approx
\frac{\mbox{MTTF}^{k+1}} { N(N-1) \ldots (N-k+1) \mbox{MTTR}^k }
\end{eqnarray}

Chen's formula assumes that the repair rate remains fixed 
regardless of the number of failed disks and one disk is repaired at a time.
With two failed disks in RAID6 corresponding strips in both disks are reconstructed together. 


Given $N$ disks and with $i \leq k$ failed disks we have the transitions:
\vspace{-1mm}
$${\cal S}_{N-i} \rightarrow {\cal S}_{N-(i+1)}: \hspace{2mm} (N-i)\delta \hspace{5mm}
{\cal S}_{N-(i+1)} \leftarrow {\cal S}_{N-i}: \hspace{2mm} \mu $$

A repair rate proportional to the number of disks is postulated in Angus 1988 \cite{Angu88}. 
and furthermore the number of failed disks is allowed to exceed $n-k$.
\vspace{-1mm}
$${\cal S}_{N-i} \rightarrow {\cal S}_{N-(i+1)}: \hspace{2mm} (N-i)\delta \hspace{5mm}
{\cal S}_{N-(i+1)} \leftarrow {\cal S}_{N-i}: \hspace{2mm} i \mu $$

Angus' MTTDL formula is as follows:

\vspace{-2mm}
\begin{eqnarray}
MTTDL_{Angus} = \frac{ \mbox{MTTF}^{n-k+1} }{k \binom{n}{k} \mbox{MTTR}^{n-k} }
\times \sum_{i=0}^{n-k} \binom{n}{i} ( \frac{MTTR}{MTTF} )^i \approx 
\frac{ \mbox{MTTF}} { k \binom{n}{k} } 
( \frac {\mbox{MTTF} }{ \mbox{MTTR}} )^{n-k}.
\end{eqnarray}

It can be observed from Table \ref{tab:Resch} obtained by simulation in Resch and Volvivski 2013 \cite{ReVo13} 
that Chen's formula overestimates MTTDL by a factor $\approx(n-k)!$.

\begin{table}[b]
\centering
\begin{footnotesize}
\begin{tabular}{|c|c|c|c|c|c|c|c|}\hline
$n$    &$k$    &MTTF           &Simul     &Chen     & S/C    & Angus    & S/C   \\ \hline \hline
10     &10     &2000           &1.988E2   &2.000E2  & 0.994  &1.988E2   &0.994  \\ \hline
10     &9      &2000           &4.488E4   &4.444E4  & 1.010  &4.467E4   &1.005 \\ \hline
10     &8      &1500           &9.446E6   &4.688E6  & 2.105  &9.438E6   &1.001  \\ \hline
10     &7      &500            &7.786E7   &1.240E7  & 6.278  &7.591E7   &1.026 \\ \hline    
10     &6      &200            &6.407E7   &2.511E6  & 25.513 &6.441E7   &0.995  \\ \hline  
\end{tabular}
\end{footnotesize}
\caption{\label{tab:Resch}Validation of Chen and Angus MTTDL formulas versus simulation.
MTTR=1 hour and small values of MTTF are used for simulation efficiency.}
\end{table}



A RAID5 system is considered operational with no or one disk failures, 
but the two states differ substantially performance-wise, 
e.g., the maximum read bandwidth is halved when a disk fails.
In the case of readonly workload $i<k$ disk failures results 
in an $(i+1)$-fold reduction in throughput.

Performability combines reliability and performance, see Trivedi 2001 \cite{Triv01}.
In the context of disk arrays it may be defined 
as the number of I/Os after first failure to the time of data loss occurs. 
Performability of four RAID1 variants in Section \ref{sec:mirhyb} 
is obtained in Thomasian and Xu 2008 \cite{ThXu08}.


In reliability modeling RAID5 can be substituted by a single component. 

\begin{eqnarray}\label{eq:appr}
R_{\mbox{RAID5}}^{appr} = e^{-t/\mbox{MTTDL}}
\end{eqnarray}
where $\mbox{MTTDL}$ is given by Eq. (\ref{eq:R5MTTDL}).



RAID designs in the context of a 70 disk strawman array is presented in Gibson 1992 \cite{Gibs92} 
and Gibson and Patterson 1993 \cite{GiPa93}.
Assuming that there are $G_{string}$s with $\mbox{MTTF}_{string}$
taking into account component unreliabilities the overall failure rate is given as:  

$$\frac{1}{MTTDL_{RAID}} = \frac{G(N+1)(N=2)MTTR_{disk}}{MTTF^2_{disk}}
 + \frac{G_{string}}{MTTF_{string}}$$

\begin{framed}
\subsection*{Crosshatch Disk Array}
\vspace{2mm}

The crosshatch disk array is proposed in Ng 1994 \cite{NgSW94}.
It takes into account supporting hardware such as a controller or cable failures.
The array consists of 7+P RAID5s placed as an array of $N \times N =64$ disks.
Faults causing data loss for the following configuration are as follows. 

\begin{description}
\item[1. Single path horizontal:]
(1) double disk failure, 
(2) single disk and single controller failure,
(3) double controller failure, MTTDL=$2.32 \times 10^6$ 

\item[2. Dual path vertical array:] 
Disks connected to two vertical buses each with own controller constitute a RAID5.
(1) double disk failures, 
(2) double disk failures. 
MTTDL=$8.72 \times 10^7$.

\item[3. Dual path horizontal:]
(1) double disk failure,
(2) single disk failure plus double controller failures,
(3) quadruple controller failures. 
MTTDL=$2.32 \times 10^7$ 

\item[4. Dual path horizontal:] 
Same as 2 except disks in rows constitute a RAID5.
(1) double disk failures,
(2) single disk failure plus double controller failure,
(3) quadruple controller failure. 
$\mbox{MTTD}L=8.72 \times 10^7$ hours

\item[5. Crosshatch diagonal:] 
(1) double disk failures,
(2) single disk failure plus double controller failure,
(3) quadruple controller failure. 
$\mbox{MTTDL}=8.90 \times 10^7$ hours.

\end{description}

The analysis is approximate and perhaps the reason that there is a small difference in MTTDLs.
The simulation method used in Section \ref{sec:HRAID} could be used for this purpose.

\end{framed}


\subsection{System Reliability Metrics}\label{sec:metrics} 
\vspace{2mm}

The MTTDL metric has been proven useful for comparing reliability schemes 
and for estimating the effect of the various parameters on system dependability. 

It is argued in Elerath and Schindler 2014 \cite{ElSc14} that the MTTDL derived 
using CTMC models provides unrealistically high reliability estimates.
This is refuted by Iliadis and Venkatesan 2015 \cite{IlVe15} by showing that MTTDL equations 
that account for latent sector errors and scrubbing operations yield satisfactory results. 


The {\it Expected Annual Fraction of Data Loss - EAFDL} metric in conjunction with the MTTDL metric, 
for various redundancy schemes and for the Weiball and Gamma real-world distributions,
see e.g., Trivedi 2001 \cite{Triv01}, is presented by the refuting authors. 

It is shown that the declustered placement scheme offers superior reliability in terms of both metrics. 
The parameters used in the study are summarized in Table \ref{tab:IlVe}.

\begin{table}
\begin{footnotesize}
\begin{center}
\begin{tabular}{|c|c|}\hline
Parameter               &Definition                                    \\ \hline \hline
$n$                     & number of storage devices                    \\
$c$                     & amount of data stored on each devise         \\
$r$                     & replication factor                           \\
$k$                     & spread factor of data placement scheme       \\
$b$                     & reserved rebuild bandwidth per device        \\
$1/\lambda$             & mean time to failure of a storage device     \\ \hline
$U$                     &amount of user data stored in the system ($U=n c / r$) \\
$1/\mu$                 &time to read $c$ bytes at rate $b$ ($1/\mu=c/b$) \\ \hline  
\end{tabular}
\end{center}
\end{footnotesize}
\caption{\label{tab:IlVe}Systems parameters with derived parameters below the line.}
\end{table}

At any point the system is either in normal or rebuild mode. 
Deferring rebuild has its advantages, 
since some disk failures are due to transient server failures and are resolved by restarting servers. 
Following a first-device failure, rebuild operations and subsequent device failures may occur, 
which eventually lead the system either to {\it Data Loss - DL} with probability $P_{DL}$ 
or back to the original normal mode by restoring all replicas with probability $1 - P_{DL}$.
Typically, rebuild time is negligible compared to the time between failures: $E[T] = 1 / (n \delta )$.
Given that the expected number of first-device failures until data loss occurs is $1/P_{DL}$. It follows: 
\vspace{-2mm}
\begin{eqnarray}
MTTDL \approx \frac{E(T)}{P_{DL}}
\end{eqnarray}

Let $H$ denote the amount in data loss when data loss occurs and $U$ the amount of stored user data.
EAFDL is then the normalized data lost per MTTDL:
\vspace{-3mm}
\begin{eqnarray}\label{eq:EAFDL}
EAFDL = \frac{E[H]}{MTTDL \cdot U} 
\end{eqnarray}

Define $Q$ as follows: $Q=H$ if DL and $Q=0$  if no DL.
Given $E(Q)$ we can determine EAFDL as follows:
\vspace{-2mm}
\begin{eqnarray}
E(Q) = P_{DL}  \cdot E(H),\mbox{  hence: }EAFDL = \frac{E(Q)}{E(T) U}
\end{eqnarray}

\subsection{Clustered versus Declustered Data Placements}\label{sec:placement}
\vspace{2mm}

To justify the EAFDL measure consider clustered and declustered data placements.
In {\it Clustered Placement - CP} the $k$ blocks on a device are replicated in another device,
while in {\it Declustered Placement - DP} the $k$ blocks 
are distributed over $k$ blocks as shown in Fig. \ref{fig:dataplacement}.

\begin{figure}
\begin{footnotesize}
$$
\begin{array}{|c|c|c|c|c|c|}\hline
D_1  &D_2    &D_3      &D_4      &D_5        &D_6       \\ \hline
A    &A      &K        &K        &\cdot      &\cdot     \\
B    &B      &L        &L        &\cdot      &\cdot     \\
C    &C      &\cdot    &\cdot    &\cdot      &\cdot     \\
D    &D      &\cdot    &\cdot    &\cdot      &\cdot     \\
E    &E      &\cdot    &\cdot    &\cdot      &\cdot     \\
F    &F      &\cdot    &\cdot    &\cdot      &\cdot     \\
G    &G      &\cdot    &\cdot    &\cdot      &\cdot     \\
I    &I      &\cdot    &\cdot    &\cdot      &\cdot     \\
J    &J      &\cdot    &\cdot    &\cdot      &\cdot     \\ \hline
\end{array}
\hspace{5mm}
\begin{array}{|c|c|c|c|c|c|}                        \hline
D_1  &D_2    &D_3  &D_4  &D_5    &D_6            \\ \hline
A    &A      &B    &C    &D      &E              \\
B    &F      &G    &H    &I      &J              \\
C    &K      &K    &L    &\cdot  &\cdot          \\
D    &D      &\cdot    &\cdot    &\cdot &\cdot   \\
E    &E      &\cdot    &\cdot    &\cdot &\cdot   \\
F    &F      &\cdot    &\cdot    &\cdot &\cdot   \\
G    &G      &\cdot    &\cdot    &\cdot &\cdot   \\
I    &I      &\cdot    &\cdot    &\cdot &\cdot   \\
J    &J      &\cdot    &\cdot    &\cdot &\cdot   \\ \hline
\end{array}
$$
\end{footnotesize}
\caption{\label{fig:dataplacement}Clustered (left) and 
declustered (right) data placement with six devices with degree of replication $r=2$.}
\end{figure}

In both cases two node failures will lead to data loss,
but two node failures with DP result in a lower volume of data loss than CP.

Symmetric placement schemes lies between the clustered and declustered schemes. 
For each node in the system the redundancy spread factor $k$ denotes the number of nodes over 
which the data on that node and its corresponding redundant data are spread. 
In a symmetric placement scheme, the $r - 1$ replicas of the data 
on each node are equally spread across $k - 1$ other nodes, 
the $r - 2$ replicas of the data shared by any two nodes are equally spread across $k - 2$ other nodes, and so on. 
The system is effectively divided into $n/k$ disjoint groups of $k$ nodes. 
Each group contains an amount of $U/k$ user data along with 
all of the corresponding replicas that are placed in its $k$ nodes in a declustered manner. 
Clustered and declustered placement schemes are special cases of symmetric placement schemes 
with in which $k=r$ and $k=n$, respectively.

Expressions for MTTDL and EAFDL derived in Iliadis and Venkatesan 2014 \cite{IlVe14} are as follows:

\vspace{-1mm}
$$
MTTDL \approx
\begin{cases}
( \frac{b}{\delta c} )^{r -1} \frac{1}{n \delta}  &\mbox{  for CP}\\
( \frac{b}{2 \delta c} )^{r-1} \frac{(r-1)!}{n \delta}
\prod_{e=1}^{r-2} ( \frac{n-e}{r-e} )^{r-e-1} &\mbox{  for DP}
\end{cases}
\hspace{5mm}
EAFDL \approx
\begin{cases}
( \frac{\delta c}{b} )^{r-1} \lambda &\mbox{  for CP}\\
( \frac{2 \delta c}{b})^{r-1} \frac{\delta}{(r-1)!} \prod_{e=1}^{r-1}
(\frac{r-e}{n-e})^{r-e} &\mbox{  for DP}
\end{cases}
$$


In an intelligent rebuild process, the system attempts to first recover the codewords of 
the user data that have the least number of codeword symbols left or most exposed codewords. 
In blind rebuild lost codeword symbols are being recovered in any order.
i.e., not specifically aimed at recovering the codewords with the least number of surviving symbols first. 

It is demonstrated that lazy rebuild result in an order of magnitude reliability degradation Iliadis 2022 \cite{Ilia22}. 
The degradation of MTTDL due to sector errors is observed in cases that apply lazy rebuild, and those that do not.
By contrast there is no degradation for the EAFDL.
It is also shown that the declustered data placement scheme offers superior reliability.

The adverse effect of LSEs on the MTTDL and the EAFDL reliability metrics 
as closed-form expressions for the metrics in Iliadis 2023 \cite{Ilia23} 
The MTTDL and EAFDL are obtained analytically for 
(i) the entire range of BERs;   
(ii) the symmetric, clustered, and declustered data placement schemes; and 
(iii) arbitrary device failure and rebuild time distributions under network rebuild bandwidth constraints. 

For realistic values of sector error rates the MTTDL degrades, 
while EAFDL remains practically unaffected. 
In the range of typical sector error rates and powerful erasure codes, EAFDL degrades as well. 
It is also shown that the declustered data placement scheme offers superior reliability.
The actual probability of data loss in the case of correlated symbol errors is smaller 
than that obtained assuming independent symbol errors and of the same order of magnitude.

\subsection{RAID5/6 Reliability Analysis With Unrecoverable Sector Errors}\label{sec:relanal}
\vspace{2mm}

Based on the discussion in Dholakia et al. 2008 \cite{DEH+08} we extend the previous discussion 
to include the effect of unrecoverable sector failures in addition to disk failures on MTTDL.
We consider a CTMC for RAID5 with two states representing the number of failed disks
and DF and UF states representing data loss due to a disk failure (DF) and an unrecoverable sector failure (UF).

\vspace{-1mm}
$$
{\cal S}_0 \rightarrow {\cal S}_1 :    \hspace{2mm} N \delta  \hspace{3mm}
{\cal S}_1 \rightarrow {\cal S}_{UF} : \hspace{2mm} \mu_1 P_{uf}^{(1)}  \hspace{3mm}
{\cal S}_1 \rightarrow {\cal S}_{DF} : \hspace{2mm} (N-1) \delta   \hspace{3mm}
{\cal S}_1 \leftarrow {\cal S}_0 : \hspace{2mm} \mu_1 (1- P_{uf}^{(1)} 
$$

$P_{uf}^{k},k=1,2$ are given by Eq. (\ref{eq:puf}) for RAID5 and RAID6 arrays.
The state transition matrix for RAID5 is given as follows.

\vspace{-1mm}
\begin{footnotesize}
$$
{\bf Q}= 
\begin{pmatrix}
- N \delta            &N \delta             &0               &0                 \\
\mu (1 - P_{uf}^{(1)} &-\mu -(N-1) \delta   &(N-1)\delta     &\mu P_{uf}^{(1)}  \\
0                     &0                    &0               &0                 \\
0                     &0                    &0               &0                 \\
\end{pmatrix}
$$
\end{footnotesize}

Since there are no transitions from the DF and UF states, 
the submatrix ${\bf Q}_T$ representing states ${\cal S}_0$ and ${\cal S}_1$ is: 

\begin{footnotesize}
\vspace{-1mm}
$$
{\bf Q}_T=
\begin{pmatrix}
- N ]\delta              &N \delta            &0                  &0                 \\
\mu (1 - P_{uf}^{(1)}    &-\mu                &-(N-1) \delta      &\mu P_{uf}^{(1)}  \\
\end{pmatrix}
$$
\end{footnotesize}

The vector of the average time spent in the transient states before the CTMC enters 
either one of the absorbing states DF and UF is obtained in Trivedi 2001 \cite{Triv01} 
by solving the following linear equations:

\begin{footnotesize}
\vspace{-1mm}
$$\tau {\bf Q}_T = {\bf P}_T(0)\mbox{  where }\tau-[\tau_0, \tau_1]\mbox{ and }{\bf P}_T (0) = [1,0.0]$$ 
\vspace{-1mm}
$$\tau_0 = \frac{ (N-1) \delta + \mu}
{ N(N-1) \delta +\mu P_{uf}^{(1)} } \hspace{5mm} 
\tau_1 = \frac{1}{(N-1) \delta + \mu P_{uf}^{(1)}}
$$

\vspace{-2mm}
\begin{eqnarray}\label{eq:MTTDLRAID5}
\mbox{MTTDL}_{RAID5} = \tau_0 + \tau_1 = 
\frac
{ (2N-1)\delta + \mu}
{N \delta [(N-1)\delta +\mu P_{uf}^{(1)} ] }.
\end{eqnarray}
\end{footnotesize}

Numerical results for typical parameters have shown that RAID5 plus IDR attains the same MTTDL as RAID6.
While IDR incurs longer transfers, this overhead is negligible compared
to the extra disk access to update Q blocks required by RAID6 with respect to RAID5.

In the case of RAID6 the probability $P_{recf}$ that a given segment of the failed disk 
cannot be reconstructed is upper-bounded by the probability that two 
or more of the corresponding segments residing in the remaining disks are in error. 
As segments residing in different disks are independent, the upper bound to $P_{recf}$ is given by:
\vspace{-1mm}
$$P^{UB}_{recf} = \sum_{j=2}^{N-1} \binom{N-1}{j}P_{seg}^j (1- P_{seg}^{N-1-j} \approx \binom{N-1}{2} P^2_{seg}.$$

Given $N_d$ segments per disk drive their reconstruction 
is independent of the reconstruction of the other disk segments.
The probability that an unrecoverable failure occurs, 
because the rebuild of the failed disk cannot be completed is given by:

\vspace{-1mm}
$$P_{uf}^r = 1 - (1- p^{UB}_{recf} )^N_d$$

It is assumed that rebuild time in the degraded and critical mode 
are exponentially distributed with rate $\mu_1$ and $\mu_2$.

Let ${\cal S}_i, i=0,1,2$ denote the operating states of RAID6 with $i$ failed disks,
${cal S}_{DF}$ and ${\cal S}_{UF}$ failed states due to disk failure and unreadable sectors.
We have the following transitions due to failures and repairs.

\vspace{-1mm}
$$
{\cal S}_0 \rightarrow {\cal S}_1 :    \hspace{2mm} N \delta  \hspace{3mm}
{\cal S}_1 \rightarrow {\cal S}_2 :    \hspace{2mm} (N-1) \delta   \hspace{3mm}
{\cal S}_1 \rightarrow {\cal S}_{UF} : \hspace{2mm} \mu_1 p_{uf}^{(2)}  \hspace{3mm}
{\cal S}_2 \rightarrow {\cal S}_{DF} : \hspace{2mm} (N-2) \delta   \hspace{3mm}
{\cal S}_2 \rightarrow {\cal S}_{UF} : \hspace{2mm} \mu_2 P_{uf}^{(2)} 
$$
\vspace{-1mm}
$$
{\cal S}_1 \leftarrow {\cal S}_0 : \hspace{2mm} \mu_1 (1- P_{uf}^r ) \hspace{3mm}
{\cal S}_2 \leftarrow {\cal S}_0 : \hspace{2mm} \mu_2 (1- P_{uf}^{(2)} ).
$$

Similarly to RAID5 we have the following submatrix ${\bf Q}_T$ for states (0,1,2): 
The rebuild rates with one and two disk failures are $\mu_1$ and $\mu_2$
In fact two failed disks in RAID6 are rebuild concurrently.

\begin{footnotesize}
\vspace{-1mm}
$$\tau {\bf Q}_T = {\bf P}_T(0)\mbox{  with  }{\bf P}_T (0) = [1,0.0]$$ 
\vspace{-1mm}
$$
{\bf Q}_T=
\begin{pmatrix}
-N \delta                 & N \delta     & 0                      \\
\mu_1 (1-P_{uf}^{(r)})    &-(N-1)\delta  &- \mu (N-1)\delta       \\ 
\mu_2(1- P_{uf}^{(2)})    &0             &-(N-2)\delta -\mu_2     \\  
\end{pmatrix}
$$

$$
\tau_0 = \frac{ [(N-1) \delta +\mu_1] [(N-1) \delta +\mu_2]}{N \delta V}, \hspace{3mm}
\tau_1 = \frac{(N-2)\delta +\mu_2 }{V}, \hspace{3mm} 
\tau_2 = \frac{N-1)\delta }{V}
$$

\vspace{-1mm}
$$V = [(N-1)\delta+ 
\mu P_{uf}^{(r)}][(N-2)\delta + 
\mu_2 P_{uf}^{(r)}]  + 
\mu_1 \mu_2 P_{uf}^{(r)} (1- P_{uf}^{(2)} ) $$

\vspace{-1mm}
$$MTTDL_{RAID6} = \tau_0 + \tau_1+ \tau_2$$
\end{footnotesize}


\subsection{Disk Scrubbing and IntraDisk Redundancy to Deal with LSEs}\label{sec:DEH}
\vspace{2mm}

The main cause of rebuild failures are LSEs rather than whole disk failures.
{\it Intra-Disk Redundancy - IDR} in Dholakia et al. 2008 \cite{DEH+08} and 
disk scrubbing are two less costly methods to deal with LSEs, rather than RAID6. 
The probability of uncorrectable disk failures ($P_{uf}$) due to segment errors is used in \cite{DEH+08} 
to extend RAID5 reliability analysis in Section \ref{sec:RAIDrel} to take into account LSEs.

IDR divides strips or SUs into segments with $n$ data and $m$ checks sectors per segment,
so that there are $\ell = n+m$ sectors per segment.
There are $S$ bits per sector and the probability that a bit is in error is $P_{bit}$.
Given that the probability of an uncorrectable sector error is $P_s$ 
the probability that a segment is in error when no coding is applied ($m=0$) is $P_{seg}$ 
\vspace{-1mm}
$$P_S = 1 - (1-P_{bit})^S, \hspace{5mm} P_{seg}^{none} = 1 - (1 - P_S )^\ell .$$

A single parity sector can be used to correct a single sector in error, 
since the sector in error is identified with a {\it Cyclic Redundancy Check - CRC}.    
\footnote{\url{https://en.wikipedia.org/wiki/Cyclic_redundancy_check}}

There are $n_d = C_d / (\ell S) $ segments on a disk with capacity $C_d$.
The probability of an uncorrectable failure for a RAID$(4+k)$ with $k$ failed disks,
so that it in critical mode:

\vspace{-2mm}
\begin{eqnarray}\label{eq:puf}
P_{uf}^{(k)} = 1 - ( 1- P_{seg} ) ^ { \frac{ (N-k) C_d } {\ell S} }
\end{eqnarray}
Note that the exponent denotes the number of disk segments in critical rebuild mode.

Two categories of sector errors are considered: independent and correlated.
In the case of independent errors there is a probability $P_s$ 
that a sector has an {\it Unrecoverable Failure - UF}.
Correlated errors have an average burst length $\bar{B}$ and there are $\bar{I}$ successive error-free sectors.  
Both are modeled with a geometric distribution:                                          
\vspace{-1mm}
$$a_j = P[I=j] =\alpha (1-\alpha)\alpha^{j-1},\mbox{   hence  }\bar{I}=1/\alpha$$          

We also have $P[B=j]=b_j$ and $\bar{B}= \sum_{j \geq 1} j b_j$.
Utilized in further discussions is the following parameter.
\vspace{-1mm}
$$G_n \overset{\Delta}{=} P[\mbox{burst\_length}\geq n]= \sum_{j \geq n} b_j$$ 

The independent model is a special case of correlated model where: 
\vspace{-1mm}
$$b_j=(1-P_s)P_s^{j-1}\mbox{ and }\bar{B}=1/(1-P_s)$$

The probability $P_s$ that an arbitrary sector has an unrecoverable error is: 
\vspace{-1mm}
$$ P_S = \frac{ \bar{B} }{\bar{B}+\bar{I}}\mbox{ it follows }
\alpha = \frac{P_s}{\bar{B}(1-P_s)} \approx \frac{P_s}{\bar{B}} 
\mbox{  hence }P_s \leq \frac{\bar{B}}{\bar{B}+1}$$ 

For independent errors:
\vspace{-2mm}
\begin{eqnarray}
P{seg}^{none} = 1- (1- P_s)^\ell \approx \ell P_s, \hspace{5mm}
P_{seg}^{none} \approx (1+\frac{\ell-1}{\bar{B}})P_s
\end{eqnarray}
and for bursty errors
\vspace{-2mm}
\begin{eqnarray}
P_{seg}^{none} = 1 - (1-P_s)(1- \frac{1}{\bar{B}} ) \approx (1+\frac{\ell-1}{\bar{B}})P_s
\end{eqnarray}

The probability that a segment is in error can be specified with the summation below,
but only one term need to be considered for small $P_s$ and the correlated model: 
\vspace{-1mm}
$$P_{seg} = \sum_{i \geq 1}c_i P_s^i  \hspace{5mm}P_{seg} = c_1 P_s + O(P_s^2)$$

\begin{framed}
\subsection{Schemes to Implement IntraDisk Redundancy - IDR}\label{sec:IDRschemes}
\vspace{2mm}

Three IDR methods are proposed and evaluated in Dholakia et al. 2008 \cite{DEH+08}.

\subsubsection*{Reed-Solomon Check}\label{sec:RS}
\vspace{1mm}

The {\it RS - Reed-Solomon} code is discussed in Fujiwara 2006 \cite{Fuji06}. 
With RS coding an error occurs when over $m$ bits among $\ell=n+m$ bits are in error.

\vspace{-2mm}
\begin{eqnarray}
P_{seg}^{RS} = \sum_{j=m+1}^ell P_s^j (1-P_s)^{\ell-j} \approx \binom{\ell}{m+1} P_s^{m+1}.
\end{eqnarray}

In the case of a correlated model for small values of $P_s$.

\vspace{-2mm}
\begin{eqnarray}
P_{seg}^{RS} \approx \left[ 1 + \frac{1}{\bar{B}} 
\left( (\ell - m - 1)G_{m+1} - \sum_{j=1}^m G_j \right) \right] P_s.
\end{eqnarray}

This is the best that can be done since for a code with $m$ parity symbols 
for a codeword of $n$ symbols any $m$ erasures can be corrected,
but RS codes are computationally expensive.

\vspace{-2mm}
\begin{eqnarray}
P_{seg}^{RS} = 
\sum_{j=m+1}^\ell \binom{\ell}{j} P_s^j  (1- P_s)^{\ell-j} \approx \binom{\ell}{m+1} P_s^{m+1}.
\end{eqnarray}
In the case of correlated model for small values of $P_s$ results in Appendix B of the paper yield:

\vspace{-2mm}
\begin{eqnarray}
P_{seg}^{RS} \approx \left[ 1 + \frac{ (\ell -m -1)G_{m+1} - \sum_{j=1}^m G_j }{\bar{B}} \right] P_s. 
\mbox{where }\bar{B}=\sum_{j=1}^\infty G_j, \hspace{2mm} b_j = G_j - G_{j-1}. 
\end{eqnarray}

\subsubsection*{Single Parity Check - SPC}
\vspace{1mm}

This is the simplest coding scheme in which a single parity sector is computed by
XORing $\ell - 1$ data sectors to form a segment with $\ell$ sectors. 
The independent model yields:

\vspace{-2mm}
\begin{eqnarray}
P_{seg}^{SPC} = \sum_{j=2}^{\ell} \binom{\ell}{j} P_s^j (1-P_s)^{\ell - j} \approx \frac{\ell(\ell-1)}{2} P_s^2 
\end{eqnarray}

In the case of correlated model for small values of $P_s$.

\vspace{-2mm}
\begin{eqnarray}
P_{seg}^{SPC} \approx
\left[ 1 + \frac{1}{\bar{B}} \left(  (\ell -2)G_2 - 1 \right)\right] P_s. 
\end{eqnarray}

\subsubsection*{Interleaved Parity Check - IPC} 
\vspace{1mm}

In this scheme $\ell - m$ contiguous sectors are placed in a rectangle with $m$ columns. 
The XOR of the sectors in a column is the parity sector which is placed in an additional row.
This scheme with $\ell/m$ sectors per interleave can correct a single error per interleave,
but a segment is in error if there is at least one interleave with two uncorrectable segment errors.

An IPC scheme with $m$ ($m \leq \ell/2$) interleaves per segment, 
i.e., $\ell/m$ sectors per interleave, has the capability of correcting a single error per interleave. 
Consequently, a segment is in error if there is at least one interleave 
in which there are at least two unrecoverable sector errors. 
Note that this scheme can correct a single burst of $m$ consecutive errors occurring in a segment. 
However, unlike the RS scheme, it in general does not have 
the capability of correcting any $m$ sector errors in a segment.
The probability of an interleave error is:
\vspace{-1mm}
$$P_{interleave\_error} = 
\sum_{j=2}^{\ell/m} \binom{\ell/m}{j} P_s^j (1-P_s)^{\ell/m-j}
\approx \frac{\ell(\ell-m)}{2m^2} P_s^2$$

\vspace{-2mm}
\begin{eqnarray}
P_{seg}^{IPC} = 1  - (1-P_{interleave\_error})^m \approx \frac{\ell (\ell- m)}{2m} P_s^2
\end{eqnarray}

In the case of correlated model:

\vspace{-3mm}
\begin{eqnarray}
P_{seg}^{IPC} \approx \left[ 1 + \frac{1}{\bar{B}} 
\left( (\ell-m-1) G_{m+1} - \sum_{j=1}^m G_j \right) \right]  P_s.
\end{eqnarray}

It follows $P_{seg}^{RS} \approx P_{seg}^{IPC}$,
but IPC is preferable since it is easier to implement.

\end{framed}

\subsection{Effect of IDR on Disk Performance}\label{sec:IDRperf}
\vspace{2mm}

The main focus in Dholakia et al. 2008 \cite{DEH+08} is to study the effect of IDR on RAID5 and RAID6 reliability, 
but performance is also investigated using simulation and {\it IO Equivalents - IOEs} Hafner et al. 2004 \cite{HDKR04}.

\vspace{-2mm}
\begin{eqnarray}\label{eq:IOE}
IOE(k) = 1 + 
\frac{\mbox{time to transfer 4 KB}}{\mbox{average time per 4 KB}}= 
1 + \frac{\mbox{4 KB/1024}}
{\mbox{Avg. media transfer rate in MB/sec}} \times \mbox{Avg IO/sec.}
\end{eqnarray}

\begin{table}[h]
\caption{Results for IOE.\label{tab:IOE}}
\begin{footnotesize}
\begin{center}
\begin{tabular}{|c|c|c|c|} \hline 
                          &10K RPM              &15K RPM             & Units    \\ \hline \hline
IOs/sec.                  &250                  &333                 &IO/sec.   \\ \hline
Avg. media transfer rate  &53.25                &62                  & MB/sec.  \\ \hline
IOE for k KB              &$1+k/55$             &$1+k/48$            &IOE       \\ \hline
\end{tabular}
\end{center}
\end{footnotesize}
\end{table}
The $IOE(k) \approx 1+ k/50$ was used in the study.

When a single-sector {\bf A} is written using the IPC scheme, for example,
it is written by a single I/O request also containing the corresponding intra-disk parity sector {\bf PA}.
The average length of an I/O request is minimized 
when the intra-disk parity sectors are placed in the middle of the segment.
The analysis in Section 8.2 of the paper determines 
the increase in IOE as 8.5\% for small and 3.6\% for long writes.

The following parameters were used by the simulator: 
$1/\delta$=500K hours, disk capacity $S=300$ GB, $P_{bit}=10^{-14}$,
number of disks $N=8/16$ for RAID5/6, rebuild time $1/\mu=17.8$ hours, 
sector size $S=512$ bytes (4KB sectors became prevalent in 2010),
$\ell = 128$ sectors, $m=8$ interleaves.

Both random number driven and trace-driven simulations results are reported.
Other scheduling policies besides FCFS such as SSTF, Look, C-Look are tried 
and shown to have a small effect on the relative performance due to IDR.  
Look is a variation of SCAN that reverses if there are no additional requests in the direction of the SCAN 
and C-look scans in one direction only Thomasian 2011 \cite{Thom11}.

\subsection*{Numerical Results}
\vspace{2mm}

A SATA drive with a capacity $C_d = 300$ GB, $P_{bit}=10^{-14}$ is considered,
so that 512 B sectors and $P_s = 4096 \times P_{bit} = 4.096 \times 10^{-11}$.
The segment length was set to $\ell=128$ sectors with $m=8$ interleaving,
burst length distribution with at most 16 sector and $\bar{B}= 1.0291$ 
\vspace{-1mm}
\begin{footnotesize}
$$\underline{b}= [0.9812, 0.016, 0.0013(2), 0.0003(2), 
0.0002 0.0001(2), 0, 0.0001, 0, 0.0001(2), 0, 0.0001(2)]$$
\end{footnotesize}
$P_{seg}$ and $P_{uf}$ with this parameter are given below.

\begin{table}
\centering
\begin{footnotesize}
\parbox{0.45\linewidth}{
\begin{tabular}{|c|c|c|} \hline
Errors     & Independent    & Correlated                              \\ \hline \hline
None       & $5.2 \times 10^{-9}$  & $5   \times 10^{-9}$             \\ \hline
RS         & $6.2 \times 10^{-81}$ & $2.5 \times 10^{-12}$            \\ \hline
SPC        & $1.3 \times 10^{-17}$ & $8.5 \times 10^{-14}$            \\ \hline
IPC        & $1.6 \times 10^{-18}$ & $9.5 \times 2.5 \times 10^{-12}$ \\ \hline
\end{tabular} 
\caption{\label{tab:pseg}Approximate $P_{seg}$ for two error models.}
}
\hspace{2.5mm}
\parbox{0.475\linewidth}{
\begin{tabular}{|c|c|c|} \hline
Scheme   &Independent               & Correlated               \\ \hline \hline
None     &   $1.5  \times 10*{-1}$  &  $1.5 \times 10^{-1}$    \\ \hline
RS       &   $2    \times 10^{-73}$ &  $7.9  \times 10^{-5}$   \\ \hline
SPC      &   $4.3  \times 10^{-10}$ &  $3.1 \times 10^{-3}$    \\ \hline
IPC      &   $6.1  \times 10^{-11}$ &  $7.95 \times 10^{-5}$   \\ \hline
\end{tabular}  
\caption{\label{tab:Puf}Approximate $P_{uf}$ for RAID5 with $N=8$.}
}
\end{footnotesize}
\end{table}

\subsection{Comparison of Disk Scrubbing and IDR with IPC}\label{sec:comparison}
\vspace{2mm}


Closed-form expressions are derived for operation with random, uniformly distributed I/O requests, 
for the probability of encountering unrecoverable sector errors, 
the percentage of unrecoverable sector errors that disk scrubbing detects and corrects, 
the throughput and the minimum scrubbing period achievable Illiadis et al. \cite{IHHE11}. 

The probability of an error due to a write is $P_w$
and writes constitute a fraction $r_w$ of disk accesses with rate $h$.
The probability of error in reading a sector without scrubbing is $P_e = r_w P_w$.
The probability of sector failure for deterministic scrubbing 
and random (exponentially distributed) scrubbing period are:
\vspace{-1mm}
$$P_s^{(det)} = [1 - (1- e^{- h T_s})/(h T_S)] P_e,\hspace{5mm}P_s^{exp)} =  [ h T_S / (1+ h T_S)] P_e,$$
where $T_S$ denotes the mean scrub period.
Given that $h T_s \ll 1 $ we have the following approximations for deterministic and exponential scrubbing:
\vspace{-1mm}
$$ P_s^{(det)}  \approx P_e h T_s, \hspace{5mm} P_s^{(exp)} \approx \frac{1}{2} P_s P_e h T_s $$
Random scrubbing has double the value for $P_s$, hence deterministic scrubbing is preferable. 



For heavy-write workloads and for typical scrubbing frequencies and loads, 
the reliability improvement due to disk scrubbing does not reach 
the level achieved by the IPC-based IDR scheme, which is load insensitive. 
Note that IPC-based IDR for SATA drives achieves the same reliability 
as that of a system with no sector errors, with a small (about 6\%) increase in capacity. 

The penalty of disk scrubbing on I/O performance can be significant, 
whereas that of the IPC-based IDR scheme is minimal, some disk space overhead.

Given IOE(k) in Eq. (\ref{eq:IOE}) the maximum rate of read/write operations to the disk is:
\vspace{-1mm}
$$\hat{\sigma}_{max}  = [IOE(K) t_{seek} ] ^{-1}.$$

Given a small write scenario with $k$ sectors per chunk, $S_D$ sectors per disk, 
with $G_S$ sectors scrubbed a time, and $p=1$ for RAID5 and $p=2$ for RAID6 

\vspace{-2mm}
\begin{eqnarray}
\sigma \leq \sigma_{max} (k, T_S) = 
\frac { \hat{\sigma}_{max} (k,T_S ) }{ 1+ (1+2p) r_w ] }
\end{eqnarray}

The scrubbing $T_S$ period should not be smaller than a critical value $T^*_S$ given below:

\vspace{-2mm}
\begin{eqnarray}
T_S \geq T_S^* =  \frac{ S_D IOE(G_S) }
{ [ 1 + (1+2p) r_w) ] G_S  IOE(k) [\sigma_{max} (k) - \sigma ]}
\end{eqnarray}

Fig. 6 in Iliadis et al. 2011 \cite{IHHE11} shows that in the area of practical interest 
the reliability offered by scrubbing is inferior to that achieved by the IDR scheme.
That IDR is preferable to disk scrubbing contradicts that of Schroeder et al. 2010 \cite{ScDG10},
on the other hand IDR does not seem to be implemented while scrubbing is common practice.

\subsection{Combining Scrubbing with SMART and Machine Learning}\label{sec:combined}
\vspace{2mm}

{\it Self-Monitoring And Reporting Technology - SMART} technology has been applied to predict HDD failures  
Hughes et al. 2002 \cite{HMKE02}, Eckart et al. 2008 \cite{ECHS08}.
\footnote{\url{https://en.wikipedia.org/wiki/Self-Monitoring,_Analysis_and_Reporting_Technology}}
Once a disk is determined to be failing its contents are copied onto a spare 
with possibly occasional assist from erasure coding to restore unreadable sectors.  
A disk detected to be failing by simply copying it, which is referred to as smart rebuild. 
\footnote{\url{https://community.ibm.com/community/user/storage/blogs/brian-kraemer1/2021/06/11/ds8000-smart-rebuild}}

SMART monitors 30 disk Attributes corresponding to different measures of performance and reliability.   
\footnote{\url{https://www.linuxjournal.com/article/6983}}
Attributes have a one-byte normalized value ranging from 1 to 253, 
e.g., SMART 187 is the number of read errors that could not be recovered using hardware ECC. 
If one or more of the Attribute values less than or equal to its corresponding threshold, 
then either the disk is expected to fail in less than 24 hours or it has exceeded its design or usage lifetime.         
\footnote{\url{https://www.backblaze.com/blog/hard-drive-smart-stats/}}

Western Digital {\it Network Attached Storage  - NAS} HDDs are automatically given a warning label 
after being powered on for three years by Synology's {\it DiskStation Manager -DSM}. 
{\it Western Digital Device Analytics - WDDA} and Seagate's IronWolf are similar offerings. 
\footnote{\url{https://arstechnica.com/gadgets/2023/06/clearly-predatory-western-digital-sparks-panic-anger-for-age-shaming-hdds/}}

HPE's Nimble predictive storage makes 86\% of problems disappear.
A 6-nines (99.9999\%) availability is guaranteed over the installed base
and 54\% of problems resolved are determined to be outside storage.    
\footnote{\url{https://m.softchoice.com/web/newsite/documents/partners/hpe/Nimble-Six-Nine-White-Paper.pdf}} 
Nimble's InfoSight applies data science to identify, predict, and prevent problems across infrastructure layers.               
\footnote{\url{https://en.wikipedia.org/wiki/Data_science}}


{\it Machine Learning - ML} methods is used by Murray et al. 2005  \cite{MuHK05}
to predict HDD failures by monitoring internally the attributes of individual drives.  
An algorithm based on the multiple-instance learning framework and 
the naive Bayesian classifier (mi-NB) are specifically designed for the low false-alarm case.  
Other considered methods are {\it Support Vector Machines - SVMs},
\footnote{\url{https://en.wikipedia.org/wiki/Support_vector_machine}}
unsupervised clustering,  
\footnote{\url{https://www.section.io/engineering-education/clustering-in-unsupervised-ml/}}
and non-parametric statistical tests (rank-sum and reverse arrangements).   
\footnote{\url{https://www.ncbi.nlm.nih.gov/pmc/articles/PMC4754273/}}                            
The failure-prediction performance of these methods is considerably better 
than the threshold method currently implemented in drives, while maintaining low false alarm rates.   

Goldszmidt 2012 \cite{Gold12} characterizes and evaluates {\it Disk Failure Advance Detector- D-FAD} 
of {\it Soon-To-Fail - STF} disks.   
The input to D-FAD is a single signal from every disk, 
a time series containing in each sample the {\it Average Maximum Latency - AML}  
and the output is an alarm according to a combination of statistical models. 
The parameters in these models are automatically trained from a population 
of healthy and failed disks ML techniques.
Results from an {\it Hidden Markov Model - HMM}
\footnote{\url{https://en.wikipedia.org/wiki/Hidden_Markov_model}}
and a threshold (peak counter) were fused using logistic regression to sound alarms. 
\footnote{\url{https://en.wikipedia.org/wiki/Logistic_regression}} 
When applied to 1190 production disks D-FAD predicted 15 out of 17 failed disks 
(88.2\% detection rate), with 30 false alarms (2.56\% false positive rate). 

Amvrosiadis et al. 2012 \cite{AmOS12} state that the goal of a scrubber 
is to minimize the time between the occurrence of an LSE 
and its detection/correction or {\it Mean Latent Error Time - MLET}, 
since during this time the system is exposed to the risk of data loss. 
In addition to reducing the MLET, a scrubber must ensure 
to not significantly affect the performance of foreground workloads.

Staggered scrubbing provides a lower MLET exploiting the fact that LSEs occur in temporal and spatial bursts.
Rather than sequentially reading the disk from beginning to end staggered scrubbing 
quickly probes different regions of a drive hoping to achieve quicker detection 
of error bursts son that they are scrubbed immediately.
Staggered scrubbing potentially reduces scrub throughput used by foreground workloads. 
Larger request sizes lead to more efficient use of the disk, 
but also have the potential of bigger impact on foreground traffic.




EMC's RAIDShield characterizes, monitors, and proactively protects 
against disk failures eliminating 88\% of triple disk failures in RAID6,
which constituted 80\% of all RAID failures Ma et al. 2015 \cite{Ma++15}.
The method uses joint failure probability to quantify and predict vulnerability to failure.
Simulation shows that most vulnerable RAID6 systems can improve the coverage to 98\% of triple errors.

The use of ML to make storage systems more reliable by detecting sector errors in HDDs and SSDs 
is explored by Mahdisoltani et al. 2017 \cite{MaSS17} 
Exploration of a widerange of ML techniques shows that sector errors can be predicted ahead of time with high accuracy. 
Possible use cases for improving storage system reliability through 
the use of sector error predictors is provided.
The mean time to detecting errors can be greatly reduced by adapting 
the speed of a scrubber based on error predictions.

A scrub unleveling technique that enforces a lower rate scrubbing to healthy disks 
and a higher rate scrubbing to disks subject to LSEs is proposed by Jiang et al. 2019 \cite{JiHZ19}.
A voting-based method ensures prediction accuracy. 
Experimental results demonstrate that the proposed approach achieves a lower scrubbing cost together 
with higher data reliability than traditional fixed-rate scrubbing methods on read-world datasets. 
Compared with the state-of-the-art method the same level of {\it Mean-Time-To-Detection - MTTD}. 
with almost 32\% less scrubbing is achieved. 

The first experimental comparison of sequential versus staggered scrubbing in production systems 
determined that staggered scrubbing implemented with the right parameters 
can achieve the same (or better) scrub throughput as a sequential scrubber, 
without penalizing foreground applications. 
A detailed statistical analysis of I/O traces is used 
to define policies for deciding when to issue scrub requests, 
while keeping foreground request slowdown at a user-defined threshold. 
The simplest approach, based on idle waiting and 
using a fixed scrub request size outperforms more complex statistical approaches.


Scrubbing schemes to deal with LSEs have several limitations according to Zhang et al. 2020 \cite{ZWW+20}.
(1) schemes use ML to predict LSEs, but only a single snapshot of training data for prediction; 
ignoring sequential dependencies between different statuses of a HDD over time. 
(2) accelerating scrubbing at a fixed rate based on the results 
of a binary classification model results in unnecessary increases in scrubbing cost; 
(3) naively accelerating scrubbing over full disks neglects partial high-risk areas; 
(4) they do not employ strategies to scrub these high-risk areas in advance based on I/O accesses patterns, 
in order to further increase the efficiency of scrubbing.

{\it Tier-Scrubbing - TS} scheme combines a {\it Long Short-Term Memory  - LSTM} based
\footnote{\url{https://en.wikipedia.org/wiki/Long_short-term_memory}} 
{\it Adaptive Scrubbing Rate Controller - ASRC}, 
module focuses on sector error locality to locate high-risk disk areas 
and a piggyback scrubbing strategy to improve the reliability of a storage system. 
Realistic datasets and workloads evaluated from two data centers demonstrate 
that TS can simultaneously decreases {\it Mean-Time-To-Detection - MTTD} by about 80\% 
and scrubbing cost by 20\%, compared to a state-of-the-art scrubbing scheme.



Disk fault detection models based on SMART data with ML algorithms 
require a large amount of disk data to train the models, 
but the small amount of training data in traditional ML algorithms greatly increases the risks of overfitting,  
which weaken the performance of the model and seriously affect the reliability.

A novel Small-Sample Disk Fault Detection optimizing method, 
with synthetic data using {\it Generative Adversarial Networks - GANs} \cite{Goo+14} is proposed in Wang 2022 \cite{Wang22}.
The approach generates failed disk data conforming to the failed disk data distribution.
Disk SMART attributes vary with the usage and are time-related; 
LSTM is adopted since it is good at learning the characteristics of time series data 
as the GAN generator to fit the distribution of SMART data, 
and use the multi-layer neural network as the discriminator to train the GAN-model to generate realistic failed disk data. 
With sufficient generated failed disk data samples, 
ML algorithms can detect disk faults more precisely than before with the small original failed disk data samples.

\subsection{Fail-Slow Faults in Drives and Their Detection}\label{sec:failslow}  
\vspace{2mm}

A major study of fail-slow or gray failures based on reports of such incidents from twelve large-scale clusters
are summarized in Table \ref{tab:gunawi2} from Gunawi et al. 2018 \cite{Gun+18}. 

\begin{table}[t]
\centering
\begin{scriptsize}
\begin{tabular}{|c|c|c|c|c|c|c||c|c|c|c|}\hline
Root   &SSD   &Disk  &Mem   &Net   &CPU     &Total  &Perm &Trans. &Partial &Tr. stop. \\ \hline \hline
ERR    &10    &8     &9     &10    &3       &40     &19   &8      &7       &6         \\
FW     &6     &3     &0     &9     &2       &20     &11   &3      &1       &4         \\   \hline
TEMP   &1     &3     &0     &2     &5       &11     &6    &2      &1       &2         \\
PWR    &1     &0     &1     &0     &6       &8      &3    &2      &1       &2         \\ 
ENV    &3     &5     &2     &4     &4       &18     &11   &3      &3       &1         \\
CONF   &1     &1     &0     &2     &3       &7      &6    &1      &0       &0         \\  \hline
UNK    &0     &3     &1     &2     &2       &8      &5    &1      &0       &2         \\  \hline \hline
Total  &22    &23    &13    &29    &25      &112    &-    &-      &-       &-         \\  \hline
\end{tabular}
\end{scriptsize}
\caption{\label{tab:gunawi2}Root causes of fail-slow problems across hardware types.
Fail-slow symptoms across root causes are given in four rightmost columns of 
as Permanent, Transient slowdown, Partial slowdown, Transient stoppage.
Abbreviations: FW=FirmWare, CONF=configuration, UNK=UNKnown.}
\end{table}

Perseus developed at Alibaba deals with {\it Fail-Slow Detection - FSD}
of components functioning with degraded performance Lu et al. 2023 \cite{LXZ+23}.
Within a ten month period 304 slow cases were found among 248K drives.
Design goals for Perseus are: 
(1) {\bf non-intrusive}, 
(2) {\bf fine-grained}, 
(3) {\bf accurate}, 
(4) {\bf general}, both SSD and HHD based block, object, and file/database storage.

Four methodologies are used:
{\bf 1. Threshold filtering.}
Setting up a threshold on drive latency to identify slow drives based on {\it Service Level Objectives - SLO}.
To account for legitimate temporary events such as {\it Garbage Collection - GC} in SSDs a minimum slowdown is also specified.
As shown in Fig. 4 and stated in the paper the method has low accuracy.
The spikes in the figure are ascribed to increased throughput rather than slowdown.
{\bf 2. Peer Evaluation.}
Drives on the same node receive equal loads due to load balancing.
The median delay over normal drives which constitute the majority is used to determine abnormal drives,
i.e., those that are twice slower for more than half the time.
{\bf 3. IASO-based model.} IASO is a fail-slow detection and mitigation
framework for distributed storage services Panda et al. 2019 \cite{PSK+19}.
The accuracy, as defined below, was only 0.48.

Perseus was developed since these attempts were either too labor-intensive or ineffective.
The following questions are raised:                                  
{\bf Q1: Reads or writes.} Read requests, 
which are satisfied with the response from one replica are used in the study.      
{\bf Q2: Modeling workload pressure.} 
Spearman's Rank Correlation Coefficient 
\footnote{\url{https://en.wikipedia.org/wiki/Spearman}}          
shows that throughput (bytes per second) is a better metric than IOPS.    
{\bf Q3: Selecting adaptive thresholds}.
Given service-, cluster-, and node-wise latency (L) versus throughput (T) (LvT) node-wise was selected.
Regression models are used to define a statistically normal drive and 
use its upper latency bound as the adaptive threshold.                     
{\bf Q4: Fail slow identification criterion}. Instead of a binary result a likelihood is expressed.

Perseus applies polynomial regression on the node-level LvT distribution 
to automatically derive an adaptive threshold. 
A scoreboard mechanism to single out the drives with severe fail-slow failures. 
The five entries of the time-series dataset are: 
(avg\_latency, avg\_throughput, drive\_ID, node\_UID, timestamp).

Perseus worklflow has the following steps:
{\bf 1. Outlier detection.}
Collect entries at each node and use {\it Principal Component Analysis - PCA}, 
to transform a large set of variables into a smaller set with those with small variance removed.
\footnote{\url{https://en.wikipedia.org/wiki/Principal_component_analysis}}
Outliers are discarded using the DBSCAN density-based clustering algorithm Ester et al. 96 \cite{EKS+96}. 
{\bf 2. Building polynomial regression model.}
The prediction upperbound is a fail-slow detection threshold,
which is then applied to the raw dataset to identify out-of-bound entries and mark them as slow.
{\bf 3. Identifying fail-slow events.}
Use is made of a sliding window and a {\it Slowdown Ratio - SR} defined as drive latency by the upper bound entry,
to identify consecutive slow entries and formulate corresponding fail-slow events.
{\bf 4. Evaluating risk.}
A risk-score mechanism estimates the duration and degree of fail-slow events 
and assigns each drive a risk score based on daily fail-slow events.   
\footnote{\url{https://en.wikipedia.org/wiki/Risk_score}}
Cases are investigated based on severity.
Perseus risk levels based on daily slowdown events specified as SRs are given in Table \ref{tab:LZX}.

\begin{table}[h]
\begin{center}
\begin{footnotesize}
\begin{tabular}{|c|c|c|c|}\hline
Slowness                   &Long-term,     &Moderate      &Temporal  \\
(minutes)                  &$\geq 120$     &[60,120)      &[30,60)   \\ \hline
Severe (SR $\geq$ 5)       &Extreme        &High          &Moderate  \\ \hline
Moderate (SR $\in$ [2,5))  &High           &Moderate      &Low       \\ \hline
Mild (SR $\in$[1,2))       &Moderate       &Low           &Minor     \\ \hline      
\end{tabular}
\end{footnotesize}
\end{center}
\caption{Perseus risk levels based on SR.\label{tab:LZX}}
\end{table}

\clearpage
\section{Appendix III: RAID Reliability Simulation}\label{sec:simul}
\vspace{3mm}

\begin{small}
\ref{sec:perfsimul} Simulation studies of RAID performance.                      \newline
\ref{sec:uber} Simulation study of a digital archive.                            \newline 
\ref{sec:HRAIDsimul} Simulation of hierarchical RAID reliability.                \newline 
\ref{sec:Proteus} Proteus open-source simulator.                                 \newline 
\ref{sec:ATT} CQSIM\_R Tool Developed at AT\&T.                                  \newline 
\ref{sec:SIMEDC} SIMedc simulator for erasure coded data centers.                \newline  
\end{small}

\subsection{Simulation Studies of RAID Performance}\label{sec:perfsimul}
\vspace{2mm}


The {\it Parallel Data Laboratory- PDL} founded by G. Gibson at CMU
initially solely concentrated on storage research.
RAIDframe was developed to assist in the implementation and  evaluation  
of new  RAID architectures by rapid prototyping disk arrays Courtright et al. 1997 \cite{CGH+97}.
DiskSim with its origin at Univ. of Michigan was further developed at PDL by G. Ganger.
\footnote{\url{https://www.pdl.cmu.edu/DiskSim/index.shtml}}                  
DiskSim relies on extracted disk drive parameters by DIXTRAC. 
\url{https://www.pdl.cmu.edu/Dixtrac/index.shtml}                                
which provides disk specifications until 2007.                                   
\footnote{\url{https://www.pdl.cmu.edu/DiskSim/diskspecs.shtml}}                      

The Pantheon simulator was developed at HP Labs by J. Wilkes et al. 1995 \cite{Wilk95},
which was used in several projects, most notably AutoRAID discussed in Section \ref{sec:autoRAID}
In evaluating the performance of IDR in Section \ref{sec:IDRperf}
IBM researchers developed their own simulator rather than relying on DiskSim or Pantheon. 

Disk and RAID simulators were developed at NJIT's Integrated Systems Lab (2000-07) by 
the author and his students.

\begin{framed}
\subsubsection{Disk Scheduling Policies}\label{sec:sched}
\vspace{1mm}

There have been numerous applications of simulation 
in evaluating the performance of disk scheduling policies.
Disk access time can be  reduced by disk scheduling by processing pending I/O requests out-of-order, 
i.e., non-FCFS scheduling, 
such as {\it Shortest Seek Time First - SSTF} and SCAN Denning 1967 \cite{Denn67}
and {\it Shortest Access Time First - SATF} Jacobson and Wilkes 1991 \cite{JaWi91}. 
Disk scheduling policies are compared via simulation in Worthington, Ganger, and Yale 1994 \cite{WoGP94}, 
Hsu and Smith 2005 \cite{HsSm04}. 

Thomasian and Liu \cite{ThLi02} extended SATF with lookahead and priorities.
An empirical equation for SATF mean disk service versus queue-length ($q$) is given in Thomasian 2011 \cite{Thom11}.
\vspace{-1mm}
$$\bar{x}_{SATF} (q) = \bar{x}_{FCFS} / q^p \mbox{  where  }p \approx 1/5$$  
i.e., disk service time is halved for $q=32$.
Increase in throughput versus $q$ is given in Table 2 in Anderson et al. 2003 \cite{AnDR03}
{\it Complete Fairness Queueing - CFQ} provides fair allocation 
of the disk I/O bandwidth for all the processes which request an I/O operation. 
\footnote{\url{https://www.kernel.org/doc/Documentation/block/cfq-iosched.txt}}
Disk scheduling mainly based on Thomasian 11 \cite{Thom11} 
is discussed in Section 3.3 in Thomasian 2021 \cite{Thom21}.

There have been several studies to reorganize data on disk to improve access times,
which are reviewed in Hsu et al. 2005 \cite{HsSY05} 
{\it Automatic Locality Improving Storage - ALIS} scheme based 
on analyzing common sequences of disk requests is proposed and evaluated in this work. 

Hsu and Smith 2004 \cite{HsSm04} is a comprehensive study using server and PC workloads 
to analyze the performance impact of I/O optimization techniques, 
i.e., read caching, sequential prefetching, opportunistic prefetching, write buffering, 
request scheduling, striping, and short-stroking. 
Disk technology improvement in the form of faster seeks, 
higher RPM, linear density improvement, and increase in track density is analyzed. 
A methodology is developed for replaying real workloads that accurately models I/O arrivals
Techniques that reduce the number of physical I/Os are investigated.
Sequential prefetching and write buffering are particularly effective, 
reducing the average read and write response time by about 50\% and 90\%, respectively. 
Larger caches up to and even beyond 1\% of the storage are shown to improve performance. 
Disk technology improvement at the historical rate increases performance 
by about 8\% per year if the disk occupancy rate is kept constant, 
and by about 15\% per year if the same number of disks are used. 
Actual average seek time and rotational latency are determined to be, 
only about 35\% and 60\% of the specified values, respectively. 
Disk head positioning time far dominates the data transfer time 
suggesting that data should be reorganized such that accesses become more sequential.

\end{framed}

In what follows several simulation studies of storage systems are discussed.

\subsection{Simulation Study of a Digital Archive}\label{sec:uber}
\vspace{2mm}

Digital archives require stronger reliability measures than RAID to avoid data loss due to device failure.
Wildani et al. 2009 \cite{WSML09} consider is multi-level redundancy coding 
to reduce the probability of data loss from multiple simultaneous device failures. 
This approach handles failures of one or two devices efficiently,
while still allowing the system to survive rare-events, i.e., four or more device failures.

An uber parity is provided which is calculated from all user data 
in all disklets belonging to a stripe in the uber-group using another erasure code.
The uber parity can be stored on NVRAM or always powered-on disks to offset write bottlenecks, 
while still keeping the number of active devices low.
Calculations of failure probabilities was used to determines that 
the addition of uber parities allows the system to absorb many more disk failures without data loss.
Adding uber-groups negatively impacts performance when these groups need to be used for a rebuild,
but since rebuilds using uber parity occur rarely, the impact on system performance is minimal.
Robustness against rare events is achieved for under 5\% of total system cost.

\subsection{Simulation of Hierarchical Reliability}\label{sec:HRAIDsimul}
\vspace{2mm}

HRAID - Hierarchical RAID described in Section{sec:HRAID} has three operational options:

\begin{description}

\item[I. Intra-SN but no Inter-SN Redundancy:]
There are $N$ independent {\it Storage Nodes - SNs} which are RAID$(4+\ell)$ disk arrays.
This option serves as a baseline in reliability comparisons.

\item[Option II: Inter-SN Redundancy and Rebuild Processing:]
Inter-SN processing is invoked on demand for accessed data and for rebuild.
Rebuild is carried out by restriping Rao et al. 2011 \cite{RaHG11}.

\item[Option III: Inter-SN Redundancy, but no Rebuild Processing:] 
Rebuild may not be possible due to limited interconnection network bandwidth,
which is geared to normal processing.
In the case of SNs considered failed due to the failure of more than $\ell$ failed disks at an SN
but no controller failures, operation continues in degraded mode with on-demand reconstruction via inter-SN redundancy, 
until data loss detected.
This is strictly done to determine the MTTDL, since the system can continue operation.

\end{description}

Given that $u$ is a uniformly distributed pseudo-random numbers in the range $(0,1)$,
it is converted to (negative) exponential distribution to determine the time to next failures as follows.
\vspace{-1mm} 
$$F(t)=1- e^{-\Omega t } = u \mbox{  leads to  }t = (- 1 / \sum_{\forall i} \delta_i) \times ln(1-u)\mbox{  or just  } 
t= - \mbox{MTTF} \times \mbox{ln}(u).$$

More complex methods for obtaining distributions are available.
The accuracy of simulation results is specified by the confidence interval at a given confidence level.
The two topics are discussed in Chapters 5 and 6 in Lavenberg 1983 \cite{Lave83}.

\begin{framed}
{\bf Procedure: Hierarchical RAID Simulator to Determine MTTDL}

\begin{description}
\item[\bf Inputs:]

\noindent
$N$: The number of SNs.                                                             \newline
$M$: The number of disks per SN.                                                    \newline
$D = N \times M$: Total number of disks.                                            \newline
$k$: InterSN redundancy level ($k=0$ for Option I).                                 \newline
$\ell$: IntraSN redundancy level, $\ell=0$ for RAID0, else RAID($4+\ell)$.          \newline
$\delta$: Disk failure rate, $\delta = 10^{-6}$, so that {\it Mean Time To Failures - MTTF}$=1/\delta = 10^6$ hours. 
\newline
$\gamma$: Controller failure rate, multiple of $\delta$).

\item
[\bf Outputs] (with initializations):

\noindent
$F_{cd}=0/1$ data loss occurred due to controller/disk failure.                    \newline
$N_c = 0$: number of failed controllers.                                           \newline
Initialize SN/controller state: $C[n]=1, \forall{n}$.                   \newline
Initialize number of failed disks at $n^{th}$ SN: $F[n]=0,\forall{n}$. 

\item
[\bf Simulation Variables:]

\noindent
$Clock=0$: Simulation time initialized.                                      \newline
State 0: Failed. State 1: Operational.                                       \newline
$C[0:N-1]$ the state of controllers.                                         \newline
$Disks[0:N-1,0:M-1]$ the state of $D= N \times M$ disks.                     \newline
$(Disks[n,m]=1,\forall{n},\forall{m}$. /* initial disk states*/              \newline
$F[0:N-1]$ \# of failed disks at node $n$                                    \newline
$\Omega = \sum_{\forall (i)} \delta_i$ sum of component failures rates.      \newline
$T_{NF}:$ Time to next failure.                                              \newline
$N_c=0$: Number of failed controllers.                                       \newline
$N_n=0:$ Number of failed nodes due to controller or $>\ell$ disk failures   \newline
$N_d=0$: number of failed disks.                                             \newline
$u_i$: $i^{th}$ uniformly distributed pseudo-random variable in $(0,1)$.

\end{description}

{\bf Simulation steps:}

\begin{description}

\item[\bf Step\_1:]
Total failure rate since all failures exponential:       
{\bf $\Omega = (N- N_c) \gamma + (D - N_d) \delta.$}

Compute time to next failure and increment $Clock$:              \newline
$T_{NF} = (-1/\Omega) ln(u_1)$,                                  \newline
$Clock +=  T_{NF}$. /* advance simulation clock */

\item[\bf Step\_2:]
Probability a controller failed: $p=(N-N_c)\gamma/\Omega$.       \newline
If $u_2 \leq p$ then controller failure: goto Step 3.            \newline
Else disk failure: goto Step 4.

\item[\bf Step\_3:]
Determine failed controller: $n= \lfloor N \times u_3 \rfloor $.     \newline
If $C[n]==0$ then regenerate $u_3$ and recompute $n$,                 \newline
else \{ $C[n] = 0$, $N_c++$, $N_n++$, $D[n,m]=0, 0 \leq m \leq M-1$, \newline
$t=M-F[n]$ disk inaccessible, so $N_d = N_d + t$. \}                 \newline
If $N_n > k$ then \{$F_{cd}=0$, goto Step 6\}, else goto Step\_1.

\item[\bf Step\_4:]
$t=\lfloor D \times u_4 \rfloor$. /* index of failed disk  */        \newline
SN number: $n= \lfloor t / M \rfloor$;                               \newline
Disk number: $m = \mbox{mod} (t, M)$.                                    \newline
If $Disks[n,m]==0$ (disk already failed) resample $u_4$ and recompute $n$ and $m$, \newline
/*  disks attached to failed controller are considered failed */    \newline
else \{$ Disks[n,m]==0$, $N_d++$, $F[n]++$]\}.                     \newline 
If $F[n] \leq \ell$ goto Step\_1.                                   \newline
else \{ C[n]=0, $t=M-F[n]$, $N_d = N_d + t$, $N_n++$ \}.           \newline
If $N_n \leq k $ then go to Step\_1, else $F_{cd}=1$.

\item[\bf Step\_5:]
Return($Clock$, $F_{cd}$, $N_c$, $N_{df}$).

\end{description}

The simulation is repeated to obtain confidence intervals 
at a sufficiently high confidence level.

\end{framed}

HRAID is conceptually similar to IBM Intelligent Bricks project Wilcke et al. 2006 \cite{WGF+06}. 
It is shown by the shortcut reliability analysis method Thomasian 2006 \cite{Thom06} 
and simulation results in Thomasian et al. 2012 \cite{ThTH12}
that for the same total redundancy level a higher MTTDL is attained 
by associating higher reliability at the intraSN rather than interSN level.
In other words to optimize apportionment of three check strips P, Q, and R, 
it is preferable to use P and Q for intra-SN and R for inter-SN recovery. 
This is contradictory to the two preferred configurations for Intelligent Bricks. \newline
(1) 1DFT (RAID5) at the the node level and 2NFT at the node level provides 40-75\% storage efficiency.\newline
(2) 0DFT in bricks and 3NFT at brick level provides 50-75\% storage efficiency.\newline


\subsection{Proteus Open-Source Simulator}\label{sec:Proteus}

Proteus was developed at {\it Storage Systems Research Center - SSRC} at {\it U. Calif. at Santa Cruz - UCSC}
to predict the risk of data loss in various RAID levels and two-dimensional arrays Kao et al. 2013 \cite{KPSL13}. 
Proteus was used to learn that there is no measurable difference 
between values obtained assuming deterministic versus exponential repair times.
The latter are required by Markov chain modeling, 
an issue raised in RAID5 reliability analysis in Gibson 1992 \cite{Gibs92}. 

\subsection{CQSIM\_R Tool Developed at AT\&T}\label{sec:ATT}
\vspace{2mm}

CQSIM\_R is a tool suited for predicting the reliability 
of large scale storage systems at AT\&T Hall 2016 \cite{Hall16}. 
It includes direct calculations based on an only-drives-fail failure model 
and an event-based simulator for predicting failures.
These are based on a common combinatorial framework for modeling placement strategies. 

CQSIM\_R models common storage systems, including replicated and erasure coded designs. 
New results, such as the poor reliability scaling of spread-placed systems 
and a quantification of the impact of data center distribution 
and rack-awareness on reliability, demonstrate the usefulness of the tool. 
Analysis and empirical studies show the tool's soundness, performance, and scalability.

\subsection{SIMedc Simulator for Erasure Coded Data Centers}\label{sec:SIMEDC}
\vspace{2mm}

The discrete-event SIMedc simulator for the reliability analysis of {\it erasure-coded data centers -edc}
was developed at {\it Chinese Univ. of Hong-Kong - CUHK} by Zhang et al. 2019 \cite{ZhHL19}.
SIMedc reports reliability metrics of an erasure-coded data center based on data center topology, 
erasure codes, redundancy placement, and failure/repair patterns 
of different subsystems based on statistical models or production traces. 

Simulation is accelerated via importance sampling assisted by uniformization Grassman 1991 \cite{Gras91}. 
It is shown that placing erasure-coded data in fewer racks generally improves 
reliability by reducing cross-rack repair traffic, 
even though it sacrifices rack-level fault tolerance in the face of correlated failures.

\clearpage
\section{Appendix IV: RAID Reliability Modeling Tools}\label{sec:tools}
\vspace{3mm}

\begin{small}
\ref{sec:ARIES} Automated Reliability Interactive Estimation System - ARIES.     \newline 
\ref{sec:SAVE} System AVailability Estimator - SAVE Project at IBM Research.     \newline 
\ref{sec:SHARPE} Symbolic Hierarchical Automated Reliability and Performance Evaluator - SHARPE.  \newline 
\ref{sec:sharpe}The mathematics behind SHARPE reliability modeling package. \newline
\ref{sec:numerical}Numerical method for computing the transient probabilities.
\end{small}

Analytic and simulation software packages developed to assess
the reliability, availability, and serviceability of computer systems 
are surveyed in Johnson and Malek 1988 \cite{JoMa88}.
Provided are the application of the tool, input, models, and model solution methods.
ARIES, SHARPE, and SAVE tools are reviewed in this section.





\subsection{Automated Reliability Interactive Estimation System - ARIES Project at UCLA}\label{sec:ARIES}
\vspace{2mm}

A. Avizienis who later joined {\it Univ. of Calif. at Los Angeles - UCLA} from {\it Jet Propulsion Labs - JPL} 
led the design and reliability analysis of fault-tolerant JPL {\it Self-Testing and Repairing - STAR} computer,
for an unmanned ten year mission to a distant planet.
The computer used {\it Trip Modular Redundancy - TMR} with a majority voter.

An example of this effort was a reliability expression derived for a 
a hybrid $k$-out-of-$n$ system with $n$ components energized reported in Mathur and Avizienis \cite{MaAv70},
which is given in Example 3.32 in Trivedi 2001 \cite{Triv01}.
Aries is a unifying method for analyzing closed systems Ng and Avizienis 1980 \cite{NgAv80}.
The parameters used in this study given in Table \ref{tab:NgAv80}>

\begin{table}[h]
\begin{footnotesize}
\begin{center}
\begin{tabular}{|c|c|} \hline
$N$          &initial number of modules in the active configuration.          \\ \hline
$D$          &number of degradations allowed in the active configuration      \\ \hline
$S$          &number of spare modules.                                        \\ \hline
$Ca$         &coverage for recovery from active module failures.              \\ \hline
$Cd$         &coverage for recovery from spare module failures.               \\ \hline
$\lambda$    &failure rate of active modules.                                 \\ \hline
$\mu$        &failure rate of spare modules                                   \\ \hline
${\bf Y}$    &sequence of allowed degradations of the active configuration.   \\ \hline
${\bf CY}$   &coverage vector for transitions into degraded configuration.    \\ \hline
\end{tabular}       
\end{center}
\end{footnotesize}                              
\caption{\label{tab:NgAv}Parameters used in ARIES.}
\end{table}

The reliability of a closed fault-tolerant system  is expressed as: 

\vspace{-3mm}
\begin{eqnarray}\label{eq:Ng}
R(t) = \sum_{\forall i} A_i e^{-\sigma_i t}
\end{eqnarray}

The analysis is based on the eigenvalues of the CTMC representing 
the failure of system components and its recovery by using spares.
Repeated roots are not considered.

Let ${\cal S}_i$ indicate a nonfailed state and $\sigma_i$ the failure rate at that state.
$A_i$ can be expressed as a function of state parameters and can be quite complex.
The MTTF easily follows from Eq.~\ref{eq:Ng}
\vspace{-1mm}
$$\mbox{MTTF} = \int_{t=0}^\infty R(t) dt = \sum_{\forall i} \frac{A_i}{\sigma_i}.$$
For closed systems $R(t)$ given by Eq. (\ref{eq:Ng}) has a specific form.

\vspace{-3mm}
\begin{eqnarray}
R(t) = X(t) {\cal A} W(t).
\end{eqnarray}

\vspace{-3mm}
\begin{eqnarray}\nonumber
X(t) = ( e^{-Y[0] \lambda t} , \ldots,  e^{-Y[D]\lambda t} ) \mbox{ with }Y[0]=N\hspace{5mm}
\mbox{ and } W(t)=(1,e^{-\mu t}, \ldots, e^{-S \mu t} ),
\end{eqnarray}

\vspace{-2mm}
\begin{eqnarray}
{\cal A} =
\begin{pmatrix}
A_{S,0}^0 & \ldots & A_{S,0}^D   \\
\vdots    & \ddots &\vdots       \\
A_{S,S}^0 & \ldots & A_{S,S}^D   \\
\end{pmatrix}
\end{eqnarray}

Two more parameters used for modeling repairable systems 
are the number of repairman or repair facilities ($M$) 
and the repair rate of one repairman ($\Psi$) 

\subsection{System AVailability Estimator - SAVE Project at IBM Research}\label{sec:SAVE}
\vspace{2mm}

An early reliability modeling effort at IBM defined the coverage factor ($c$), 
which is the probability that the system can successfully recover Bouricius et al. 1971 \cite{BCJ+71}.
%
%
The SAVE project was started in S. Lavenberg's Systems Modeling and Analysis Dept. at IBM Research in mid-1985. 
The effort first dealt with analytical reliability modeling,
along the lines adopted in ARIES and SHARPE Blum et al. 1994 \cite{BGH+94}. 

Realistic system models are often not amenable 
to analysis using conventional analytic or numerical methods Stewart 2021 \cite{Stew21}.
Adopting simulation as an alternative to analysis for reliability estimation 
led to a unified framework for simulating Markovian models 
of highly dependable systems Goyal et al. 1992 \cite{GSH+92},

Straightforward simulations is too costly since failures are rare.
Techniques for fast simulation of models of highly dependable systems 
are reviewed in Nicola et al. 2001 \cite{NiSN01}.
Importance  sampling is one such method and when it works well 
it can reduce simulation run lengths by several orders of magnitude.
\footnote{\url{http://jrxv.net/x/16/ism.pdf}}

The following example is provided in this study:                    
(1) two sets of processors with four processors per set,                 
(2) two sets of controllers with two controllers per set,
(3) four clusters of disks each consisting of four disks. 
The multiprocessors are connected to all controllers. 
The pair of controllers have two connections to disk clusters.
In a disk cluster, data are replicated so that one disk can fail without affecting the system.
The ID mirrored disk organization in Section \ref{sec:mirhyb} is utilized.

When a processor fails it has a 0.01 probability of causing another processor to fail.
Each unit in the system has two failure modes which occur with equal probability.
The repair rates for mode 1 and 2 failures are 1 and 0.5 hour, respectively.
The failure rates in hour$^{-1}$ are 1/1000 for processors,  
1/20,000 for controllers, and 1/60,000 for disks.
The repair rates are 1 and 1/2 for mode 1 and mode 2 failures.

Efficient simulation techniques for estimating steady-state quantities
in models of highly dependable computing systems with general component failure
and repair time distributions are developed in Nicola et al. 1993 \cite{NNHP93}.
The regenerative method of simulation for steady state estimation
can be used when the failure time distributions are exponentially distributed.
A splitting technique is used for importance sampling 
to speed up the simulation of rare system failure events during a cycle.
Experimental results show that the method is effective in practice.

\subsection{Symbolic Hierarchical Automated Reliability and Performance Evaluator - SHARPE}\label{sec:SHARPE}
\vspace{2mm}

SHARPE was developed at Duke Univ. by Sahner, Trivedi and Puliafito 1996 \cite{SaTP96}.
Several generations of K. Trivedi's students contributed to SHARPE.              
\footnote{\url{https://sharpe.pratt.duke.edu/}}

Section 4.7.2 in \cite{SaTP96} considers a CTMC with $m=2$ non-absorbing and $n=2$ absorbing states,
which have no transitions to other states Trivedi 2001 \cite{Triv01}.
In the case of RAID5 a second disk failure or unreadable sector will lead 
to data loss as discussed in Section \ref{sec:DEH}. 
Section 5.4 titled ``Imperfect Fault Coverage and Reliability'' in Trivedi 2001 \cite{Triv01}
provides a good discussion and formulas derived in Ng and Avizienis \cite{NgAv80}.
Section 8.5 titled ``Markov Chains with Absorbing States'' discusses a technique used in Sharpe..
Hierarchical reliability modeling is discussed in Chapter 16 in Trivedi and Bobbio 2017 \cite{TrBo17}.

\subsubsection{The Mathematics Behind SHARPE Reliability Modeling Package}\label{sec:sharpe}
\vspace{1mm}

The infinitesimal generator matrix with the states indexed as $(0:m-1)$ and $(m:m+n-1)$ for $m=n=2$ is given as:

\begin{footnotesize}
\vspace{-2mm}
$$
{\bf Q}= 
\begin{pmatrix}
-(\gamma+\lambda) &\gamma &\lambda &0 \\
\beta & -(\beta+\delta + \mu)    &\mu   &\delta \\
0 &0 &0 &0 \\ 
0 &0 &0 &0 \\ 
\end{pmatrix}
$$
\end{footnotesize}

Differential equations and respective LSTs are as follows where
${\bf \alpha} = {\bf \pi}(0) = \{ \alpha_0,\alpha_1,\alpha_2,\alpha_3 \}$ denotes the initial state vector.

\begin{footnotesize}
\vspace{-2mm}
\begin{align}\nonumber
\frac{d\pi_0 (t)}{dt} &= - (\gamma + \lambda) \pi_0(t) +\beta \pi_1(t), \hspace{5mm}
s L^*_0 (s) - \alpha_0 =  -(\gamma +\lambda) L^*_0 (s) + \beta L^*_1 (s) \\
\nonumber
\frac{d\pi_1 (t)}{dt} &= -(\beta +\delta +\mu) \pi_1 (t_ +\gamma \pi_0 (t),  \hspace{5mm} 
s L^*_1 (s) - \alpha_1 = -(\beta +\delta + \mu) L^*_1(s) +\gamma L^*_0 (s) \\
\nonumber
\frac{d\pi_2 (t)}{dt} &= \lambda \pi_0 (t) +\mu \pi_1 (t),   \hspace{5mm}    
s L^*_2 (s) - \alpha_2 = \lambda L^*_0 (s) +\mu L^*_1 (s) \\
\nonumber
\frac{d\pi_3 (t)}{dt} &= \delta \pi_1 (t), \hspace{5mm}   
s L^*_3 (s) - \alpha_3 = \delta L^*_1 (s)
\end{align}
\end{footnotesize}

The vector of transforms is given as ${\bf L^*} = \left( L^*_0, \ldots , L^*_{m-1} \right)$.
the vector of initial probabilities as ${\bf \alpha} = (\alpha_0, \alpha_1, \ldots, \alpha_{m-1})$ and 
${\bf T}$ is the upper $m \times m$ upper left hand corner of generator matrix ${\bf Q}$, we have:
The system of first $m$ equations in matrix form and their solution is given as follows:

\vspace{-3mm}
\begin{eqnarray}\nonumber
s {\bf L}^*  - {\bf \alpha} = {\bf L}^* {\bf T}  \Rightarrow
{\bf  L}^* ={\bf \alpha}  \left( s  {\bf I} - {\bf T} \right)^{-1}
\end{eqnarray}

We now have the LST to the transient distribution functions for nonabsorbing states
We define ${\bf T}_i$ to be the $m \times 1$ vector that is transpose of $(q_{1i}, q _{2i}, \ldots, q_{mi}) $ so:

\vspace{-1mm}
$$s L^*_2 (s) = \alpha_2 + {\bf L} {\bf T}_2 = \alpha_2 +{\bf \alpha} (s{\bf I} - {\bf T} )^{-1} {\bf T}_2$$

For the absorbing state $L^*_i$ we have:
\vspace{-1mm}
$${\bf L}^*_i (s) = \frac{1}{s} \left( \alpha_i + {\bf \alpha} 
\left( s {\bf I} - {\bf T} \right)^{-1} \right) {\bf T}_i ,$$ 

Define $T_i$ as the transpose of $\left( q_{1i}, q_{2i}, \ldots, q_{mi} \right)$. We have:

$$s L^*_2 (s) = \alpha_2 + {\bf L}{\bf T_2} = \alpha_2 + {\bf \alpha} (s {\bf I}= {bf T})^{-1} T_2$$

In general, for absorbing state $i$, we have:

$$L_i (s) = \frac{1}{s} \left( \alpha_i + {\bf \alpha} (s {\bf I}- {\bf T})^{-1} T_i \right) $$ 

The real work is inverting the LST using partial fraction expansion:
\vspace{-1mm}
$${\bf P} (s) = \frac{1}{s} \left( {\bf \alpha} \left( s {\bf I} - {\bf T} \right)^{-1} T_i \right) $$

Vector ${\bf X}(s)$ is defined as 
\vspace{-1mm}
$${\bf X}(s) = \left( s {\bf I} - {\bf T} \right)^{-1} {\bf T}_i \mbox{  then  }
\left( s {\bf I}- {\bf T} \right) {\bf X}(s) = {\bf T}_i$$

Applying Cramer's rule the $i^{th}$ element of ${\bf X}$ is:
\footnote{\url{https://en.wikipedia.org/wiki/Cramer's_rule}}
\vspace{-1mm}
$$X_i (s) = \frac{N_i(s)}{D(s)}$$

$D(s)$ is the determinant of $({\bf I} - {\bf T})$
and $N_i (s)$ is the determinant of the matrix obtained 
by replacing the $i^{th}$ column of $s{\bf I} - {\bf T}$ by the column vector $T_i$. 

Given the column vector ${\bf X}$ we have:
\vspace{-3mm}
\begin{eqnarray}\nonumber 
P(s) = \frac{1}{s} \left( {\bf \alpha} {\bf X}(s) \right) = \frac {\sum \alpha_i N_i(s) }{s D(s)}
\end{eqnarray}

To invert $P(s)$ we need to carry out a partial fraction expansion. 
The zero root of the denominator yields the steady state solution.
Given $D(s) = det(s {\bf I} - {\bf T})$ eigenvalues of ${\bf T}$ are given as sorted list of the roots
$( s_1, s_2, \ldots , s_{m+1})$, where root $s_k$ may occur $d_k$ times. 
Then after determining $a_k$ we have the inversion:

\begin{eqnarray}\nonumber 
P^*(s) = \sum_{k=1}^{m+1} \frac{a_k}{ (s-s_k)^{d_k}} \Rightarrow
F(t) = \sum_{k=1}^{m+1} \frac{a_k} {(d_k - 1)!} t^{d_k -1} e^{s_k t}
\end{eqnarray}

This solution yields transient state probabilities. 
However, the number and types of matrix manipulations make the algorithm unstable for some CTMC.

\subsubsection{Numerical Method for Computing the Transient Probabilities}\label{sec:numerical}
\vspace{1mm}

A formal solution to $\frac{d \pi(t) }{dt} = \pi (t) Q$ is:  

\vspace{-2mm}
$${\bf \pi} (t) ={\bf \pi}(0) e^{{\bf Q} t} \mbox{  where }
e^{ {\bf Q} t} = \sum_{i=0}^\infty \frac{({\bf Q}t)^i}{i!}$$

Given $q \geq \mbox{max}_i |q_{ii} |$
\vspace{-2mm}
$${\bf \pi}(t) =\sum_{k=0}^\infty \theta (k) e^{-q t} \frac{ (q t)^k}{k!}$$

where 
$\theta (0) = {\bf \pi}(0)$, ${\bf Q}^* = \frac{ {\bf Q}} {q} + {\bf I}$ and 
$\theta (k) = \theta (k-1) {\bf Q}^*$

The reader is referred to Chapter 8 in Stewart 2021 \cite{Stew21} for alternate solution methods.

\clearpage
\section{Appendix V: Mirrored and Hybrid Disk Arrays}\label{sec:mirhyb}
\vspace{3mm}

\begin{small}
\ref{sec:relcomp}. Mirrored and hybrid disk arrays description and reliability comparison.\newline
\ref{sec:shortcut}. Shortcut Method to compare RAID reliability.                          \newline
\ref{sec:multilevel}. Reliability analysis of multilevel RAID arrays.                     \newline 
\ref{sec:RAID15}. Mirrored RAID5 - RAID1/5 reliability analysis.                          \newline 
\ref{sec:RAID51}. RAID5 with mirrored disks - RAID5/1.                                    \newline 
\ref{sec:shortcut2}. Shortcut reliability analysis to compare RAID1/5 and RAID5/1.        \newline
\ref{sec:HDArel} Reliability and performance comparison of two Heterogeneous Disk Arrays
\end{small}

\subsection{Mirrored and Hybrid Disk Arrays Descriptions and Reliability Comparison}\label{sec:relcomp}
\vspace{2mm}


In addition to increased availability replication helps with meeting {\it Service Level Agreements - SLAs}.
\footnote{\url{https://en.wikipedia.org/wiki/Service-level_agreement}}
A request does not have to wait if at least one disk with utilization $\rho_i, 1 \leq i \leq m$ is idle: 
\vspace{-1mm}
$$P[\mbox{at least one disk idle out of }m]= 1 - \prod_{i=1}^m (1- \rho_i).$$

Mirrored or duplexed disks which are the simplest form of replication 
are classified as RAID1 in Patterson et al. 1988 \cite{PaGK88}.
Mirroring and duplexed processors were adopted by Tandem (now part of HP) 
which evolved to NonStop SQL system Tandem 1987 \cite{TaDG87}.
The Teradata {\it Data Base Computer - DBC}/1012  with terabyte capacity 
was first shipped in 1980s Sloan 1992 \cite{Sloa92}.
EMC's (now Dell/EMC) Symmetrix was an early raid1 which could emulate IBM DASD. 
\footnote{\url{https://en.wikipedia.org/wiki/EMC_Symmetrix}}  

Four mirrored disk organizations are described and their reliability and performance compared in 
Thomasian and Blaum 2006 \cite{ThBl06} and Thomasian and Xu 2008 \cite{ThXu08}.
Hybrid disk arrays which store XORs of multiple disks instead of just mirroring are patented Wilner 2001 \cite{Wiln01}
and evaluated in Thomasian and Tang 2012 \cite{ThTa12}

$k$-way replication provides the opportunity to reduce disk access time for read requests
by accessing the disk with the lowest seek distance, hence with minimum seek distance.
Since all disks need to be updated the maximum of $k$ uniformly distributed seek distances 
as well as the minimum is derived in Gray and Bitton 1988 \cite{BiGr88}:

\vspace{-1mm}
\begin{eqnarray}\label{eq:Ik}
E[\mbox{min of k-way seeks}] \approx \frac{C}{2k+1}, \hspace{5mm} 
E[\mbox{max of k-way seeks}] \approx C(1 - I_k),\mbox{  where:  }I_k= 
\frac{2k}{2k+1} \frac{2k-2}{2k-1} \ldots \frac{2}{3}. 
\end{eqnarray}

{\it Distributed Shortest Processing Time First - DSPTF} 
is a request distribution protocol for decentralized brick storage system with high-speed interconnects,
which dynamically selects servers with a replica to satisfy a request, 
while balancing load exploits aggregate cache capacity Lumb and Golding 2004 \cite{LuGo04} 
Compared to existing decentralized approaches,  such as hash-based request distribution,  
D-SPTF achieves up to 65\% higher throughput. 
Storage bricks are reviewed in Sections 11.12 and 11.13 in Thomasian 2021 \cite{Thom21}, 
presentation on the topic is by Gray 1992 in \cite{Gray02}.

Anticipatory arm placement places the arm at $C/2$ in disk with $C$ cylinders 
reducing the seek distance to C/4 for uniform requests over all cylinders,
while otherwise it is $C/3$ according to Eq. (\ref{eq:meanseek}).   
In mirrored disks the arms are placed at $C/4$ and $3C/4$ reducing the mean seek distance to $C/4$,
while $C/5$ is achievable by shortest seek routing according to Eq. (\ref{eq:Ik}).
Disks with hot spots (frequently accessed cylinders) and zoning 
are also considered Thomasian and Fu 2006 \cite{ThFu06}. 

Loads across BM disk pairs can be balanced via striping in the hierarchical array RAID0/1,
with RAID0 at the higher level and RAID1 at the lower level.
The load due to read requests to disk pairs is balanced via uniform routing 
(with equal probabilities) or round-robin routing.
In what follows we show that the latter exhibits shorter mean waiting times 
under the assumption that arrivals are Poisson with arrival rate $\lambda$ 
and service times exponentially distributed with mean $\bar{x}=1/\mu$.

With round robin routing the arrival process to each disk is the sum of two exponentials 
or Erlang-2 distribution with {\it Coefficient of Variation - CV}$=1/2$, 
while the CV for exponential distribution is one.
The mean waiting time on a GI/M/1 queue with Erlang-2 arrivals is Kleinrock 1975 \cite{Klei75}:
\vspace{-1mm} 
$$W_{GI/M/1} = \frac{\sigma\bar{x}}{1- \sigma} \mbox{ with } \sigma= {\cal A}^* (\mu - \mu \sigma),\mbox{ where }
{\cal A}^* (s) = [\frac{2 \lambda } {s+2 \lambda } ]^2 $$  
\vspace{-1mm}
$$\sigma^2 - (1+4 \rho)\sigma + 4 \rho^2 =0, \hspace{3mm} \sigma = \frac{1}{2}(1+4 \rho - \sqrt{1+8 \rho} )$$
$W_{GI/M/1} < W_{M/M/1}$ since it can be shown that $\rho > \sigma$.
A smaller CV does not insure a smaller mean waiting time for $W_{GI/M/1}$ 
as shown by a counterexample in Thomasian 2014b \cite{Thom14b}.

RAID1 is especially suited in doubling disk bandwidth 
for random updates of small blocks as in OLTP, while RAID5 incurs SWP. 
RAID1 is less efficient than RAID5 in writing large blocks
or accessing large blocks of data that can be carried out in parallel in RAID5.

When one of two of disk pairs in BM fails the load of the surviving disk is doubled
and this may lead to overload or at least higher queueing delays.
This BM weakness led to several RAID1 configurations which are described here:

\begin{description}

\item[Basic Mirroring - BM:]
This is the simplest form of replication, 
where the same data appears at both disks. 
RAID0/1 is a striped array with $M=N/2$ pairs of mirrored disks.
Up to $M$ disk failures can be tolerated as long as they are not pairs.
When a disk fails BM has the worst performance in terms of unbalanced disk loads,
i.e., the read load of the surviving disk is doubled.
According to Table \ref{tab:MTTDL} BM has the highest reliability among 
the four RAID1 organizations considered here.
This is because the probability that the second disk failure 
leading to data loss by being a pair of the first failed disk is lowest.

\item[Interleaved Declustering - ID:] 
Teradata DBC/1012 partitioned its $N$ disks into $c$ clusters
with $n=N/c$ disks per cluster as shown in Fig. \ref{fig:ID}.
Each disk has a primary and a secondary area. 
The data in primary area is partitioned and placed onto $n-1$ secondary areas in the same cluster. 
A failed disk results in $n/(n-1)$ load increase at the $n-1$ surviving disks in a cluster. 
Two disk failure in a cluster will lead to data loss.
The setting $c=N/2$ or $n=2$ is tantamount to BM.
Teradata latest product is Vantage 17.20, whose {\it Native Object Store - NOS} can reside on cloud storage.
\footnote{\url{https://www.dwhpro.com/teradata-nos-native-object-store/}}

\item[Group Rotate Declustering - GRD:]
There are two mirrored RAID0 arrays with $M=N/2$ disks each, specified as RAID1/0,
but strips on the right side are rotated with respect to the strips on the left side 
as shown in Fig. \ref{fig:GRD} Chen and Towsley 1996 \cite{ChTo96}.
When a disk fails its load is evenly distributed over the disks on the other side.
Two disk failures on one side will lead to data loss.

\item[Chained Declustering- CD:]
Hsiao and DeWitt 1990 \cite{HsDe90} proposed CD in conjunction 
with the Gamma database machine project at U Wisconsin in 1990, i.e., years after the Teradata DBC/1012.
\footnote{\url{https://research.cs.wisc.edu/techreports/1990/TR921.pdf}}
Data in the primary area on the $i^{th}$ disk is replicated 
on the secondary area of the ${i+1}^{st}\mbox{mod}(N)$ disk as shown in Fig. \ref{fig:CD}
Similarly to ID primary and secondary data can be placed in outer or inner cylinders,
or upper and lower disk tracks when there are an even number of tracks per cylinder.
Two successive disk failures leads to data loss,
but the reliability is lower than BM since there are twice as many data loss opportunities as in BM:
when following the $i^{th}$ disk failure the $i \pm 1 \mbox{mod}(N)$ disk fails.

\end{description}

\begin{figure}
\centering
\begin{footnotesize}
 \begin{tabular}{|c c c c||c c c c|}
 \hline
 \multicolumn{4}{|c||}{Primary Disks} &  \multicolumn{4}{c|}{Secondary Disks}\\  \hline \hline
 $D_1$ & $D_2$ & $D_3$ & $D_4$ & $D_5$ & $D_6$ & $D_7$ & $D_8$ \\ \hline
 $A$ & $B$ & $C$ & $D$ & $A'$ & $B'$ & $C'$ & $D'$ \\  \hline
 $E$ & $F$ & $G$ & $H$ & $E'$ & $F'$ & $G'$ & $H'$ \\  \hline
 $I$ & $J$ & $K$ & $L$ & $I'$ & $J'$ & $K'$ & $L'$ \\  \hline
 $M$ & $N$ & $O$ & $P$ & $M'$ & $N'$ & $O'$ & $P'$ \\ \hline
 \end{tabular}
\end{footnotesize}
\caption{\label{fig:BM}RAID0/1 with $N=8$ disks.}
\end{figure}

\hspace{1mm}

\begin{figure}
\centering
\begin{footnotesize}
\begin{tabular}{|c c c c||c c c c|}  \hline
\multicolumn{4}{|c||}{Primary Disks} &  \multicolumn{4}{c|}{Secondary Disks}\\  \hline \hline
 $D_1$ & $D_2$ & $D_3$ & $D_4$ & $D_5$ & $D_6$ & $D_7$ & $D_8$ \\ \hline
 $A$ & $B$ & $C$ & $D$ &   $A'$ & $B'$ & $C'$ & $D'$ \\  \hline
 $E$ & $F$ & $G$ & $H$ &   $H'$ & $E'$ & $F'$ & $G'$ \\  \hline
 $I$ & $J$ & $K$ & $L$ &   $K'$ & $L'$ & $I'$ & $J'$ \\  \hline
  $M$ & $N$ & $O$ & $P$ &  $N'$ & $O'$ & $P'$ & $M'$  \\ \hline
 \end{tabular}
\end{footnotesize}
\caption{\label{fig:GRD}Group Rotate Declustering with $N=8$ disks.}
\end{figure}

\hspace{1mm}

\begin{figure}
\centering
\begin{footnotesize}
 \begin{tabular}{|c c c c||c c c c|} \hline
 \multicolumn{4}{|c||}{Cluster 1} &  \multicolumn{4}{c|}{Cluster 2}\\  \hline
 $D_1$ & $D_2$ & $D_3$ & $D_4$ & $D_5$ & $D_6$ & $D_7$ & $D_8$ \\ \hline \hline \hline
 $~A~$ & $~B~~$ & $~C~~$ & $~D~~$ & $~E~~$ & $~F~~$ & $~G~~$ & $~H~~$ \\  \hline
 $b_3$ & $a_1$ & $a_2$ & $a_3$ & $f_3$ & $e_1$ & $e_2$ & $e_3$ \\  \hline
 $c_2$ & $c_3$ & $b_1$ & $b_2$ & $g_2$ & $g_3$ & $f_1$ & $f_2$ \\  \hline
 $d_1$ & $d_2$ & $d_3$ & $c_1$ & $h_1$ & $h_2$ & $h_3$ & $g_1$ \\ \hline
 \end{tabular}
\end{footnotesize}
\caption{\label{fig:ID} Interleaved Declustering with $N=8$ disks, $c=2$ clusters, and $n=4$ disks per cluster.
Capital letters denote primary data and small letters subsets of secondary data.}
\end{figure}

\hspace{1mm}

\begin{figure}[h]
\centering
\begin{footnotesize}
 \begin{tabular}{|c c c c c c c c|}  \hline
 $D_1$ & $D_2$ & $D_3$ & $D_4$ & $D_5$ & $D_6$ & $D_7$ & $D_8$ \\ \hline \hline
 $\frac{1}{2}A$ & $\frac{1}{2}B$ & $\frac{1}{2}C$ & $\frac{1}{2}D$ & $\frac{1}{2}E$ & $\frac{1}{2}F$ & $\frac{1}{2}G$ & $\frac{1}{2}H$ \\  \hline
 $\frac{1}{2}h$ & $\frac{1}{2}a$ & $\frac{1}{2}b$ & $\frac{1}{2}c$ & $\frac{1}{2}d$ & $\frac{1}{2}e$ & $\frac{1}{2}f$ & $\frac{1}{2}g$ \\ \hline
 \end{tabular}
\end{footnotesize}
\caption{\label{fig:CD} Chained declustering with $N=8$ disks.
Primary (resp. secondary) blocks are in capital (resp. small) letters.
The read load is evenly distributed among the primary and secondary copies.}
\end{figure}

Let $I$ denote the maximum number of disk failures that can be tolerated without data loss 
For BM with $N=2M$ disks $I=M$ and $I=c$ for the ID organization.

Setting $r=r(t)$ to simplify the notation RAID1 reliability expressions can be expressed as follows:

\vspace{-2mm}
\begin{eqnarray}\label{eq:rel}
R_{RAID} (N) = \sum_{i=0}^I A(N,i) r^{N-i} (1-r)^i .
\end{eqnarray}
$A(N,0)= 1$ by and $A(N,i)=0$ for $i > M$.

In the case of BM up to $M$ disk failures can be tolerated, as long as one disk in each pair survives, hence:

\vspace{-2mm}
\begin{eqnarray}\label{eq:BM}
A (N,i) = \binom{M}{i} 2^i , \hspace{2mm} 0 \leq i \leq M.
\end{eqnarray}

In the case of ID with $c$ clusters and $n=N/c$ disks per cluster,
there can only be one disk failure per cluster.

\vspace{-2mm}
\begin{eqnarray}\label{eq:ID}
A (N,i) = \binom{c}{i} n^i , \mbox{ where }n=N/c, \hspace{2mm} 0 \leq i \leq c.
\end{eqnarray}



In the case of GRD up to $M$ disks on each side all $M$ disks 
can fail as long as they are all on one side.

\vspace{-2mm}
\begin{eqnarray}\label{eq:GRD}
A (N,i) = 2 \binom{M}{i}, \hspace{2mm} 0 \leq i \leq M.
\end{eqnarray}

The expression for $A(N,i)$ for CD is derived in Thomasian and Blaum 2006 \cite{ThBl06} as:

\vspace{-2mm}
\begin{eqnarray}\label{eq:CD}
A (N,i) = \binom{N-i-1}{i-1} + \binom{N-i}{i}, \hspace{2mm} 1 \leq i \leq M.
\end{eqnarray}



Hybrid disk arrays combine mirroring with parity coding.
LSI RAID has disks holding the XOR of neighboring disks according to Wilner 2001 \cite{Wiln01}.
\vspace{-1mm}
$$D_A, (D_A \oplus D_B), D_B, (D_B \oplus D_C), D_C, (D_C \oplus D_D), D_D, (D_D \oplus D_A)$$ 
Three consecutive disk failures can be tolerated as long as the middle disk is not a data disk.

SSPiRAL is a generalization to three disks participating in parity computation Amer et al. 2008 \cite{ASPL08} 
Given four data disk $D_A, D_B, D_C, D_D$ the four parities are computed as follows:

$$(D_A \oplus D_B \oplus D_C), \hspace{3mm}
(D_B \oplus D_C \oplus D_D), \hspace{3mm}
(D_C \oplus  D_D \oplus D_A), \hspace{3mm}
(D_D \oplus D_A \oplus D_B).$$ 

Up to four disk failures can be tolerated as long as it is not a data disk 
and the three parities it participates in.
\footnote{A minor correction to the reliability expression in SSPiRAL paper is made in Thomasian and Tang 2012 \cite{ThTa12}.}


Duplexed disks in Tandem computers came with two cross-connected processors, 
instead of processor disk pairs.
Given that $R_c$ and $R_d$ are the CPU and disk reliability, 
the former configuration has a higher reliability.
\vspace{-1mm}
$$R_{cross-connected} = 1 - (1- R_c R_d)^2 \approx 2 R_c R_d - 4 R_c^2 R_d^2,
\hspace{5mm} R_{dedicated} = [1- (1-R_c)^2][1-(1-R_d)^2] \approx 4 R_c R_d - 2 R_c R_d^2 - 2 R_d R^2_cd$$

\subsection{AutoRAID Multilevel RAID}\label{sec:autoRAID}
\vspace{2mm}

AutoRAID is a two level memory hierarchy with RAID1 acting as a cache holding hot data 
and cold data is held in a RAID5/LSA array Wilkes et al. 1996 \cite{WGSS96}.
Data is initially written onto RAID1 and as RAID1 disks fill their contents are moved to RAID5/LSA.

AutoRAID assumes that the working set for RAID1's active data changes slowly.
A trace-driven analysis of storage working set sizes was conducted by Ruemmler and Wilkes 1993 \cite{RuWi93}.
{\it Relocation Blocks - RBs} are promoted to RAID1 when they are updated.   

\subsection{Shortcut Method to Compare the Reliability of Mirrored  Disks and RAID(4+$\ell$}\label{sec:shortcut}

The shortcut method allows the comparison of the reliability of various RAID configurations,
which does not require the transient analysis or
plotting disk reliability $R(t)$ versus $t$ Thomasian 2006 \cite{Thom06}. 
This is accomplished by expressing disk reliability as $r = 1 - \varepsilon$,
e.g., for MTTF=$10^6$ hours or 114 years, 
a good approximation for relatively short time $(t_s)$ 
\vspace{-1mm}
$$R(t_s) = e^{-\delta t_s} \approx 1 - \delta t_s 
\mbox{ after three years }R(3)=1 -3/114=0.975, \mbox{ hence :} \varepsilon=0.025.$$

RAID with $n$-way replication fails with $n$ disk failures and 
this is indicated by the power $n$ in the reliability expression.
In other words the number of failures tolerated is $n-1$ (we use $\bar{r}=1-r$).
\vspace{-1mm}
$$ R_{n-way} = 1  - (1-r)^n = 1 - \bar{r}^n = 1- \varepsilon^n $$

The expressions for the RAID1 with BM configurations is as follows:

\vspace{-2mm}
\begin{eqnarray}\label{eq:BM2}
R_{BM} \approx r^N + N r^{N-1} \bar{r} + \binom{N}{2} r^{N-2} (\bar{r})^2 \approx 1 - 0.5 N \varepsilon^2.
\end{eqnarray}

For GRD the reliability expression can be written directly,
since it is as if we have two logical disks each comprising $M=N/2$ disks.

\vspace{-2mm}
\begin{eqnarray}\label{eq:GRD2}
R_{GRD} = 2 r^{N/2} - r^N  \approx 1 -  \frac{1}{4} N^2 \varepsilon^2.
\end{eqnarray}

\vspace{-2mm}
\begin{eqnarray}\label{eq:ID2}
R_{ID} \approx
r^N +  c (\frac{N}{c}) r^{N-1} \bar{r} + \binom{c}{2} (\frac{N}{c})^2 r^{N-2} (\bar{r})^2
= 1 - \frac{N}{2} ( \frac{N}{c} -1 ) \varepsilon^2.
\end{eqnarray}

We only need the first three terms in the reliability expression to obtain $R_{CD}$
With two disk failures there are N configurations leading to data loss:
\vspace{-1mm}
$$A(N,2) \approx \binom{N}{2} - N = \frac{1}{2} N (N-3)$$

\vspace{-2mm}
\begin{eqnarray}\label{eq:CD2}
R_{CD} \approx r^N + N r^{N-1} \bar{r} + \frac{1}{2} N(N-3) r^{N-2} (\bar{r})^2 = 1- N \varepsilon^2 .
\end{eqnarray}

We denote the reliability of RAID5 and RAID(4+$\ell$) as $R_\ell$.

\vspace{-1mm}
$$R_5 = r^N + N \bar{r} r^{N - 1}  = 
(1-\varepsilon)^N + N \varepsilon (1-\varepsilon)^{N-1} \approx 1- \frac{1}{2} 5N(N-1) \varepsilon^2$$

\vspace{-2mm}
\begin{eqnarray}
R_\ell \approx 1 - \binom{N}{\ell+1} \varepsilon^{\ell+1} + \binom{N}{\ell+2} \varepsilon^{\ell+2} - \ldots
\end{eqnarray}

A summary of shortcut reliability analysis results in \cite{ThTa12} are given in Table \ref{tab:MTTDL}.

\begin{table}[h]
\centering
 \begin{tabular}{|c|c|c|c|c|c|c|c|c|}\hline
RAID5 & BM & CD & GRD & ID & RAID6 & LSI & RAID7 & SSP\\ \hline
& $\frac{163}{280 \delta}$                       
& $\frac{379}{840 \delta}$                       
& $\frac{3}{8 \delta }$                          
& $\frac{61}{168 \delta}$                        
& $\frac{ 73}{ 168 \delta}$                      
& $\frac{82}{105 \delta}$                       
& $\frac{533}{840 \delta}$                       
& $\frac{701}{840 \delta}$ \\ \hline              
\scriptsize {$0.268 \delta^{-1}$}              
& \scriptsize {$0.582 \delta^{-1}$}              
& \scriptsize {$0.451 \delta^{-1}$}              
& \scriptsize {$0.375 \delta^{-1}$}             
& \scriptsize {$0.363 \delta^{-1}$}              
& \scriptsize {$0.435 \delta^{-1}$}              
& \scriptsize {$0.781 \delta^{-1}$}              
& \scriptsize {$0.635 \delta^{-1}$}              
& \scriptsize {$0.8345 \delta^{-1}$} \\ \hline   
$\binom{N}{2} \varepsilon^2$                     
& $\frac{N \varepsilon^2}{2}$                    
& \scriptsize{$N\varepsilon^2$}                  
&  $\frac{N(N-1) \varepsilon^2}{4}$              
& $\frac{N(N-c) \varepsilon^2}{2c}$              
& $\binom{N}{3} \varepsilon^3$ 
& $( \binom{N}{3} - \frac{N}{2})$\scriptsize{$\varepsilon^3$}     
& $\binom{N}{4} \varepsilon^4$                  
& $ \frac{1}{5}  \binom{N}{4} \varepsilon^4$  \\ \hline 
\end{tabular}
\caption{\label{tab:MTTDL}MTTDLs as a ratio and a fraction of the MTTF ($\delta^{-1}$)
and the first term in asymptotic reliability expression with $\varepsilon$
denoting the unreliability of a single disk and 
the power is the minimum number of disk failures leading to data loss.}
\end{table}

RAID6 results apply to EVENODD, X-code, and RDP, all three of which are 2DFT. 

Hybrid disk arrays which combine replication and parity coding provide 
a higher reliability at the same redundancy level.
RAID1 provides 2-way, LSI RAID 3-way, and SSPiRAL 4-way replication.
The update penalty is proportional to the number of ways.
In the case of hybrid arrays in addition to a write a RMW is required.

\subsection{Reliability Analysis of Multilevel RAID Arrays}\label{sec:multilevel}
\vspace{2mm}

Two multilevel RAID disk arrays are considered in this section:
mirrored RAID5 (RAID1/5) and RAID5 consisting of mirrored disks (RAID5/1).
RAID1/0 and RAID0/1 was discussed in the previous section.
Note that the number of disks is the same in both cases,
but one configuration provides higher reliability than the other
and this is verified with the shortcut reliability analysis method in Thomasian 2006 \cite{Thom06}.

\subsection{Reliability of Mirrored RAID5 - RAID1/5 Reliability}\label{sec:RAID15}
\vspace{2mm}

Consider mirroring of RAID5 arrays and no repair via mirroring, from one side to the other.
If more than one disk on each side fails, that side is considered failed although the remaining disks are intact. 
\vspace{-1mm}
$$R_{RAID1/5}= 2 R_5(t) - R_5^2 (t)$$

Substituting $R_{RAID5}(t) $ given by Eq. (ref{eq:RAID5t}):

\vspace{-2mm}
\begin{eqnarray}
R_{RAID1/5} = 
2 [ \frac{\zeta e^{\eta t} - \eta e^{\zeta t}}{\zeta - \eta} ]   
- [\frac{\zeta e^{\eta t} - \eta e^{\zeta t}}{\zeta - \eta}]^2 =  \\
\nonumber
2 \frac{\zeta e^{\eta t} - \eta e^{\zeta t}}{\zeta - \eta}  
- \frac{ \zeta^2 e^{2 \eta t} + \eta^2 e^{2 \zeta t} -2 \eta \zeta e^{(\eta + \zeta) t }} 
{ (\zeta - \eta)^2 } 
\end{eqnarray}

\vspace{-2mm}
\begin{eqnarray}
MTTDL_{mirrored/RAID5}= 
\frac{1}{\zeta-\eta}
[ - 2 \frac{\zeta}{\eta} + \frac{\eta}{\zeta}]          \nonumber
- \frac{1}{(\zeta - \eta)^2}                              
[ - 2 \frac{\zeta^2}{ \eta} + 2 \frac{\eta^2}{\zeta} + 2 \frac{\eta\zeta}{\eta+\zeta} ]
\end{eqnarray}

RAID5 reliability can be expressed as a single component 
whose MTTF equals the MTTDL as was given by Eq. (\ref{eq:appr}).
\vspace{-1mm}
$$R_{RAID15}^{appr} (t) = 2 R_5^{appr} (t) - [ R_5^{appr} (t) ]^2 $$ 
\vspace{-1mm}
$$MTTDL_{RAID15}^{appr}=  \frac{1.5 \mu }{N(N+1) \delta^2} = \frac{1.5 MTTF^2}{N(N+1)MTTR}$$


\subsection{Reliability of RAID5 Consisting of Mirrored Disks - RAID5/1}\label{sec:RAID51}
\vspace{2mm}

An approximate expression for mirrored RAID5 MTTDL was given in Xin et al. 2003 \cite{Xin+03}.
The contents of a failed disk are recovered by its mirror, but otherwise by invoking the RAID5 paradigm
\vspace{-1mm}
$$MTTDL_{RAID5/1} \approx \frac{\mu^3}{4N(N-1)\delta^4} = \frac{MTTF^4}{4N(N-1)MTTR^3}$$

A rigorous method to derive RAID5/1 MTTDL using the shortest path reliability model 
is presented in Iliadis and Venkatesan 2015b \cite{IlVe15b}.
The state tuples $(x, y, z)$ indicate that there are x pairs with both devices in operation, 
$y$ pairs with one device in operation and one device failed, and $z$ pairs with both devices failed.
With $D$ devices on each side and failure rate $\delta$ and repair rate $\mu$
we have the following state transitions leading to {\it Data Loss - DL}. 

\vspace{-2mm}
\begin{align}\nonumber
(D,0,0) &\xrightarrow{2 D \delta} (D-1,1.0), \hspace{5mm} \xleftarrow{\mu} (D-1,1,0) \\
\nonumber
(D-1,1,0) &\xrightarrow{\delta} (D-1,0,1), \hspace{5mm} (D-1,0,1) \xleftarrow{2\mu} (D-1,1,0) \\
\nonumber
(D-1,1,0) &\xrightarrow{2(D-1)\delta} (D-2,2.0), \hspace{5mm} (D-2,2,01) \xleftarrow{2\mu} (D-1,1,0) \\
\nonumber
(D-1,0,1) &\xrightarrow{2(D-1) \delta} (D-2,1,1), \hspace{5mm} (D-2,1,1) \xleftarrow{\mu} (D-1,0,10) \\
\nonumber 
(D-2,2,0) &\xrightarrow{2\delta} (D_2,1,1), \hspace{5mm}  (D-2,1,1) \xleftarrow{\mu} (D-2,2,0) \\
\nonumber 
D_2,1,1) &\xrightarrow{\delta} \mbox{DL} 
\end{align}

Taking into account that $\delta \ll \mu$ we get 
the following end-to-end probabilities for the upper and lower paths.
\vspace{-1mm}
$$P_u \approx \frac{\delta}{\mu} \times \frac{2(D-1)\delta}{2 \mu} \times \frac{\delta}{2 \mu} = 
\frac{(D-1) \delta^3 } {2 \mu^3} $$
\vspace{-1mm}
$$P_{ell} \approx \frac{2(D-1)\delta}{\mu} \times \frac{\delta}{\mu} \times \frac{\delta}{2 \mu}=
\frac{(D-1) \delta^3) }{\mu^3}$$
\vspace{-1mm}
$$P_{DL} =  P_u + P_\ell \approx \frac{3 (D-1) \delta^3}{2 \mu^3}$$

It is argued that the MTTDL is a product of two first device failures 
and the expected number of first device failure events:

Given that $MTTDL \approx E[T] / P_{DL} $, where $E[T] = 1/(N\delta)$.
It follows that $ MTTDL \approx 1 / (N \delta P_{DL})$ and noting that $N=2D$ we obtain:

\vspace{-2mm}
\begin{eqnarray}
MTTDL^{(approx)}_{RAID(5/1)} \approx \frac{\mu^3}{3D(D-1)\delta^4} = \frac{MTTF^4}{D(D-1)MTTR^3} 
\end{eqnarray}

\begin{framed}
\subsection{Shortcut Reliability Analysis to compare RAID1/5 and RAID5/1}\label{sec:shortcut2}
\vspace{2mm}

That RAID5/1 is more reliable than RAID1/5 as shown 
by the shortcut reliability analysis in Thomasian 2006 \cite{Thom06}.
We consider RAID with $N$ disks and denote disk reliability with $r$.

\vspace{-2mm}
\begin{eqnarray}\label{eq:15}
R_{RAID1/5}= 1 - [1- R_{RAID5}]^2 = 1- [R^N + N r^{N-1} (\bar{r})]^2 .
\end{eqnarray}

The reliability of RAID5 with mirrored disks can be expressed as:

\vspace{-2mm}
\begin{eqnarray}\label{eq:16}
R_{RAID5/1} = N R_{RAID1}^{N-1} -(N-1)R_{RAID}^N = N[1 -(1-r)^2]^{N-1} - (N-1) [ 1 -(\bar{r})^2]^N .
\end{eqnarray}

For $N=3$ it easily follows that $RAID_{5/1} > R_{RAID1/5}$, since:
$R_{RAID5/1}- R_{RAID1/5}  = 6r^2 (\bar{r})^4 > 0$.
Rather than attempting to show this for arbitrary $N$
we use the shortcut method setting $r=1-\varepsilon$ in the above equations and 
retaining only the lowest power of $\varepsilon$ we have:

\vspace{-2mm}
\begin{eqnarray}
\nonumber
R_{RAID1/5} \approx 1 -\frac{1}{4} N^2 (N-1)\varepsilon^4 , \hspace{5mm}
R_{RAID5/1} \approx 1- \frac{1}{2} N(N-1)\varepsilon^4. 
\end{eqnarray}
It is easy to see that $R_{RAID5/1} > R_{RAID1/5}$. 

\end{framed}
\subsection{Reliability and Performance Comparison of Two Heterogeneous Disk Arrays}\label{sec:HDArel}
\vspace{2mm}

We next utilize the asymptotic reliability analysis 
to compare the reliability of two HDA configurations described in Section \ref{sec:HDA}.
The total number of disks is $N=8$.

\begin{description}

\item[${\cal C}_1$: Vertical Disk Partitioning:]
$n=2$ disks are dedicated to RAID1 and $N-n=6$ disks to RAID5.

\item
[${\cal C}_2$: Horizontal Disk Partitioning:]
This is the HDA paradigm.
RAID1 is allocated over $N=8$ disks, occupying $2/N$ or 25\% of the space on each disk.
The remaining space on $N$ disks is more than adequate
to hold the RAID5 array allocated on six disks in ${\cal C}_1$,
since the wider RAID5 array allocates less space to parity.
\footnote{RAID5 data (not parity) blocks are distributed over seven disks,
utilizing 5/7 or 72\% of disk capacity, i.e., less than 75\% of available space.}

\end{description}

The reliability of RAID1/0 with $p$ disk pairs is $R_{R1}(p) = [1- (1-r)^2]^p$,
while the reliability of a RAID5 with $w$ disks is
$R_{R5}(w) = r^w + w (1-r) r^{w-1}$.
We consider highly reliable disks, so that $r=1-\epsilon$ with $\epsilon \ll 1$.

\[ R_{{\cal C}_1} =
[ 1- (1-r)^2] \times [r^6+6r^5(1-r)] \approx 1 -16\epsilon^2. \]

\[ R_{{\cal C}_2} =
\{ [1-(1-r)^2]^4 \} \times [r^8 +8r^7 (1-r)]
\approx 1-32\epsilon^2. \]
It follows that the ${\cal C}_1$ configuration is more reliable than ${\cal C}_2$,
since it tolerates two disk failures as long as long as they do not affect the same array: RAID1 or RAID5.
This reliability analysis is extensible to more complicated cases.

In evaluating HDA performance we postulate an M/M/1 queueing system \cite{Klei75}
with Poisson arrivals and exponential service times,
which has been utilized in RAID5 performance studies \cite{Meno94}.
Given an arrival rate $\lambda$ and mean disk service time $\overline{x}_d$,
the mean disk response time is $R = \overline{x}_d / (1- \rho)$,
where $\rho = \lambda \overline{x}_d$ is the disk utilization factor.

The arrival rate of read requests to RAID1 and RAID5 arrays is
$\Lambda_{R1}$ and $\Lambda_{R5}$, respectively.
The mean response time for RAID1 with ${\cal C}_1$ is:
$R_{R1} = \overline{x}_d / (1- \rho)$,
where the disk utilization is $\rho = (\Lambda_1 /2 ) \overline{x}_d$.
The mean response time for RAID5 with ${\cal C}_1$ is $R_{R5} = \overline{x}_d/(1-\rho)$,
where $\rho = (\Lambda_{R5} / 6) \overline{x}_{disk}$,

For ${\cal C}_2$ there are two components to disk utilization,
since RAID1 and RAID5 disk arrays share disk space:
$\rho' = [(\Lambda_{R1} + \Lambda_{R5}) / N] \overline{x}_d$,
so that ${R'}_{R1}  = {R'}_{R5} = \overline{x}_d / (1- \rho')$.
If $ \rho' < \rho $ or $\Lambda_{R5} < (N/2-1)\Lambda_{R1}$ then $ {R'}_{R1} < R_{R1}$.
For $N=8$ ${\cal C}_2$ will improve the RAID1 response time
with respect to ${\cal C}_1$ for $\Lambda_{R1} > \Lambda_{R5} / 3$.

RAID1 response times in ${\cal C}_2$ can be improved by processing
its accesses at a higher priority than RAID5 accesses.
${R''}_{R1} = \overline{x}_d / (1- \rho'')$,
where $\rho''= (\Lambda_{R1} / N) \overline{x}_d$,
since only the disk utilization due to RAID1 accesses affects ${R''}_{R1}$ \cite{Klei75}.
For ${\cal C}_1$ with $\rho''= (\Lambda_{R1}/ 2) \overline{x}_{disk} = 0.8$,
$R_{R1} = \overline{x}_d / (1-0.8) = 5 \overline{x}_d$
and for ${\cal C}_2$ with ${\rho''} = n \times 0.8 / N = 0.2$ and
${R''}_{R1} = \overline{x}_d / (1-0.2) = 1.25 \overline{x}_d$, i.e., a 4-fold improvement.

\clearpage
\subsection*{Abbreviations} 
\vspace{2mm}

\begin{small}
{\bf AFR} - Annual Failure Rate,
{\bf BER} - Bit Error Rate,
{\bf BIBD} - Balanced Incomplete Block Design,
{\bf CTMC} - Continuous Time Markov Chain,
{\bf DAC} - Disk Array Controller,
{\bf EAFDL} - Expected Annual Fraction of Data Loss,
{\bf ECC} - Error Correcting Code,
{\bf FB} - Fixed Block (disks),
{\bf HDFS} - Hadoop Distributed File System,
{\bf IOPS} -Input/Output Per Second,
{\bf IDR} - Intra-Disk Redundancy,
{\bf kDFT} - k Disk-Failure Tolerant,
{\bf LRC} - Local Redundancy Code.
{\bf LSE} - Latent Sector Error,
{\bf LST} - Laplace Stieltjes Transform,
{\bf MDS} - Maximum Distance Separable,
{\bf ML} - Machine Learning,
{\bf MTTDL} - Mean Time To Data Loss,
{\bf MTTF} - Mean Time To Failure,
{\bf MTTR} - Mean Tine To Repair, 
{\bf NRP} - Nearly Random Permutation,
{\bf NVMeoF} - NVMe\textregistered over Fabrics (NVMe-oF\texttrademark ),
{\bf NVRAM} - Non-Volatile Random Access Memory,
{\bf OLTP} - OnLine Transaction Processing,
{\bf PUE} - Power Usage Effectiveness,
{\bf RAID} - Redundant Array of Inexpensive/Independent Disks, 
{\bf RDP} - Rotated Diagonal Parity,
{\bf RMW} - Read-Modify Write,
{\bf RU} - Rebuild Unit,
{\bf SCM} - Storage Class Memory,
{\bf SDC} - Silent Data Corruption,
{\bf SLA/O} - Service Level Agreement/Objective.
{\bf SSD} - Solid State Disk,
{\bf SU} - Stripe Unit,
{\bf SWP} - Small Write Penalty,
{\it TCO} - Total Cost of Ownership,
{\bf UDE} - Undetected Disk Errors,
{\bf UPS} - Uninterruptible Power Supply,
{\bf VSM} - Vacationing Server Model, 
{\bf XOR} - eXclusive OR,
{\bf ZBR} - Zoned Bit Recording,
{\bf ZLA} - Zero Latency Access.
\end{small}

\end{document}